\documentclass[aps,pre,groupedaddress,longbibliography]{revtex4-1}

\usepackage{graphicx}
\usepackage{bm}
\usepackage{amsmath}
\usepackage{amssymb}
\usepackage{natbib}
\usepackage[export]{adjustbox}
\usepackage{array}

\usepackage{latexsym} 
\usepackage{mathtools} 
\usepackage{multirow} 
\usepackage{mathrsfs} 

\usepackage{color}

\newcommand{\BEQ}{\begin{equation}}
\newcommand{\EEQ}{\end{equation}}
\newcommand{\BER}{\begin{eqnarray}}
\newcommand{\EER}{\end{eqnarray}}
\newcommand{\BEQN}{\begin{equation*}}
\newcommand{\EEQN}{\end{equation*}}
\newcommand{\BERN}{\begin{eqnarray*}}
\newcommand{\EERN}{\end{eqnarray*}}

\newcommand{\vect}[1]{\mathbf{#1}}

\newcommand\oned{one-di\-men\-sion\-al}

\newcommand\twod{two-di\-men\-sion\-al}
\newcommand\twody{two-di\-men\-sion\-al\-ity}

\newcommand\threed{three-di\-men\-sion\-al}
\newcommand\lsa{lin\-ear sta\-bi\-li\-ty analy\-sis}

\newcommand\qtwod{qua\-si-two-di\-men\-sion\-al}

\newcommand\mhd{MHD}

\newcommand\RB{Rayleigh--B\'{e}nard}
\newcommand\PRB{Poiseuille--\RB}

\newcommand\Poi{Poiseuille}

\newcommand\TollSchl{Tollmien--Schlichting}

\newcommand\ie{i.e.}
\newcommand\eg{e.g.}

\newcommand\Rey{\textit{Re}}
\newcommand\Ray{\textit{Ra}}
\newcommand\Ha{\textit{Ha}}
\newcommand\HH{\textit{H}}
\newcommand\Prn{\textit{Pr}}

\newcommand\ci{\mathrm{i}}
\newcommand\ub{\bar{u}}
\newcommand\tb{{\bar{\theta}}}  
\newcommand\up{\tilde{{u}}} 
\newcommand\vp{\tilde{{v}}} 
\newcommand\tp{\tilde{{\theta}}} 
\newcommand\pp{\tilde{p}}
\newcommand\fp{\tilde{f}}

\newcommand\order{\mathscr{O}}

\newcommand{\ignore}[1]{}

\newcolumntype{M}{m{\dimexpr.5\linewidth-2\tabcolsep}} 

\begin{document}


\title{Linear stability of horizontal, laminar fully developed, quasi-two-dimensional liquid metal duct flow under a transverse magnetic field and heated from below}



\author{Tony Vo}
\affiliation{The Sheard Lab, Department of Mechanical and Aerospace Engineering, Monash University,
Victoria 3800, Australia}

\author{Alban Poth\'{e}rat}
\affiliation{Applied Mathematics Research Centre, Coventry University,
Priory Street, Coventry CV15FB, UK}

\author{Gregory J. Sheard}
\email[]{Electronic mail: greg.sheard@monash.edu}
\affiliation{The Sheard Lab, Department of Mechanical and Aerospace Engineering, Monash University,
Victoria 3800, Australia}


\date{\today}

\begin{abstract}
This study considers the linear stability of \PRB\ flows, subjected to a transverse magnetic field to understand the instabilities that arise from the complex interaction between the effects of shear, thermal stratification and magnetic damping. This fundamental study is motivated in part by the desire to enhance heat transfer in the blanket ducts of nuclear fusion reactors. In pure \mhd\ flows, the imposed transverse magnetic field causes the flow to become \qtwod\ and exhibit disturbances that are localised to the horizontal walls. However, the vertical temperature stratification in \RB\ flows feature convection cells that occupy the interior region and therefore the addition of this aspect provides an interesting point for investigation.

The linearised governing equations are described by the \qtwod\ model proposed by \citet{sommeria1982mhd} which incorporates a Hartmann friction term, and the base flows are considered fully developed and \oned. The neutral stability curves for critical Reynolds and Rayleigh numbers, $\Rey_c$ and $\Ray_c$, respectively, as functions of Hartmann friction parameter $\HH$ have been obtained over $10^{-2}\leq\HH\leq10^4$. Asymptotic trends are observed as $\HH\rightarrow\infty$ following $\Rey_c\propto\HH^{\,1/2}$ and $\Ray_c\propto\HH$. The linear stability analysis reveals multiple instabilities which alter the flow both within the Shercliff boundary layers and the interior flow, with structures consistent with features from plane \Poi\ and \RB\ flows.
\end{abstract}


\maketitle 

\section{Introduction}
The liquid lithium flowing within the ductwork of proposed tritium breeder modules of magnetic confinement fusion reactors present an exciting example of confined duct flows combining thermal and velocity shear destabilisation processes with mag\-ne\-to\-hy\-dro\-dy\-nam\-ic (\mhd) effects. In this application, the strong magnetic field is transverse to the duct and two-dimensionalises the flow, while both MHD and viscous effects act to damp fluctuations, inhibiting heat transport. While configurations for liquid metal blankets in fusion reactors considered in recent times use poloidal ducts (\eg\ \cite{abdou2015blanket}) where flows are predominantly vertical, the present study investigates the fundamental problem of the stability of horizontal \qtwod\ MHD duct flow with vertical heating - thus serving as an extension of the classical \PRB\ flow to \qtwod\ flows under the model of \citet{sommeria1982mhd}. Understanding the stability of these flows underpins endeavours to enhance heat transfer in these duct flows via convective mixing.

The primary non-dimensional parameters that characterise \PRB\ flows are the Reynolds number $\Rey$, Rayleigh number $\Ray$, and Prandtl number $\Prn$. These parameters respectively characterise the ratios of inertial to viscous forces, buoyancy to viscous and thermal forces, and momentum diffusivity to thermal diffusivity. It is well known that \RB\ flow bounded by no-slip horizontal boundaries (rigid-rigid) develops convection rolls at critical Rayleigh number $\Ray_c=1708$ (with the depth of the fluid as the characteristic length scale) \citep{drazin2004hydrodynamic}. Due to the infinite extent of the system, the convection rolls have no directional preference. However, an imposed through-flow results in the selection of longitudinal rolls as opposed to transverse rolls at onset. This result is supported by \citet{gage1968stability} who performed \lsa\ on a thermally stratified \Poi\ flow and has been observed experimentally in large aspect ratio configurations (\eg\ \cite{yasuo1966forced,akiyama1971experiments,fukui1983longitudinal}). The stability of \PRB\ flows has been studied extensively, uncovering a rich variety of thermoconvective instabilities including longitudinal and transverse rolls, mixed rolls, wavy rolls and oscillating rolls (\eg\ \cite{muller1992effect,carriere1999convective,nicolas2000linear,grandjean2009experimental,mergui2011sidewall,nicolas2012influence}).

Although longitudinal rolls are dominant in \PRB\ flows in infinite aspect ratio domains, it is possible for transverse rolls to be more unstable than longitudinal rolls in finite aspect ratio channels \citep{luijkx1981existence,nicolas2000linear}. This is relevant to applications involving a strong transverse magnetic field due to the suppression of longitudinal structures. The liquid metal flow in rectangular blanket ducts surrounding the plasma can therefore be considered as an extension of \PRB\ flow via the incorporation of \mhd\ effects. Such flows have been motivated by the need to demonstrate the viability of nuclear fusion as an energy source. Specifically, the stability (\eg\ \cite{potherat2007quasi,fakhfakh2010selective,krasnov2012numerical,priede2012linear}) and heat transfer properties of a liquid metal flow is of great importance to the future designs of experimental fusion reactors where maximising heat transfer is a key criterion. The enhancement of heat transfer and instability growth in \mhd\ flows can be achieved by implementing turbulent promoters such as bluff bodies \citep{yoon2004numerical,hussam2011dynamics,cassells2016heat} and current injection \citep{hamid2016combining}. However, physical modifiers are not always practicable and therefore the characterisation of flow instabilities is required. Although the stability results of the horizontal duct presented herein are limited in their relevance to recent poloidal liquid metal blanket duct designs (\eg\ \cite{abdou2015blanket}), they provide fundamental insights into the stability of a flow influenced by thermal and shear effects.

In \threed\ \mhd\ duct flows both the Hartmann and Shercliff boundary layers play a significant role in determining the stability of the flow. These layers form on the walls perpendicular and parallel to the magnetic field, respectively. However, the study of a \threed\ domain is computationally expensive. The system can be simplified to two-dimensions by assuming that the imposed magnetic field is sufficiently strong and therefore described adequately using a \qtwod\ model developed by \citet{sommeria1982mhd} (referred to as the SM82 model hereafter). This model considers solutions in the plane perpendicular to the magnetic field. That is, only the Shercliff layers are resolved while the frictional effects of the Hartmann layers in the out-of-plane direction are integrated across the depth of the duct and modelled through an added Hartmann friction term.

The stability of a \mhd\ duct flow to \qtwod\ perturbations under a transverse magnetic field, without consideration of thermal stratification, has been investigated previously by \citet{potherat2007quasi}. It was found that the critical modes of linear instability are \TollSchl\ waves. An asymptotic regime was observed for $\HH\gtrsim200$ with neutral stability curves described by critical Reynolds number and streamwise wavenumber, $\Rey_c=4.83504\times10^4\HH^{\,1/2}$ and $k_c=0.161532\HH^{\,1/2}$, respectively. The same study also investigated the stability of the system through an energy analysis. A lower threshold of stability defined by $\Rey_c=65.3288\HH^{\,1/2}$ with corresponding $k_c=0.863484\HH^{\,1/2}$ was established. Thus, according to both analyses the stability of the system in the limit of $\HH\rightarrow\infty$ is determined only by the thickness of the Shercliff layer, which scales according to $\delta_{S} \propto 1/\HH^{\,1/2}$. That is, the stability of \qtwod\ \mhd\ flows are governed solely by the Shercliff layers.

Besides advantages in terms of simplicity and computational costs, the physical representation based on the two-dimensional Navier-Stokes equation with linear friction extends the relevance of our analysis to a number of other physical problems described by this equation: rapidly rotating flows \citep{fruh2003origin,vo2015non}, plane flows with parabolic profiles subjected to Rayleigh friction \citep{duran2010dynamics}. In the case of MHD flows, high friction parameters are relevant to ducts under strong magnetic fields (as in fusion applications), whereas low friction parameters could be experimentally achieved in thin quasi-two dimensional layers of fluids (such as Hele-Shaw cells submitted to high spanwise magnetic field, in the spirit of \citet{krasnov2008optimal}).

\citet{zikanov2013natural} studied \mhd\ flow in a pipe with the lower half of the wall heated and subjected to a transverse magnetic field. Linear stability analysis and direct numerical simulations confirm the existence of convection structures at high $\HH$. This verifies experimental observations of temperature fluctuations being suppressed as $\HH$ is increased, but re-emerging as $\HH$ is further increased due to secondary flows \citep{genin2011temperature,belyaev2015temperature}. These results are not limited to the pipe geometry as they have also been detected in duct flows (\eg\ \cite{vetcha2013study,zhang2014mixed}). Recently, \citet{zhang2014mixed} used direct numerical simulations to observe two types of secondary flows in horizontal duct flows; one which is dominated by \qtwod\ spanwise rolls and another which characterised by a combination of streamwise and spawnwise rolls.

The focus of this paper is on the stability of flows through electrically insulated ducts subjected to a transverse magnetic field and a vertical heating gradient, as modelled by the SM82 model with a Boussinesq approximation. Linear stability analysis on the basic velocity and temperature solutions over a large range of $\Rey$, $\Ray$ and $\HH$ are reported. Thus, this may be considered an extension of the research of \citet{potherat2007quasi} through the introduction of natural convection effects, and as an extension of classical \PRB\ instability via incorporation of \mhd\ effects through the SM82 model. The remaining sections of this paper are organised as follows. The methodology is presented in Sec.~\ref{sec:methodology} which includes a description of the system, the governing equations and parameters, the \lsa\ solver and its validation. Results are discussed in Sec.~\ref{sec:results} with attention to neutral stability curves for various fixed $\Rey$ flows and fixed $\Ray$ flows separately. Lastly, the key conclusions of this study are outlined in Sec.~\ref{sec:conclusions}.

\section{Methodology}\label{sec:methodology}
\subsection{Problem formulation}

\begin{figure}
  \begin{center}
    \begin{tabular}{c}
        \includegraphics[width=0.6\columnwidth]{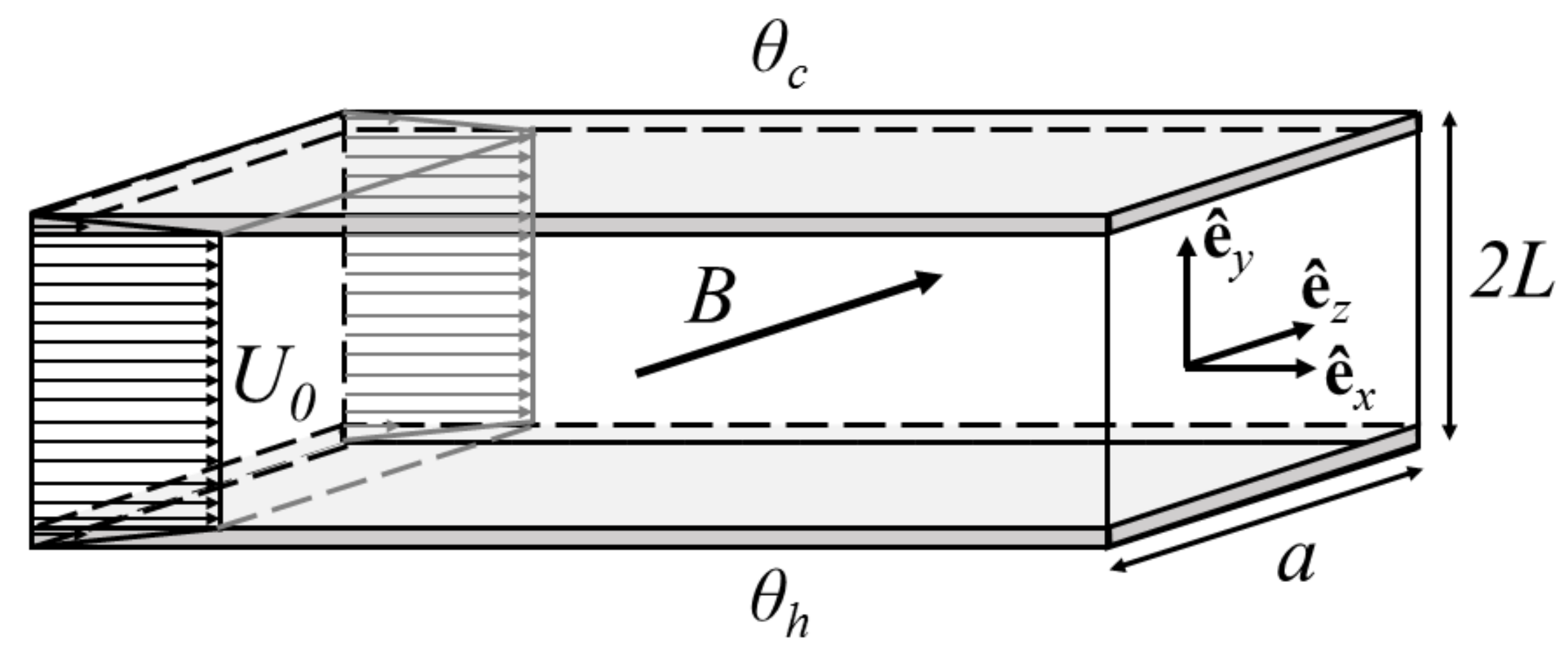}
    \end{tabular}
        \caption{A schematic diagram of the system under investigation. The key dimensions are the duct height $2L$, duct width $a$, magnetic field $B$ and the maximum velocity $U_0$. The variables ($\mathbf{\hat{e}}_x, \mathbf{\hat{e}}_y, \mathbf{\hat{e}}_z$) represent unit vectors in the ($x,y,z$) directions, respectively. The liquid metal flows in the positive-$x$ direction while the magnetic field acts in the $z$ direction. Temperatures $\theta_h$ (hot) and $\theta_c$ (cold) are imposed on the bottom and top duct walls, respectively. Shaded regions represent the Shercliff boundary layers.}\label{fig:schematic}
  \end{center}
\end{figure}

The system studied in this paper represents an electrically conducting fluid with kinematic viscosity $\nu$, thermal diffusivity $\kappa$, volumetric expansion coefficient $\alpha$, density $\rho$, and electrical conductivity $\xi$, flowing through a horizontal rectangular duct of height $2L$ and width $a$, exposed to a transverse magnetic field of strength $B$ and a vertical temperature gradient. The schematic of the system is shown in figure~\ref{fig:schematic}. The duct walls are electrically insulated. Provided that the imposed transverse magnetic field is sufficiently strong relative to the through-flow, the flow solution can be described accurately by a \qtwod\ model \citep{sommeria1982mhd}. This is a result of the magnetic field suppressing motions and gradients parallel to the magnetic field in the interior of the flow. It should be noted that the Hartmann boundary layers that develop along the out-of-plane duct walls can become unstable despite the effects of a strong transverse magnetic field. Linear stability analyses have revealed that the stability of Hartmann layers can be described by a Reynolds number based on the Hartmann layer thickness (\ie\ $\Rey_\Ha=\Rey/\Ha$). Numerous different critical $\Rey_{\Ha}$ have been obtained in previous studies with values ranging from $\Rey_{\Ha}=48311.016$ \citep{takashima1996stability}, to $\Rey_{\Ha}=50000$ \citep{lock1955stability}. However, destabilisation of the Hartmann layer is known to be subcritical and transition to turbulence in fact takes place at $\Rey_\Ha\simeq380$ \citep{moresco2004experimental}. This study assumes that the Hartmann layers are laminar and therefore can be described by the SM82 model. This model is particularly well suited to linear stability analyses as it achieves maximum precision in the limit of vanishing inertia \citep{potherat2005numerical}. Derivation and details of the SM82 model can be found in \citet{sommeria1982mhd}.

\subsection{Governing equations and parameters}\label{subsec:governing}
The SM82 equations are coupled with a thermal transport equation through a Boussinesq approximation to describe the \mhd\ duct flow with vertical thermal stratification (\ie\ $\Rey>0$, $\Ray>0$). These equations are given in the non-dimensional form as
\begin{subequations}\label{eq:nondimensional_SM82}
\begin{align}
\dfrac{\partial \mathbf{u}}{\partial t} + (\mathbf{u} \cdot \nabla)\mathbf{u} & = -\nabla p + \dfrac{1}{\Rey}{\nabla}^2\mathbf{u} - \dfrac{\HH}{\Rey}\mathbf{u}+\dfrac{\Ray}{\Prn\Rey^2}\theta\mathbf{\hat{e}}_y,\\
\dfrac{\partial \mathbf{\theta}}{\partial t} +(\mathbf{u} \cdot \nabla)\mathbf{\theta} & = \dfrac{1}{\Prn\Rey}{\nabla}^2\mathbf{\theta}\\
\nabla \cdot \mathbf{u} & = 0,
\end{align}
\end{subequations}
where $\mathbf{u}$ is the velocity vector, $t$ is time, $p$ is pressure, and $\mathbf{g}$ is the gravitational acceleration acting in the negative $y$ direction. These equations are obtained by normalising lengths by $L$, velocity by $U_0$ the maximum velocity of the base flow profile, time by $L/U_0$, pressure by $\rho U_0^2$ and temperature by $\Delta\theta$. Here, $\mathbf{\hat{e}}_y$ is a unit vector in positive $y$ direction. The non-dimensional parameters $\Rey$ (Reynolds number), $\Ray$ (Rayleigh number), $\HH$ (modified Hartmann number), and $\Prn$ (Prandtl number) are respectively defined as
\begin{subequations}
\begin{align}
\label{eq:Rey} \Rey &= \dfrac{U_0 L}{\nu},\\
\label{eq:Ray} \Ray &= \dfrac{\alpha gL^3\Delta\theta}{\nu\kappa},\\
\label{eq:HH}  \HH  &= n\Ha {\left(\dfrac{L}{a}\right)}^2,\\
\label{eq:Prn} \Prn &= \dfrac{\nu}{\kappa},
\end{align}
\end{subequations}
where $\Delta\theta$ is the temperature difference between the bottom (hot) and top (cold) walls ($\theta_h-\theta_c$), $n=2$ the number of Hartmann layers on out-of-plane walls imparting friction on the \qtwod\ flow, and $\Ha$ is the Hartmann number $\Ha = aB\sqrt{{\xi}/{(\rho\nu)}}$ whose square describes the relative influence of magnetic to viscous forces on the flow. The Prandtl number (ratio of momentum diffusivity to thermal diffusivity) is fixed at $\Prn=0.022$ to represent the eutectic liquid metal alloy Galinstan (GaInSn) that is used in a number of modern \mhd\ experiments \citep{morley2008gainsn}. It should be mentioned that the SM82 model is typically valid for $\Ha\gg1$, $N=\Ha^2/\Rey\gg1$ and $\Rey/\Ha\lesssim380$ \citep{sommeria1982mhd,sommeria1988electrically,krasnov2004numerical,moresco2004experimental}.

The present study includes analysis conducted at small $H$ where the SM82 model may be invalid. Generally speaking, the SM82 model describes a \twod\ incompressible fluid flow with a linear friction term. Beyond the \mhd\ applications satisfying the SM82 model, other flows may adopt the same model form where an additional forcing term describes the out-of-plane effects imparted onto the \twod\ flow. One example is the quasi-geostrophic model which incorporates a forcing term to describe the frictional effects induced by the Ekman layers (\eg\ \cite{fruh2003origin},\cite{vo2015non}) of the form $\tau_E=(a^2/\nu)E^{1/2}$, where the Ekman number $E$ represents the ratio of Coriolis to viscous forces. The linear term can also represent the Rayleigh friction in plane shallow water flows with a parabolic velocity profile between the planes, of dimensional characteristic friction time $\tau_F=4a^2/(\pi^2\nu)$. Interestingly, this case is recovered both from the quasi-geostrophic model and from the SM82 model for moderate Reynolds numbers and in the limits $\Ha\rightarrow0$ and $E\rightarrow\infty$ where the flows lose quasi their \twody\ under the effect of viscous friction. In all these cases, although the most obvious relevance of the SM82 model is to \qtwod\ flows at large $H$, it must be kept in mind that since $H$ is of the form $H=n Ha(L/a)^2$, moderate values of $H$ still satisfy $Ha\gg1$ in Hele-Shaw cell geometries with a spanwise magnetic field (\ie\ $(L/a)\gg1$). In particular, the stability properties of the SM82 model at low $H$ in \citet{potherat2007quasi} provide evidence of transient growth in channels with spanwise magnetic fields in the range of $Ha$ that is computationally difficult to reach with a \threed\ approach \citep{krasnov2008optimal}. This geometry lends itself well to experiments in large solenoidal magnets. This possibility remains to be explored and could elucidate the open problem of the transition to turbulence in MHD channel flows with spanwise magnetic fields. For these reasons the full range of values of $H$ from zero to infinity deserves to be explored, even though our prime motivation remains for regimes of high $H$ relevant to fusion.


The horizontal walls ($y=0,2L$) adopt no-slip conditions and are electrically insulated. Under these conditions, the base flow solutions for velocity $\vect{\bar{u}}=\bar{u}(y)\vect{\hat{e}}_x$ and temperature $\tb(y)$ are fully developed and are expressed as
\BEQ\label{eq:ub}\begin{array}{c}
\ub(y) =  \begin{dcases}  1-y^2 & \,\textrm{for }\,\,  \Rey>0, \HH = 0, \\
\left(\dfrac{\cosh(\sqrt{\HH})}{\cosh(\sqrt{\HH})-1}\right)\left(1-\dfrac{\cosh(\sqrt{\HH}y)}{\cosh(\sqrt{\HH})}\right) & \,\textrm{for }\,\,  \Rey>0, \HH>0,
\end{dcases}\end{array}\EEQ
\BEQ\label{eq:tb}
\tb(y) =  \dfrac{1-y}{2} \,\textrm{ for }\,\,  \Ray > 0.
\EEQ
The base velocity and temperature profiles are zero everywhere for $\Rey=0$ and $\Ray=0$, respectively. The plane \Poi\ flow and the classical \RB\ problems are recovered for $\HH=0$, $\Ray=0$ and $\HH=0$, $\Rey=0$, respectively. The governing equations and corresponding scales for these cases are presented in Appendix~\ref{appsec:governing_eqs}.

\subsection{Linear stability analysis}\label{subsec:linear}
The governing equations are linearised by decomposing velocity, temperature and pressure solutions into the mean (base flow denoted by an overbar) and fluctuating components (perturbation denoted by a prime). Due to the translational invariance of the problem in the $x$ direction, the perturbations are waves travelling in the $x$ direction:
\begin{subequations}\label{eq:modal_pert}
\begin{align}
f^\prime &= \delta\fp(y)\textrm{e}^{\ci(kx - \omega t)},
\end{align}
\end{subequations}
where $f$ is any of $u$, $v$, $p$ or $\theta$. Here, $\delta$ is taken to be a small parameter, $k$ is the streamwise wavenumber, $\omega$ is the complex eigenvalue and the tilde components ($\up,\vp,\pp,\tp$) are eigenfunctions. Substituting these expressions into equation~(\ref{eq:nondimensional_SM82}) (for $\Rey>0$, $\Ray>0$) and retaining the terms up to order $\order(\delta)$ at most yields
\begin{subequations}\label{eq:linearised}
\begin{align}
  \begin{split}
    \dfrac{1}{\Rey}\left(\mathrm{D}^2 - k^2\right)^2\vp +\ci k\ub^{\prime\prime}\vp - \ci k\ub\left(\mathrm{D}^2-k^2\right)\vp \\
    -\dfrac{\HH}{\Rey}\left(\mathrm{D}^2-k^2\right)\vp -\dfrac{\Ray}{\Rey^2\Prn}k^2\tp &= -\ci\omega\left(\mathrm{D}^2 - k^2\right)\vp,
  \end{split}\\
      -\tb^\prime\vp - \ci k\ub\tp + \dfrac{1}{\Rey\Prn}\left(\mathrm{D}^2 - k^2\right)\tp & = -\ci\omega\tp,
\end{align}
\end{subequations}
where $\mathrm{D}$ is the differentiation operator with respect to $y$. A zero Dirichlet boundary condition is imposed on the perturbation fields (directly on $\vp$ and $\tp$ as part of the eigenvalue problem, and on $\up$ and $\pp$ during its reconstruction). The linearised equations are limited to transverse rolls since longitudinal rolls are outside the scope of the SM82 model and therefore not considered in this study.

\subsection{Numerical procedure and validation}\label{subsec:numerical}
The code we use has been successfully implemented for \oned\ stability analysis of a flow driven by horizontal convection \citep{tsai2016origin} and is based on several numerical methodologies described in \citet{trefethen2000spectral}, \citet{weideman2000matlab} and \citet{schmid2001stability}, and briefly recalled this section.

The flow solutions are discretised in the $y$ direction using Chebychev collocation points. To ensure that the Shercliff layers are sufficiently resolved, testing revealed that the number of points needed to exceed $N=\textrm{max}(100,50\HH^{\,1/4})$. This criterion ensures that the  solution is grid-independent and maintains at least $10$ collocation points in each Shercliff layer, consistent with \citet{potherat2007quasi} for the range of $\HH$ investigated here. Thus, the condition of $N\geq\textrm{max}(100,50\HH^{\,1/4})$ is adopted throughout this study. For example, $N=[100,159,282,500]$ for $\HH=[0,10^2,10^3,10^4]$, respectively.

The linearised governing equations are treated as an eigenvalue problem with the eigenvectors
\begin{align}
\pmb{\phi}=
\begin{bmatrix}
\vp\\
\tp
\end{bmatrix}.
\end{align}
The corresponding instability modes are recovered using equation~(\ref{eq:modal_pert}) while the functions $\up$ and $\pp$ can be recovered from the continuity and momentum equations, respectively. The complex eigenvalues of the problem are represented by $\omega$.

The imaginary component of leading eigenvalue $\omega$ represents the growth rate $\sigma$, of the instability for a specific streamwise wavenumber $k$. A MATLAB eigenvalue solver is used to obtain the leading eigenvalues and corresponding eigenvectors for equation~(\ref{eq:linearised}) ($\Rey>0$, $\Ray>0$). The outputs from the solver were found to be more consistent and generated smaller errors when the generalised eigenvalue problem was posed in the standard form
\begin{align}\label{eq:standard_form}
\left(\mathcal{B}^{-1}\mathcal{A}\right)\pmb{\phi} = \omega\pmb{\phi},
\end{align}
following \citet{mcbain2009numerical}. The leading eigenvalue recovered by the present code for $\Rey=10^4$, $\Ray=0$, $\HH=0$ and $k=1$ agrees very well with the studies tabulated in \citet{mcbain2009numerical} to at least $8$ significant figures.

\begin{figure}
  \begin{center}
     \begin{tabular}{ccc}
     \multicolumn{1}{l}{(\textit{a})} & \multicolumn{1}{l}{(\textit{b})} & \multicolumn{1}{l}{(\textit{c})} \\
     \includegraphics[width=0.32\columnwidth]{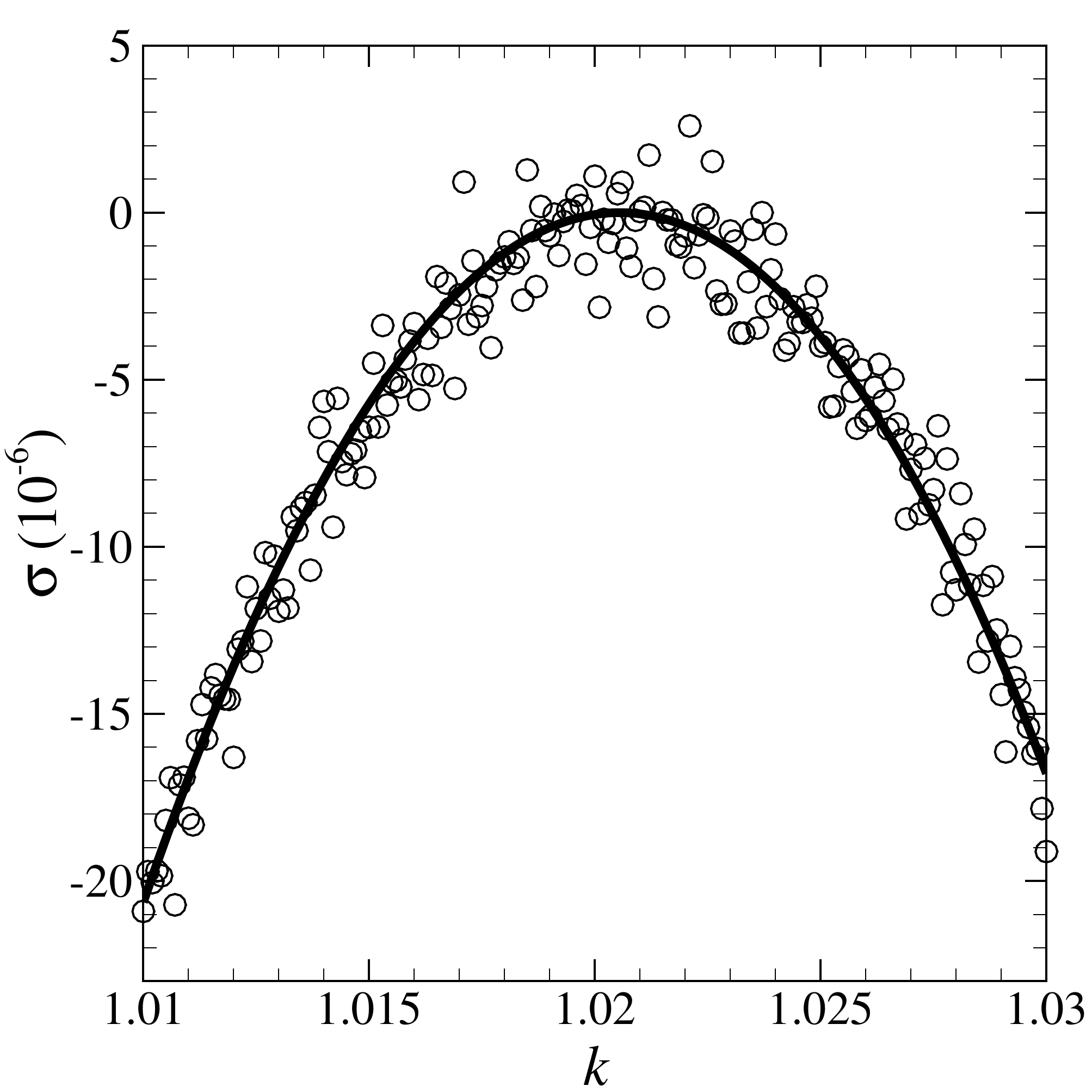}&
       \includegraphics[width=0.32\columnwidth]{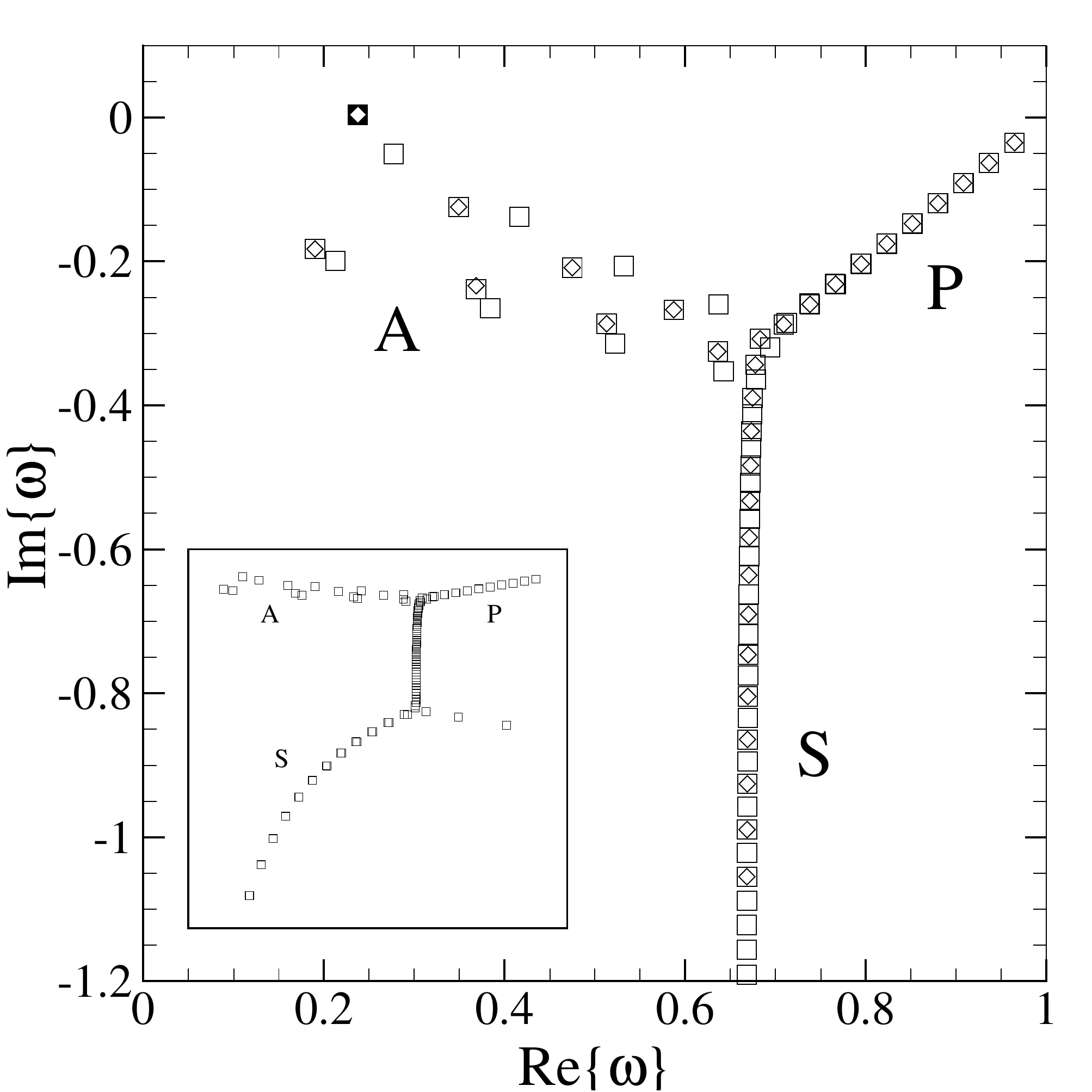}&
       \includegraphics[width=0.32\columnwidth]{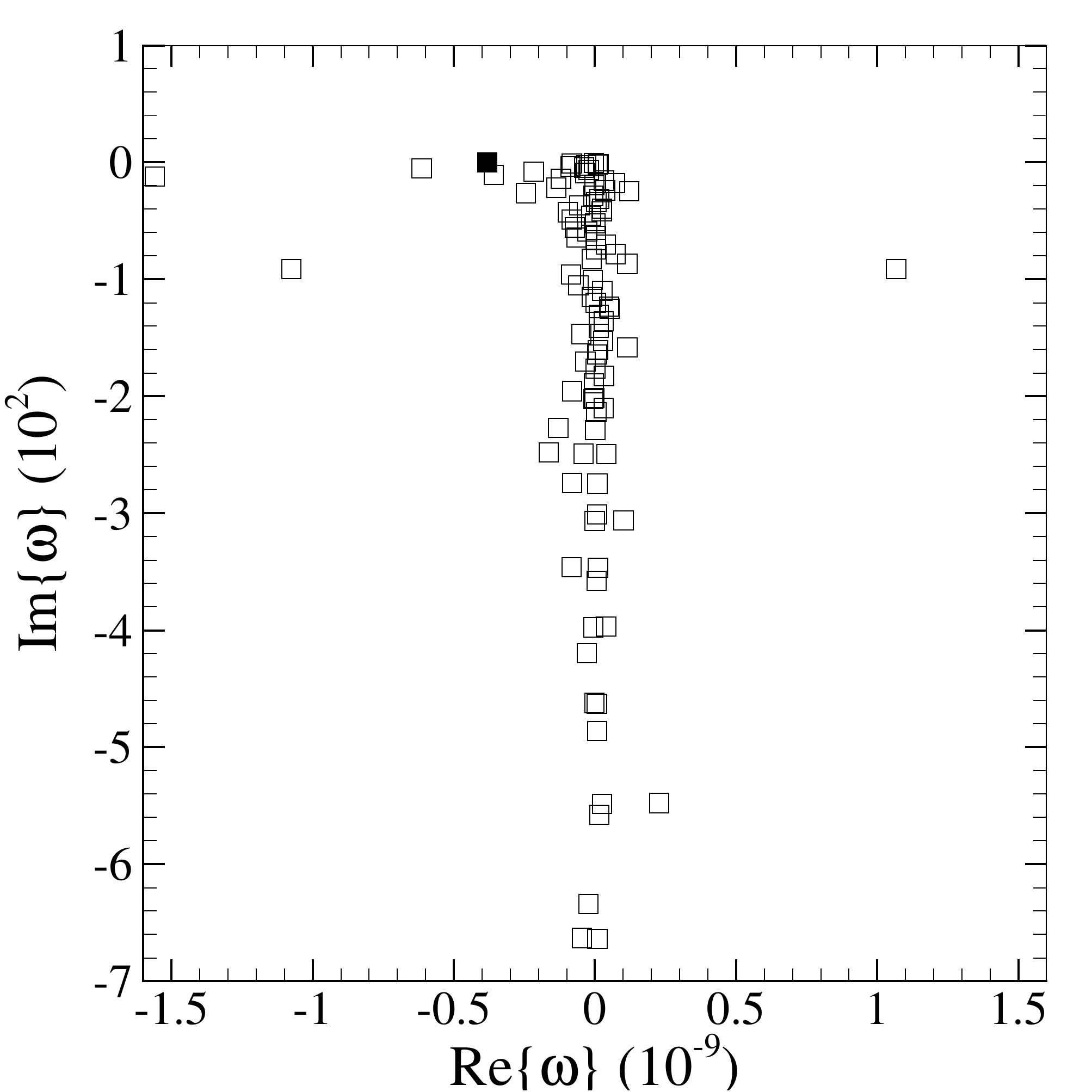}
     \end{tabular}
  \caption{(\textit{a})~Growth rates as a function of wavenumber at $\Rey=5772$ solved in the standard form (piecewise line segments connecting each computed point) and the generalised form ($\bigcirc$). The same set of wavenumbers are solved in both cases. (\textit{b})~Eigenvalue spectrum for \Poi\ flow ($\Rey=10^4$, $\Ray=0$, $\HH=0$) displaying a comparison between the eigenvalues obtained for the present study ($\square$) and \citet{mack1976numerical} ($\diamond$). Note that additional eigenvalues are displayed for the present study. The inset displays the same eigenvalue spectrum with the inclusion of smaller $\textrm{Im}\{\omega\}$ eigenvalues. (\textit{c})~Eigenvalue spectrum for \RB\ flow ($\Rey=0$, $\Ray=250$, $\HH=0$). The solid symbol represents the leading eigenvalue.}\label{fig:Re5772_growth_rate_noisy}
  \end{center}
\end{figure}

The growth rates over $1.01\leq k\leq1.03$ for plane \Poi\ flow at the critical Reynolds number $\Rey_c=5772.22$ have been obtained using generalised and standard forms, and are plotted in figure~\ref{fig:Re5772_growth_rate_noisy}(\textit{a}). In addition to its superior accuracy, solutions using the standard form converged with a much shorter compute time than the generalised form. Thus, all results presented in this study are obtained by solving the eigenvalue problem in standard form unless otherwise specified.

We also recovered the three-branch structure of the corresponding eigenvalue spectrum, in excellent agreement with \citet{mack1976numerical} (see figure~\ref{fig:Re5772_growth_rate_noisy}\textit{b}). The branches are labelled A, P and S \citep{schmid2001stability}. The S branch is found to bifurcate at lower $\textrm{Im}\{\omega\}$ values as illustrated in the inset panel. This behaviour was also identified by \citet{potherat2007quasi} for \mhd\ flows. The leading eigenvalue is highlighted by the solid symbol and is observed on the A branch. The A and P modes are sometimes referred to as the wall and center modes, respectively \citep{schmid2001stability}. Thus, the unstable mode for plane \Poi\ flow is a wall mode described by a \TollSchl\ wave.

The critical conditions for \RB\ flow is found to be $\Ray_c=213.47$ (equivalent to $\Ray_c=1707.76$ using the traditional scaling where the full duct height is used as the length scale as opposed to half the duct height used in the present study) with a corresponding $k_c=1.5582$ (equivalent to $k_c=3.1164$ based on full duct height scaling). These values are in excellent agreement with $\Ray_c=1708$ and $k_c=3.117$, which are often quoted to $4$ significant figures in previous literature \citep{reid1958some,chandrasekhar1961hydrodynamics,drazin2004hydrodynamic}. The eigenvalue spectrum for \RB\ flow is shown in figure~\ref{fig:Re5772_growth_rate_noisy}(\textit{c}) and exhibits a vertical branch situated at very small phase velocities, values less than the precision of the present code and therefore considered to be zero. This suggests that the instability is non-propagating, which is consistent with the steady (non-oscillatory) modes predicted at the onset of \RB\ convection \citep{chandrasekhar1961hydrodynamics}.

In the present study, there are three governing parameters ($\Rey$, $\Ray$, $\HH$). Therefore, to begin searching for the critical conditions, two of the three governing parameters are fixed while the third parameter is varied to seek $\sigma=\textrm{Im}\{\omega\}=0$. The flow condition is considered to be critical once the varying parameter (either $\Rey_c$, $\Ray_c$ or $\HH_c$) and corresponding growth rate have converged to at least $5$ significant figures.

\section{Results and discussion}\label{sec:results}
This section discusses the neutral stability curves obtained for $\Rey_c$ and $\Ray_c$ as a function of $\HH$, for $10^{-2}\leq\HH\leq10^{4}$. Neutral stability curves for the special cases with $\Rey=0$ and $\Ray=0$ are presented first, and $\Ray_c$ and $\Rey_c$ are subsequently sought for positive $\Rey$ and $\Ray$, respectively.

\subsection{Stability for $\Rey=0$ or $\Ray=0$}\label{subsec:Re0_Ra0_flows}
\begin{figure}
  \begin{center}
     \begin{tabular}{cc}
       \multicolumn{1}{l}{(\textit{a})} & \multicolumn{1}{l}{(\textit{b})}\\
       \includegraphics[width=0.475\columnwidth]{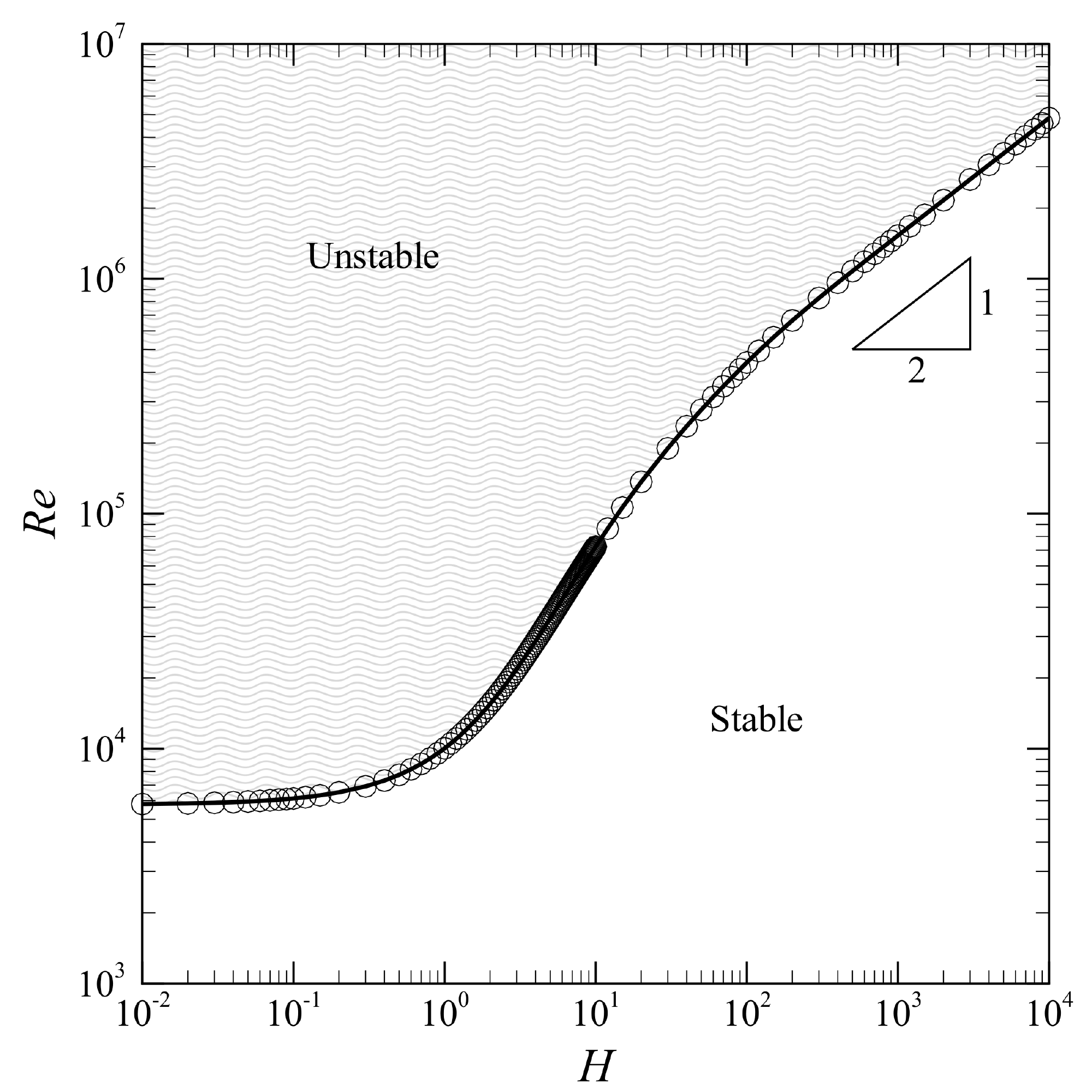}&
       \includegraphics[width=0.475\columnwidth]{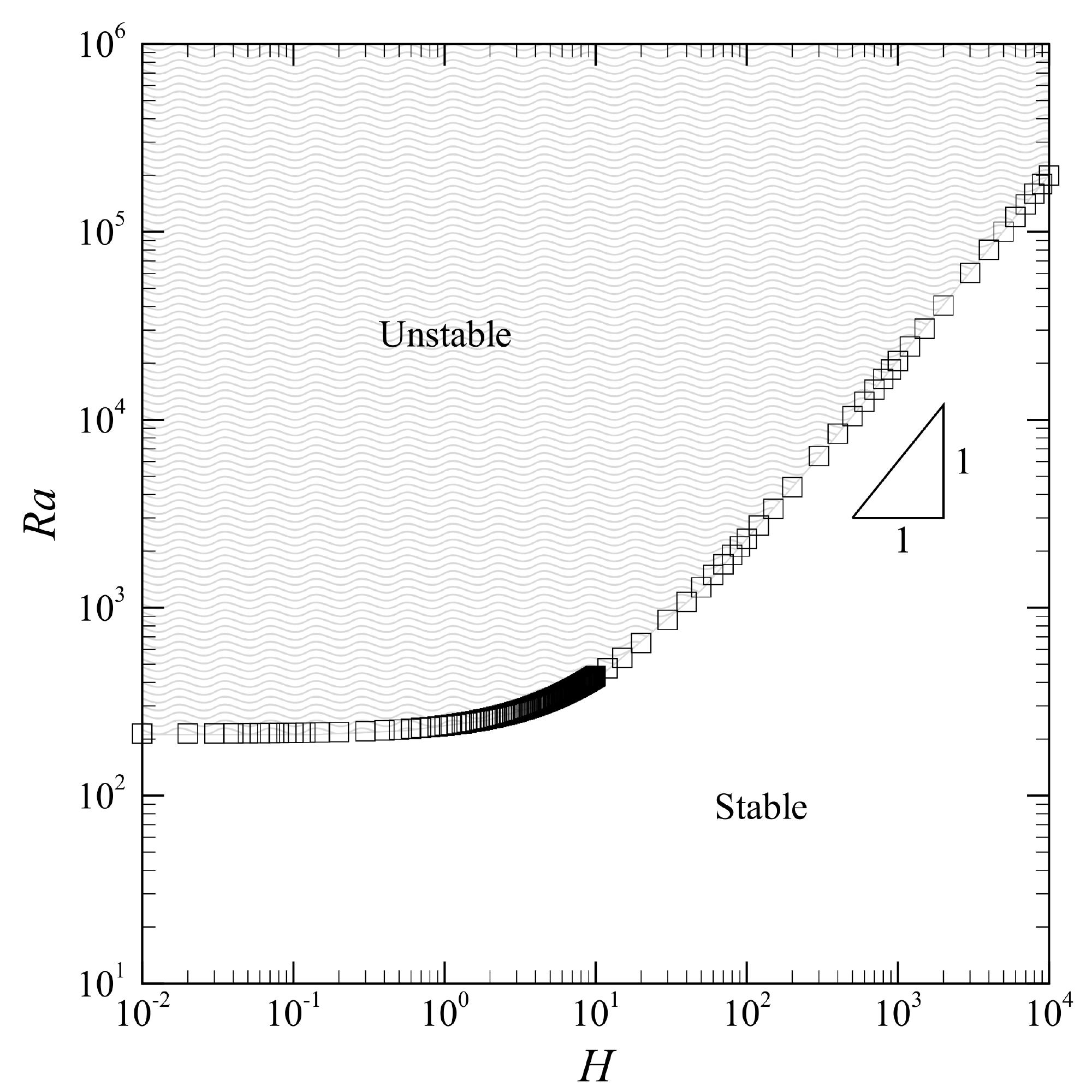}  \\
       \multicolumn{1}{l}{(\textit{c})} & \multicolumn{1}{l}{(\textit{d})}\\
       \includegraphics[width=0.475\columnwidth]{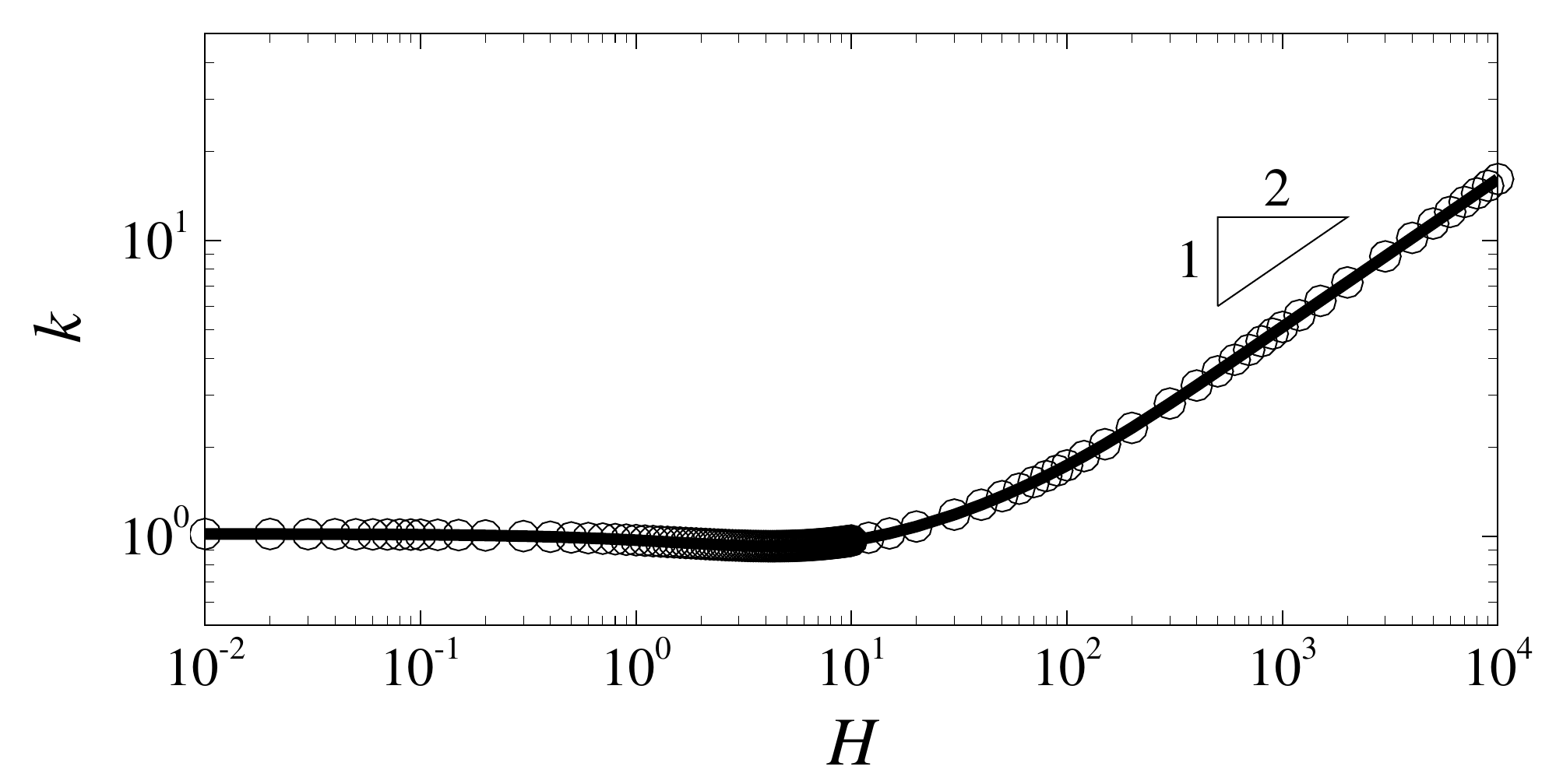}&
       \includegraphics[width=0.475\columnwidth]{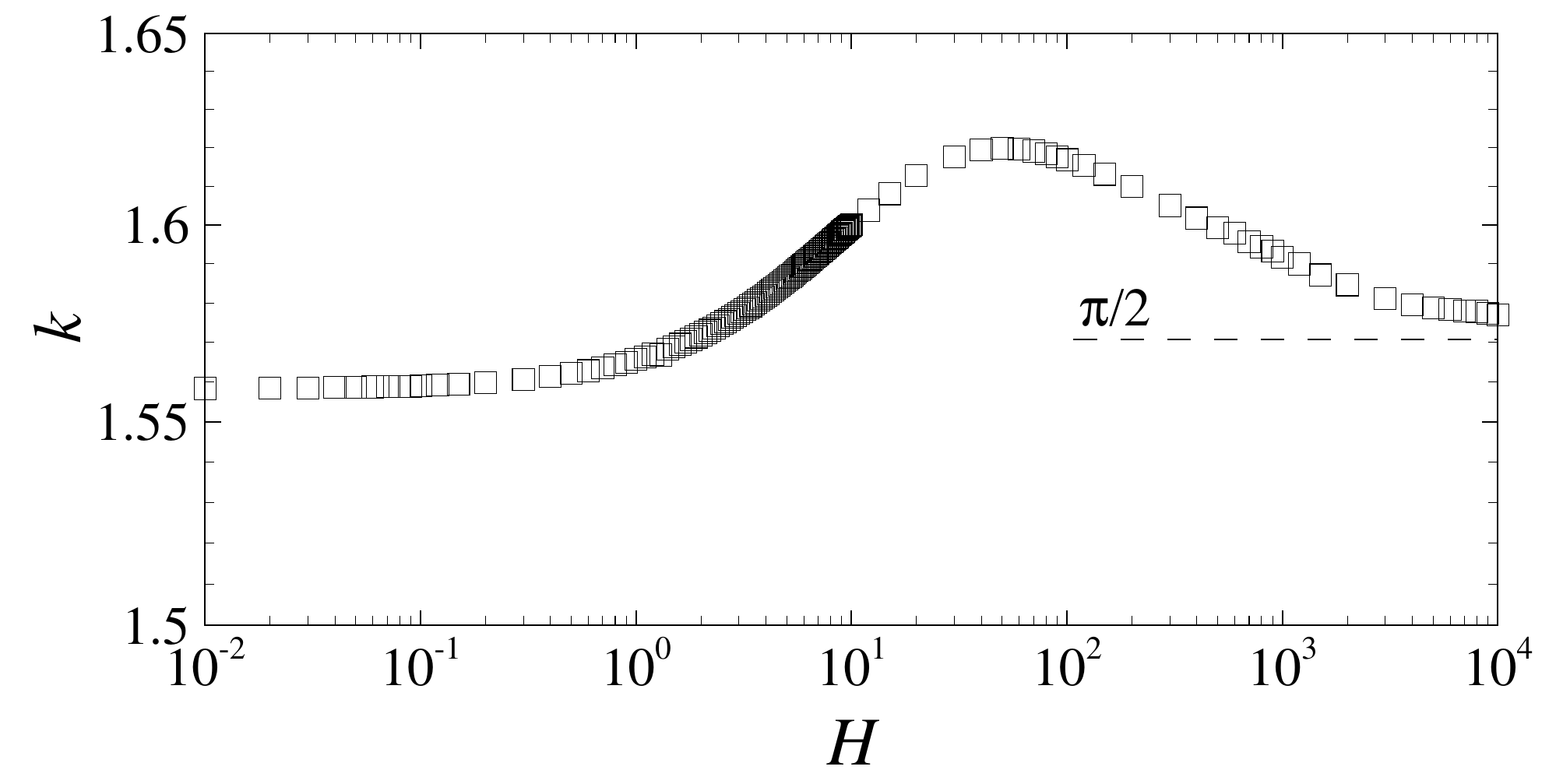}  \\
     \end{tabular}
  \caption{Neutral stability curves of (\textit{a})~$\Rey_c$, (\textit{b})~$\Ray_c$ and (\textit{c})~$k_c$ as a function of $\HH$ for $\Ray=0$~($\circ$) and $\Rey=0$~($\square$). The solid lines in (\textit{a},~\textit{c}):~\citet{potherat2007quasi} and the dashed line in (\textit{d}):~$k=\pi$/2. Shaded regions represent the unstable flow conditions.}\label{fig:Rec_Ra0}
  \end{center}
\end{figure}

The neutral stability curve for $\Ray=0$ is shown in figure~\ref{fig:Rec_Ra0}(\textit{a}). Above and below the curve are unstable and stable flows, respectively. An excellent agreement in $\Rey_c$ values between the present study and \citet{potherat2007quasi} can be seen, with a maximum percentage error of less than $0.1\%$ across the range of $10^{-2}\leq\HH\leq10^4$. The curve demonstrates that $\Rey_c$ increases monotonically with increasing $\HH$ which is due to a stronger through-flow being required to counteract the increased damping induced by the increased magnetic field strength. As $\HH\rightarrow0$, the flow becomes purely hydrodynamic and achieves the expected instability to \TollSchl\ waves at $\Rey_c=5772.22$. As $\HH\rightarrow\infty$, the onset of instability is described by an asymptotic trend regime $\Rey_c=48347\HH^{\,1/2}$ for $\HH\gtrsim200$ governed by the Shercliff layers.

Similar to the $\Ray=0$ case, the neutral stability curve for the $\Rey=0$ case of natural convection with no through-flow is unique and continuous as shown in figure~\ref{fig:Rec_Ra0}(\textit{b}). As $\HH\rightarrow0$, the flow reverts to classical plane \RB\ flow and the perturbations are controlled by the balance between buoyancy and viscous dissipation, with $\Ray_c=213.47$ ($\HH=0$). However, as $\HH\rightarrow\infty$ the neutral stability curve approaches a scaling of $\Ray_c=(21.672\pm0.212)\HH^{\,(0.991\pm0.001)}$ for $2000\leq\HH\leq10^4$. The uncertainties presented throughout this paper correspond to the standard error (estimated standard deviation) of the least-squares estimates using linear regression for the coefficients and exponents, and are only provided if the uncertainty is greater than $0.01\%$. It is expected here that the exponent $0.991\pm0.001$ will approach unity as $\HH\rightarrow\infty$. It may be understood as follows: in the limit $\HH\rightarrow\infty$, Hartmann friction dominates and has to be balanced by buoyancy for convection to set in. Hence, at the onset the balance between these forces implies $\Ray\sim\HH$, which scales with the duct height $2L$. This scaling is in agreement with the \lsa\ performed by \citet{burr2002rayleigh} who studied \RB\ convection in liquid metal layers under the influence of a horizontal magnetic field. This linear scaling was also obtained by \citet{mistrangelo2016magneto} who studied a the stability of a motionless basic steady-state solution induced by a parabolic temperature distribution.

The critical wavenumber $k_c$ for both $\Ray=0$ and $\Rey=0$ are shown in figures~\ref{fig:Rec_Ra0}(\textit{c,d}), respectively. For $\Ray=0$, $k_c=1.02$ at low $\HH$ and adopts an asymptotic trend at high $\HH$ of the form $k_c=0.1615\HH^{\,1/2}$. In contrast, the critical wavenumber for $\Rey=0$ remains relatively constant across all $\HH$. As $\HH\rightarrow0$, $k_c\rightarrow1.5582$ and as $\HH\rightarrow\infty$, the critical wavenumber appears to approach $k_c=\pi/2$. These values are in agreement with the asymptotic values from \citet{burr2002rayleigh}. These authors also observed an unexpected maximum at $\HH\approx60$. It is unclear as to why there is an increase in wavenumber at the onset of \RB\ convection cells in this intermediate regime where the flow transitions from a buoyancy-viscous dissipation balance to a buoyancy-Hartmann friction balance. However, the increase in $k$ with increasing $\HH$ has been observed experimentally using shadowgraph visualisation \citep{andreev2003visualization}. The streamwise wavenumber scaling for both $\Rey=0$ and $\Ray=0$ cases as $\HH\rightarrow\infty$ support the view that the instability scales with the Shercliff layer thickness and channel height, respectively.

\begin{figure}
  \begin{center}
  \begin{tabular}{cc}
     \begin{tabular}{c}
     \multicolumn{1}{l}{(\textit{a})} \\
     \includegraphics[width=0.48\columnwidth, trim=0 0 2mm 0, clip=true]{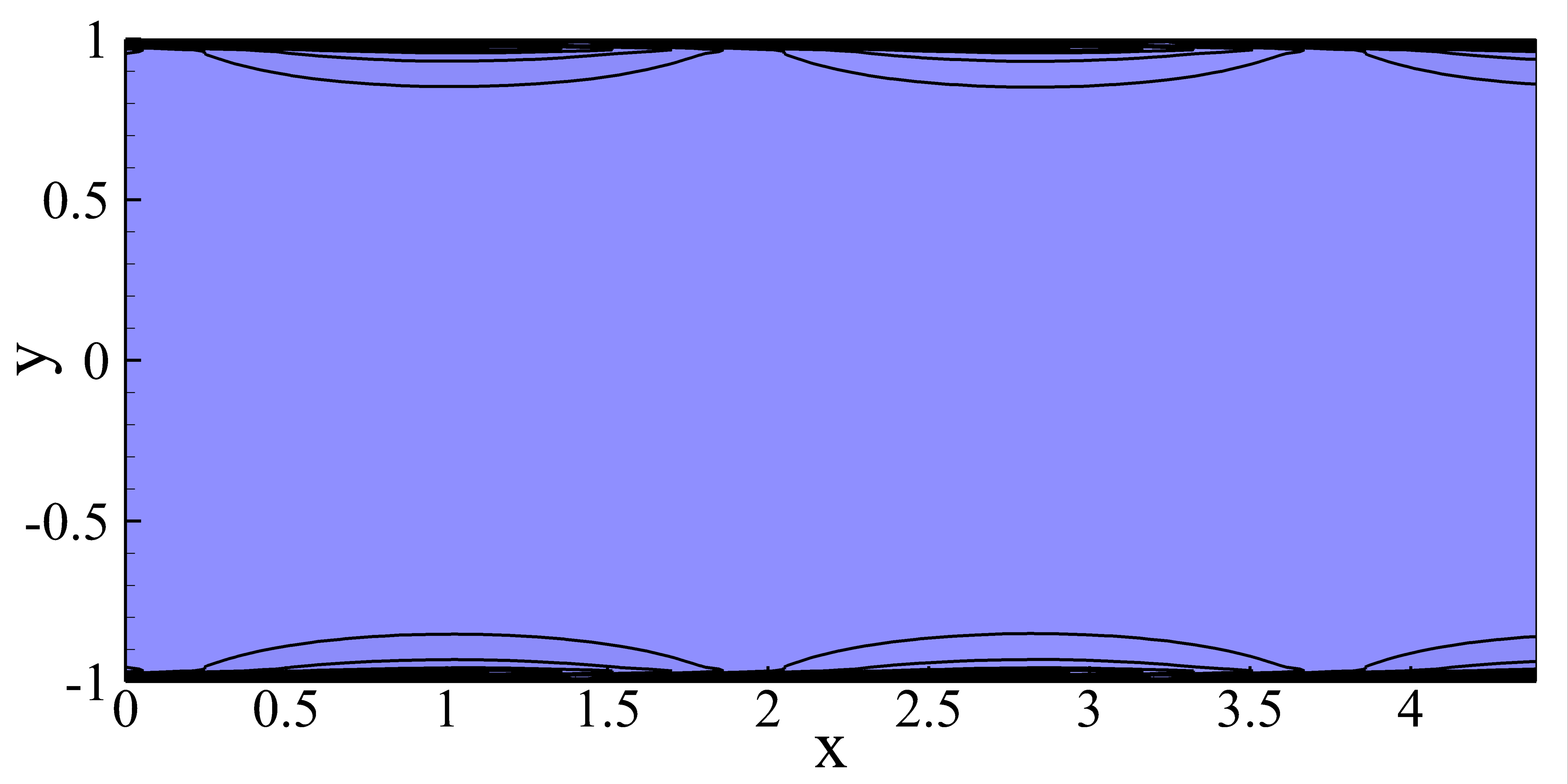}
     \end{tabular}

     \begin{tabular}{c}
     \multicolumn{1}{l}{(\textit{b})} \\
     \includegraphics[width=0.48\columnwidth]{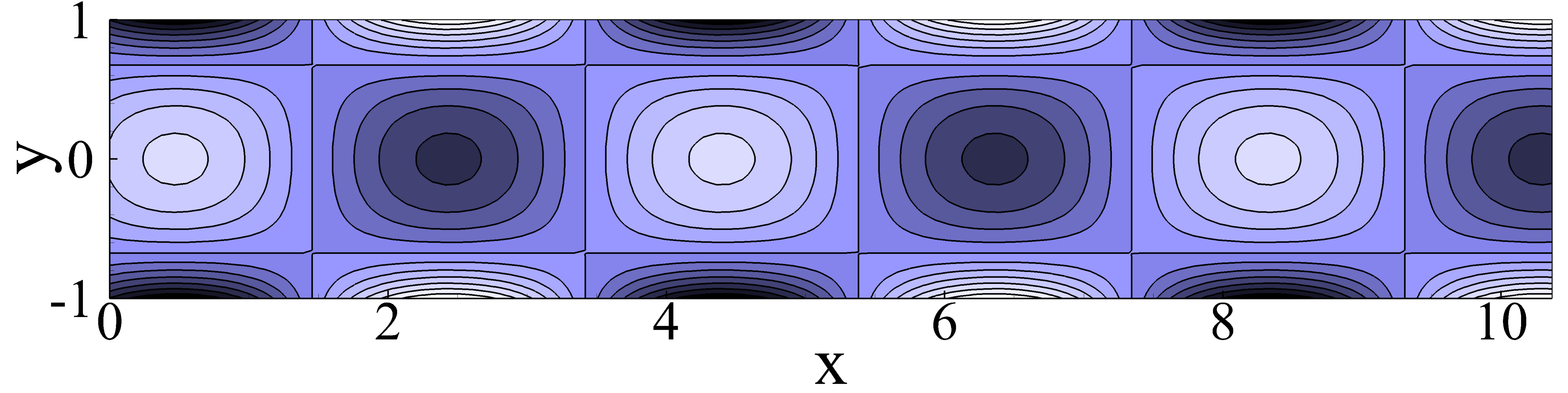}\\
     \includegraphics[width=0.48\columnwidth]{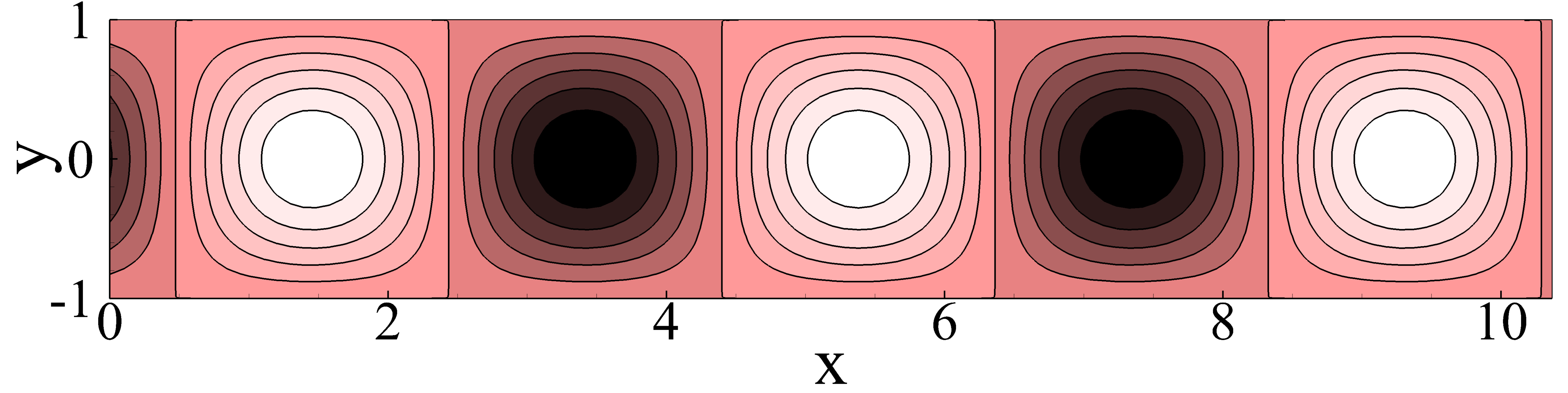}
     \end{tabular}\\

     \begin{tabular}{cc}
     \multicolumn{1}{l}{(\textit{c})} & \multicolumn{1}{l}{(\textit{d})}\\
     \includegraphics[width=0.48\columnwidth]{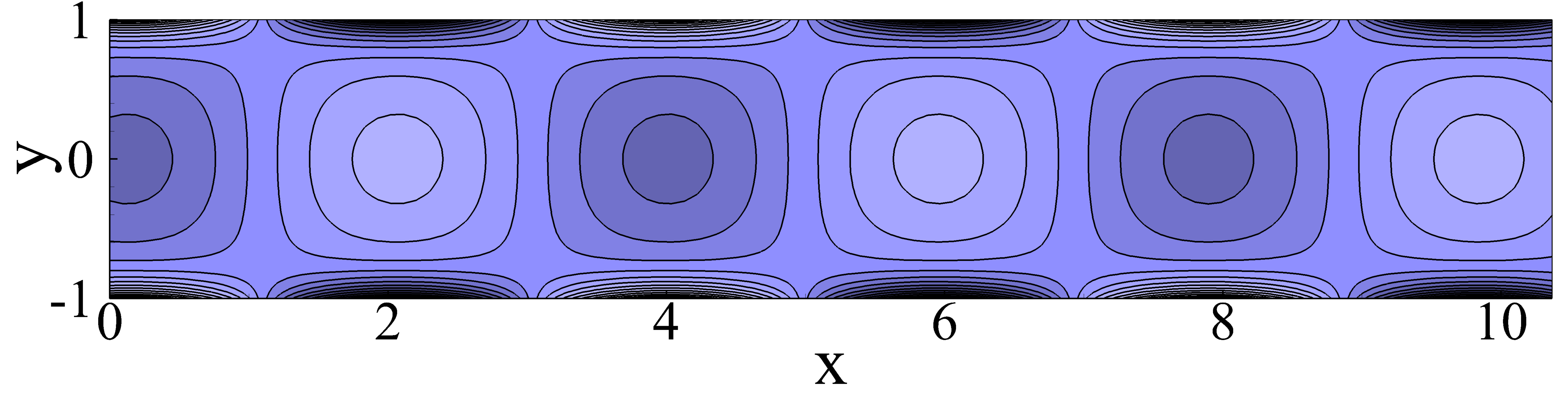}&
     \includegraphics[width=0.48\columnwidth]{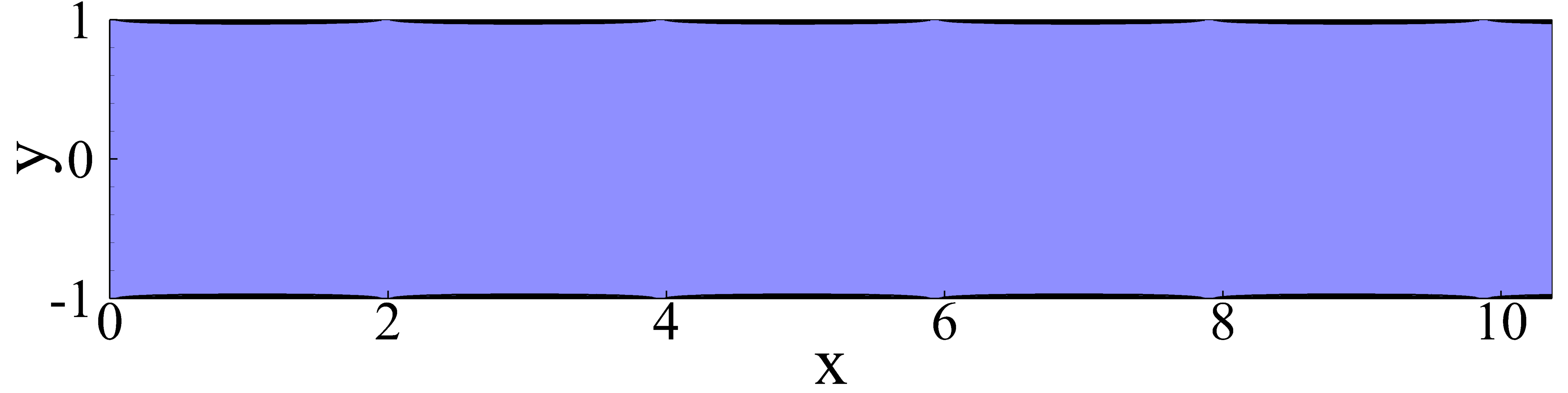} \\
     \includegraphics[width=0.48\columnwidth]{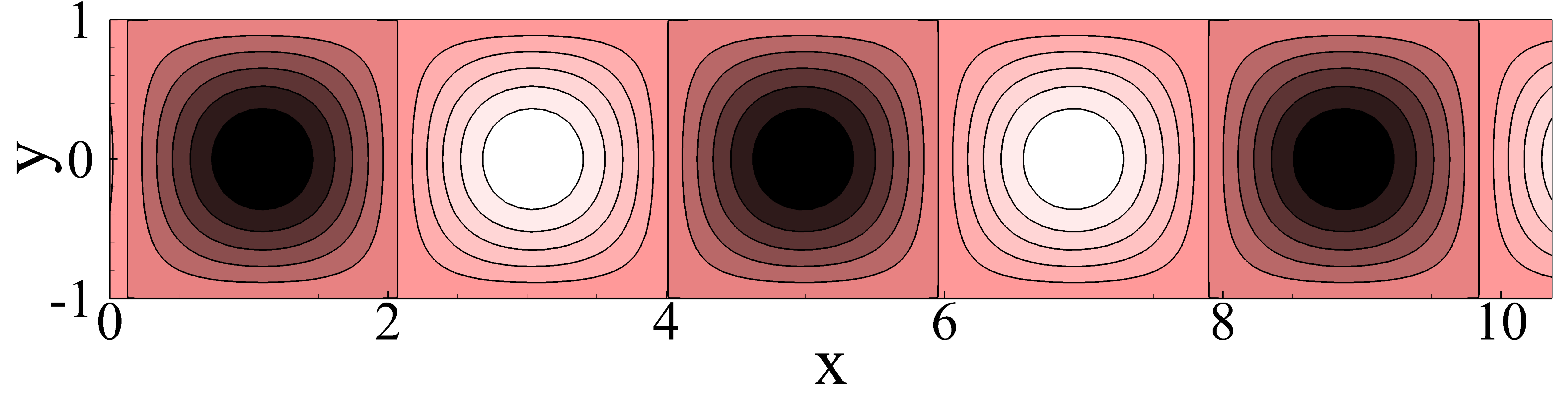}&
     \includegraphics[width=0.48\columnwidth]{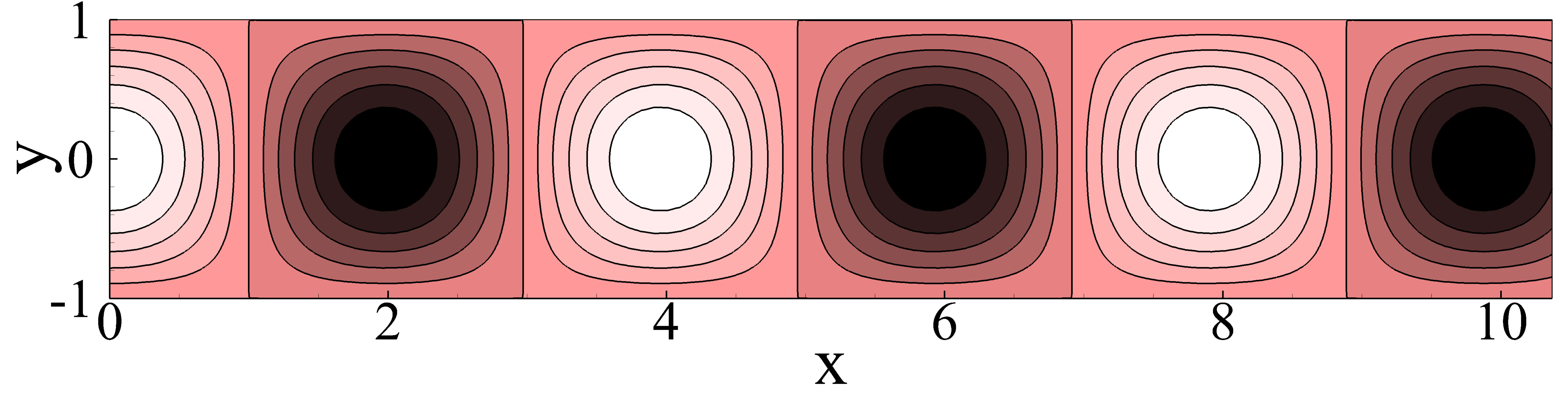}
     \end{tabular}
  \end{tabular}
  \caption{(\textit{a})~Vorticity in the eigenvector field of $\Rey_c=440223$, $\Ray=0$, $\HH=100$ for $k_c=1.7393$. Vorticity (top panel) and temperature (bottom panel) in the eigenvector fields of (\textit{b})~$\Rey=0$, $\Ray_c=432.79$, $\HH=10$ for $k_c=1.6001$, (\textit{c})~$\Rey=0$, $\Ray_c=2322.66$, $\HH=100$ for $k_c=1.6169$, and (\textit{d})~$\Rey=0$, $\Ray_c=199513.4$, $\HH=1\times10^4$ for $k_c=1.5929$. Dark and light shading represent negative and positive values, respectively.}\label{fig:H1_Ra0_Re10033d15_vort}
  \end{center}
\end{figure}

A typical plot of vorticity in the eigenvector field for $\Rey_c$ with $\Ray=0$ and $\HH=100$ is illustrated in figure~\ref{fig:H1_Ra0_Re10033d15_vort}(\textit{a}). The instability caused by the shearing through-flow is localised along the horizontal walls in the boundary layers. Further increasing $\HH$ only results in thinner Shercliff layers and therefore shrinking the region occupied by perturbation (figure~\ref{fig:H1_Ra0_Re10033d15_vort}\textit{d}). For all $\HH$, the leading eigenvalue was consistently found to lie on the A branch, in agreement with \citet{potherat2007quasi}. For the case of heating and no through-flow (\ie\ $\Ray_c$ with $\Rey=0$), the vorticity perturbations occupy both regions along the walls and throughout the interior of the duct as shown in figure~\ref{fig:H1_Ra0_Re10033d15_vort}(\textit{b})(top panel) for $\HH=10$. The interior vorticity disturbances are evident only at relatively low $\HH$ as the strength of the counter-rotating \RB-like cells are comparable to the wall disturbances. The vorticity and thermal perturbation fields at low, moderate and high $\HH$ are shown in figures~\ref{fig:H1_Ra0_Re10033d15_vort}(\textit{b},\textit{c},\textit{d}), respectively. Despite the disappearance of vorticity disturbances in the interior, the structure of the thermal disturbances appear to depend little on $\HH$. The eigenvalue spectrum for each case is very similar to that shown in figure~\ref{fig:Re5772_growth_rate_noisy}(\textit{c}) since the same instability is induced at onset.

\subsection{Critical Rayleigh numbers at finite Reynolds numbers}\label{subsec:fixed_Re_flows}
In this section, the linear stability of the system is investigated for fixed, non-zero Reynolds numbers with natural convection. The results are presented in a progressive manner for $0\leq\Rey\leq10^5$. The regimes $\Rey\leq300$ and $\Rey\geq350$ are discussed separately as these ranges correspond to observations of single and multiple instability modes, respectively.

\subsubsection{Critical Rayleigh numbers for $0<\Rey\leq300$}\label{subsubsec:fixed_Re300_flows}

\begin{figure}
  \begin{center}
     \begin{tabular}{cc}
     \multicolumn{1}{l}{(\textit{a})} & \multicolumn{1}{l}{(\textit{b})}\\
       \includegraphics[width=0.475\columnwidth]{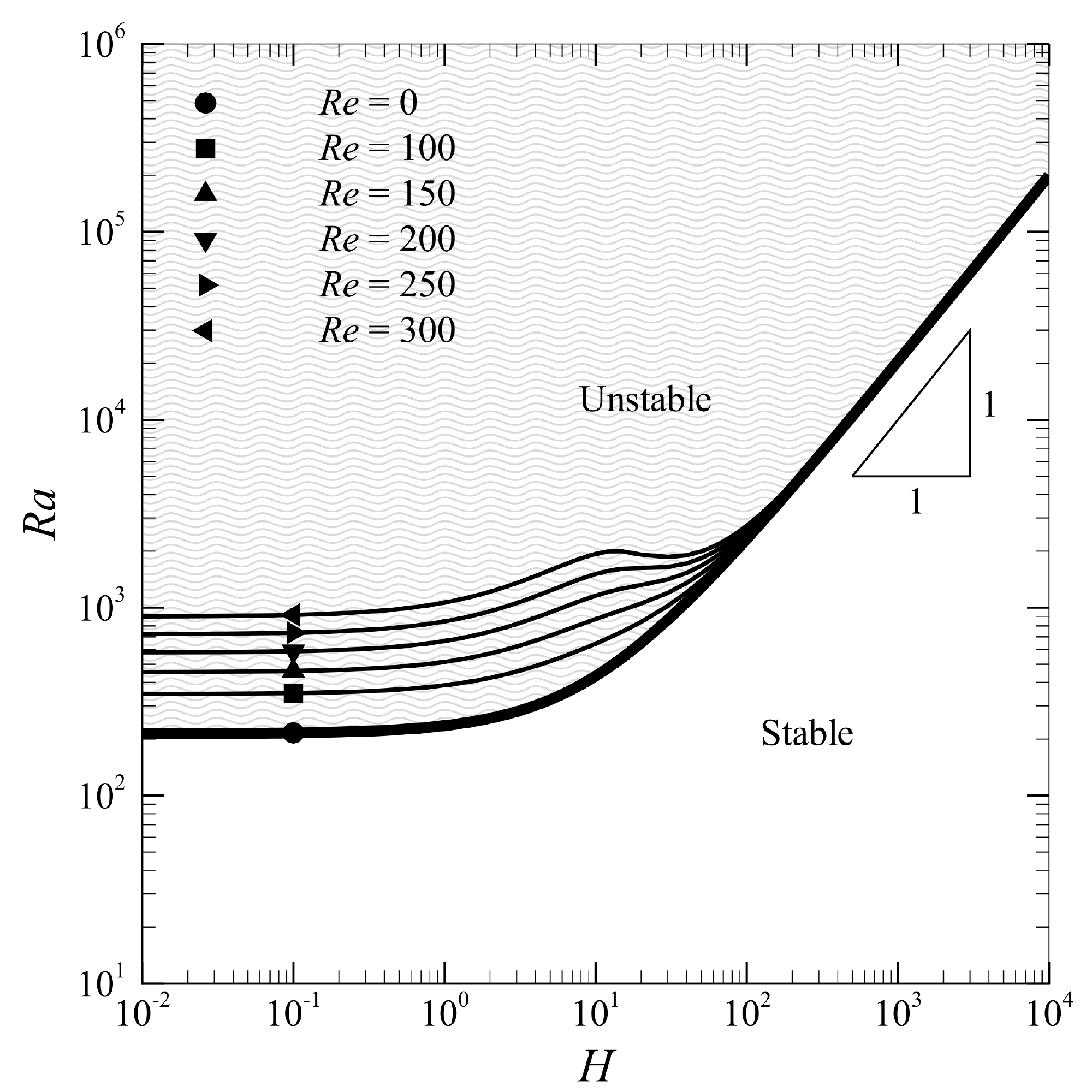}&
       \includegraphics[width=0.475\columnwidth]{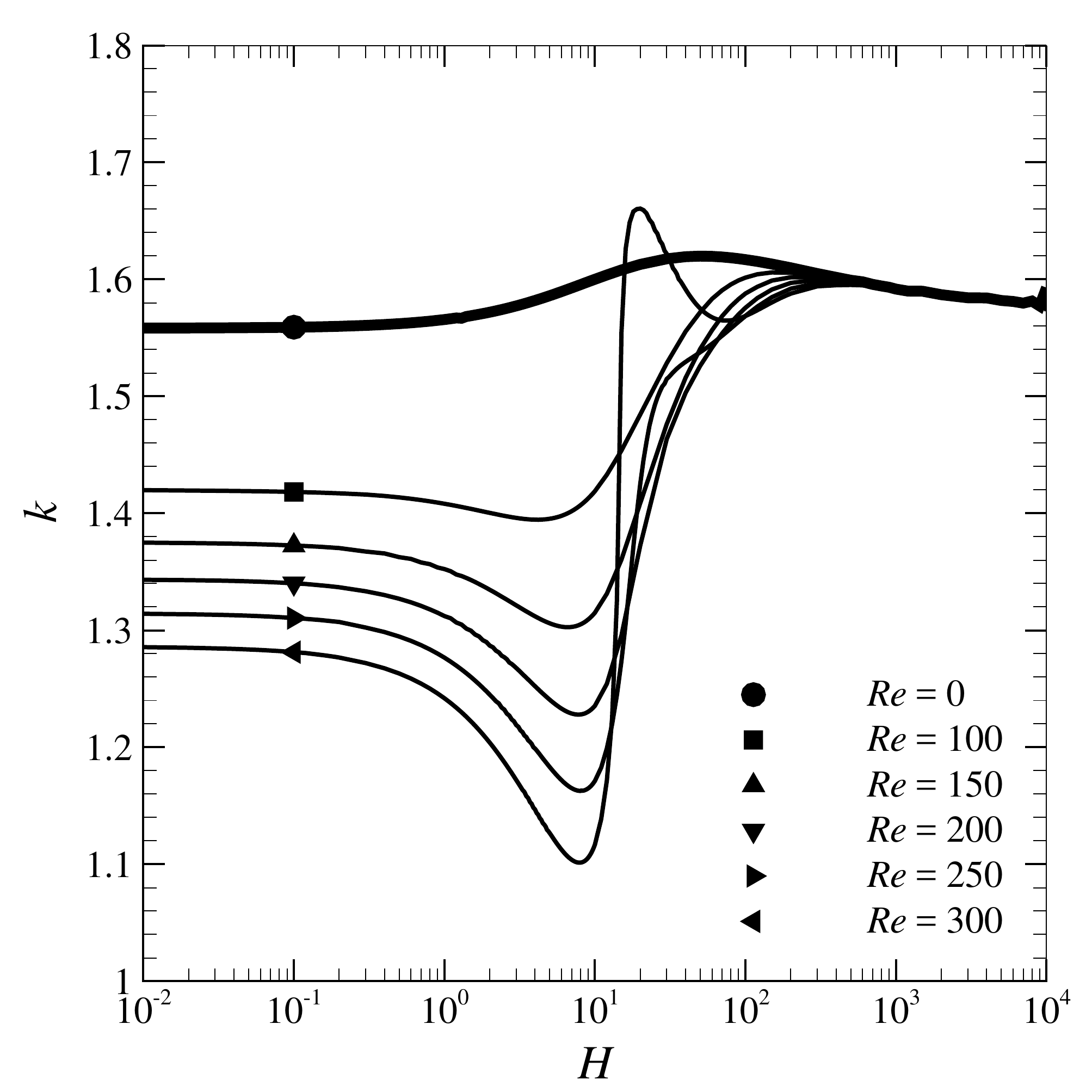}  \\
     \end{tabular}
  \caption{Neutral stability curves of (\textit{a})~$\Ray_c$ and (\textit{b})~$k_c$ as a function of $\HH$ for fixed Reynolds numbers $0\leq\Rey\leq300$. Thick solid lines represent $\Rey=0$ while thinner solid lines represent $100\leq\Rey\leq300$. The symbols identify the Reynolds number for each curve. The flows are unstable above each curve. Shaded region represents the unstable flow conditions for $\Rey=0$.}\label{fig:Rac_all_sub300}
  \end{center}
\end{figure}

Figure~\ref{fig:Rac_all_sub300}(\textit{a}) shows neutral stability curves for several Reynolds numbers over $0\leq\Rey\leq300$. Throughout this range of $\Rey$, the neutral stability curves maintain their continuous profiles with slight changes with increases in $\Rey$. These curves reveal that $\Ray_c$ is Reynolds number-dependent in the limit $\HH\rightarrow0$, and follows a Reynolds-number-independent linear scaling $\Ray_c=(21.672\pm0.212)\HH^{\,(0.991\pm0.001)}$ as $\HH\rightarrow\infty$. As $\HH$ increases, the Hartmann friction becomes the dominant damping process over viscous dissipation acting on the instabilities, and therefore at sufficiently large $\HH$, the instability threshold becomes insensitive to the Reynolds number. However, for low to moderate values of $\HH$, $\Ray_c$ increases with $\Rey$. This suggests that viscous shear flow inhibits the thermal instability: because of the disruption of recirculating convection cells caused by the shear, a stronger thermal gradient is required to overcome the through-flow. This is a well known result \citep{nicolas2000linear,nicolas2012influence}, demonstrated by experiments by \citet{luijkx1981existence}. More recently, this stabilising effect was also observed by \citet{hattori2015stability} who investigated the stability of a laterally confined fluid layer induced by absorption of radiation influenced by horizontal through-flow.

The curves of critical wavenumber for $\Rey\leq300$ are shown in figure~\ref{fig:Rac_all_sub300}(\textit{b}). Increasing $\Rey$ demonstrates a decrease in $k_c$, especially at low $\HH$ where the onset of instability is governed by a balance between buoyancy and viscous dissipation. The decrease in critical wavenumber is caused by the stronger shear which elongates convection cells. A decrease in critical wavenumber is also observed with increasing $\HH$ up to $\HH\approx1$, beyond which $k_c$ increases and approaches an asymptotic curve as $\HH\rightarrow\infty$. The increase in $k$ at intermediate values of $\HH$ is related to the change in force balances at the onset of convection cells. At higher $\HH$, the instability is controlled solely by the Hartmann layers and therefore $k_c$ is independent of $\Rey$. Furthermore, the eigenvector field between $\Rey=0$ and $300$ are indistinguishable at $\HH=10^{4}$ which confirms that the viscous dissipation has a negligible effect on the instability under a sufficiently strong magnetic field.

\begin{figure}
  \begin{center}
     \begin{tabular}{cc}
       \multicolumn{1}{l}{(\textit{a})} & \multicolumn{1}{l}{(\textit{b})}\\
       \includegraphics[width=0.475\columnwidth]{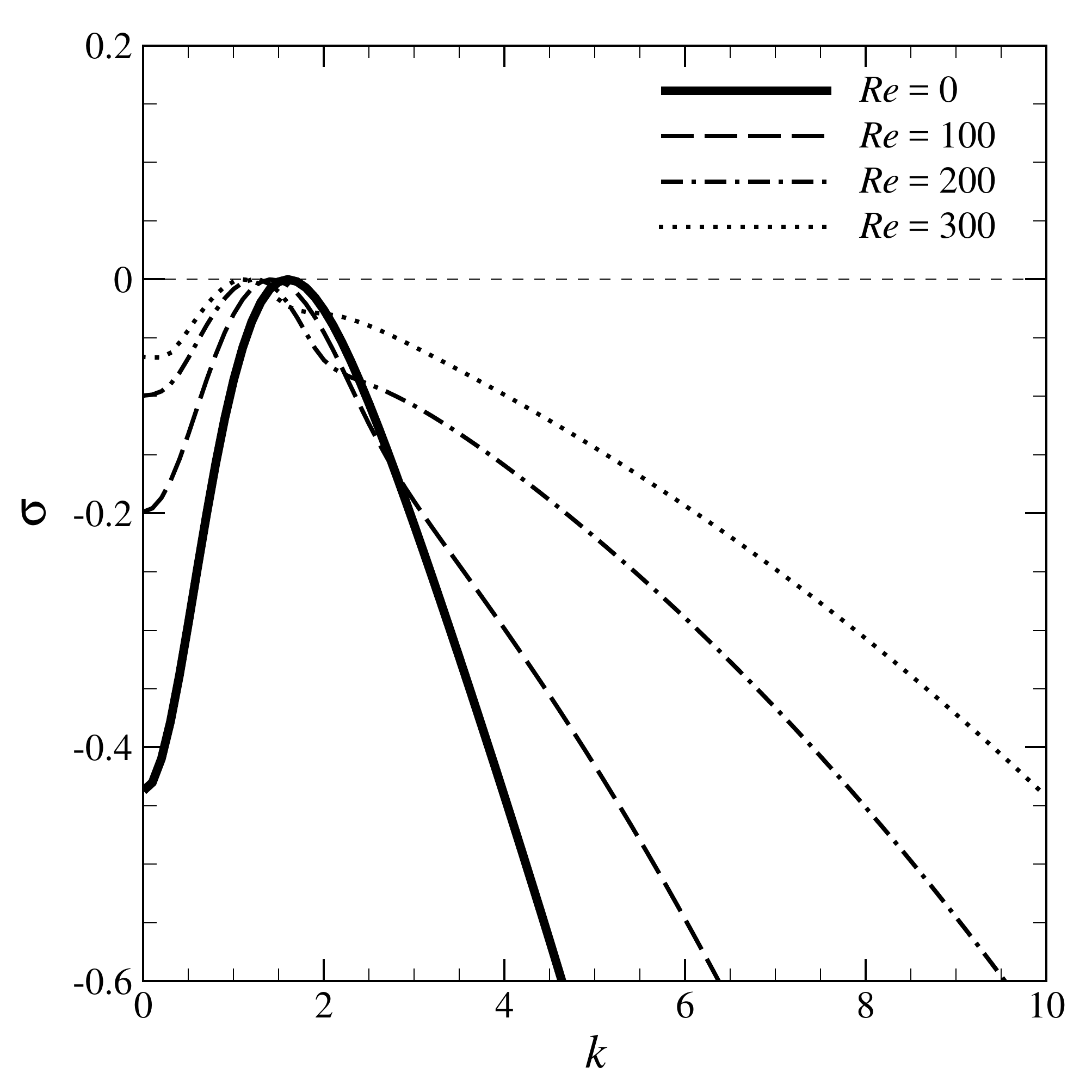}&
       \includegraphics[width=0.475\columnwidth]{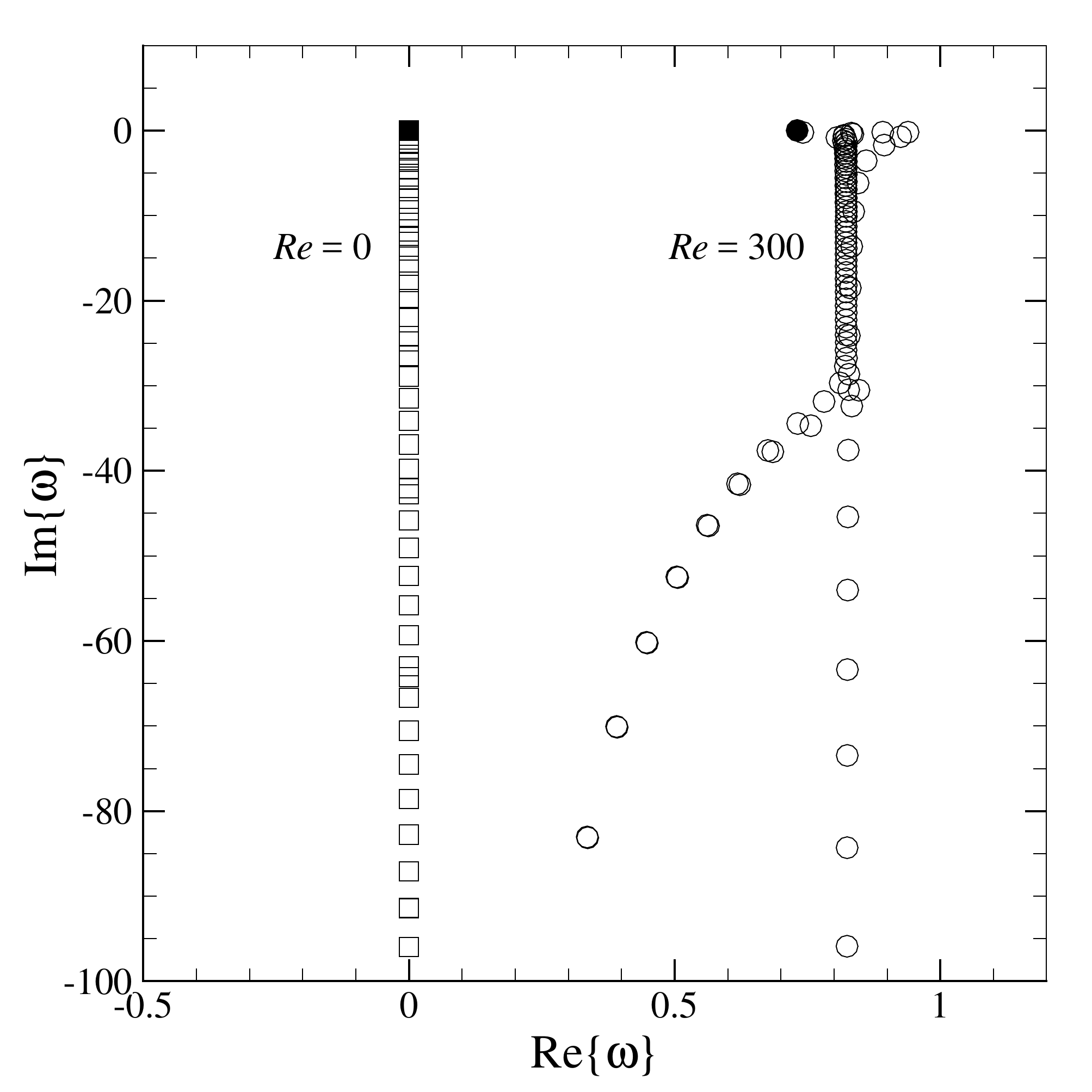}
     \end{tabular}
  \caption{(\textit{a})~Growth rate profiles and (\textit{b})~eigenvalue spectra of the wavenumber of peak growth rate for $\HH=10$ and various $\Rey$ with corresponding $\Ray_c$. The eigenvalue spectra are shown for $\Rey=0$~($\square$) and $300$~($\circ$). The solid symbols denote the leading eigenvalue.}\label{fig:H1_Rac_Re0to300_growth_lsa}
  \end{center}
\end{figure}

The streamwise wavenumber dependence of the critical growth rate with fixed $\Rey\leq300$ and $\HH=10$ is shown in figure~\ref{fig:H1_Rac_Re0to300_growth_lsa}(\textit{a}). The $\Rey=0$ data exhibits a single maximum. With increasing $\Rey$, a secondary maximum begins to emerge at higher wavenumbers. However, this secondary maximum never becomes unstable in the range of $0\leq\Rey\leq300$ as increasing $\Ray$ beyond $\Ray_c$ results in the lower-wavenumber mode absorbing the secondary, higher-wavenumber maximum. Hence, the neutral stability curves for $\Rey\leq300$ are described exactly by the onset of the single instability of the mode $k=k_c$, not by the onset of multiple instabilities. The critical wavenumber decreases slightly with the Reynolds number, which is consistent with figure~\ref{fig:Rac_all_sub300}(\textit{b}).

\begin{figure}
  \begin{center}
     \begin{tabular}{cc}
       \multicolumn{1}{l}{(\textit{a})~$\HH=10$}\\
       \includegraphics[width=0.475\columnwidth]{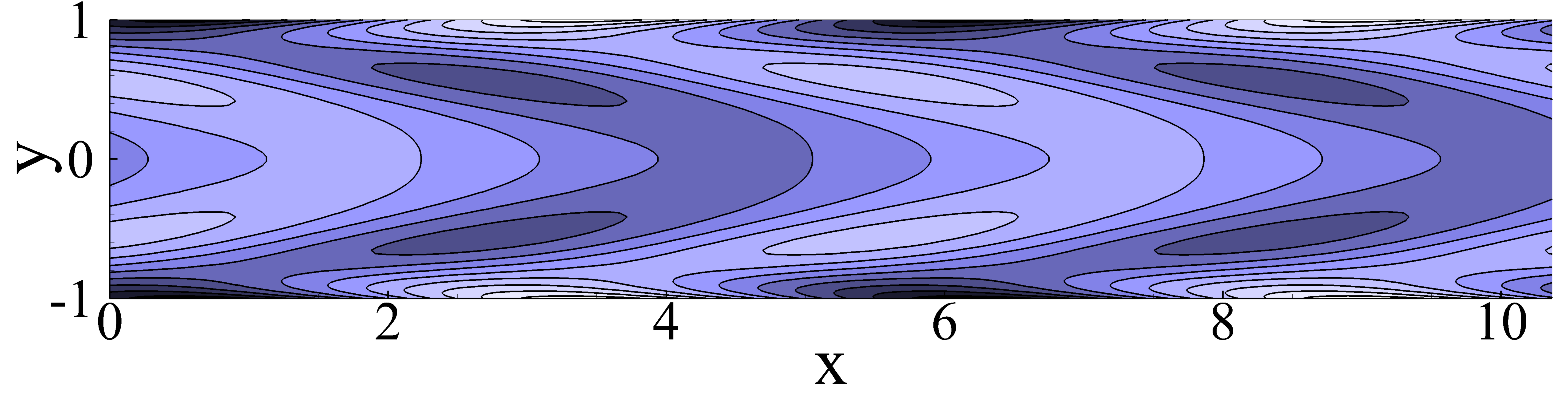}&
       \includegraphics[width=0.475\columnwidth]{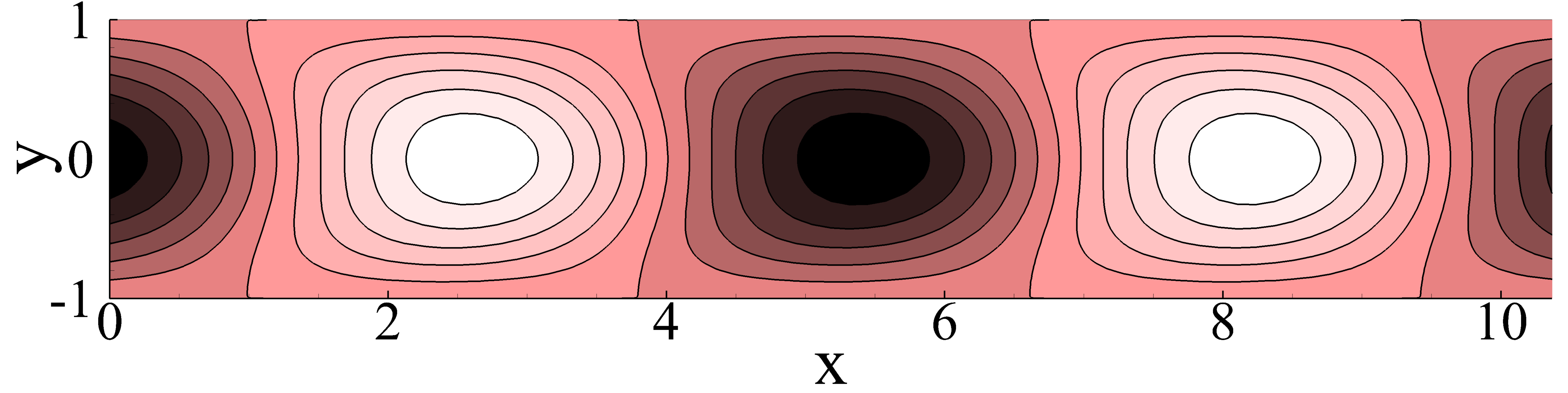}  \\
       \multicolumn{1}{l}{(\textit{b})~$\HH=100$}\\
       \includegraphics[width=0.475\columnwidth]{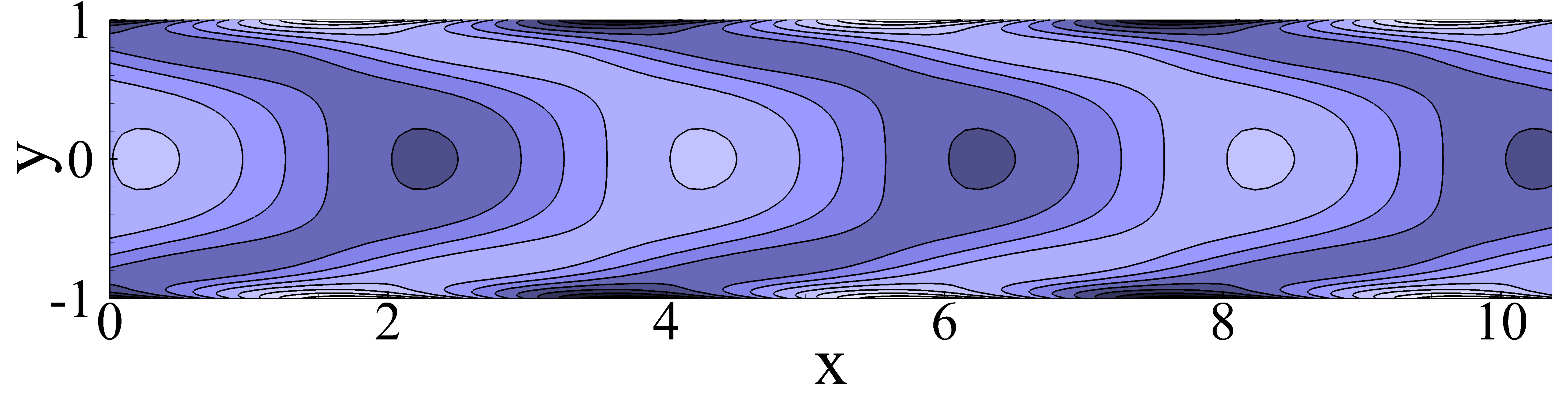}&
       \includegraphics[width=0.475\columnwidth]{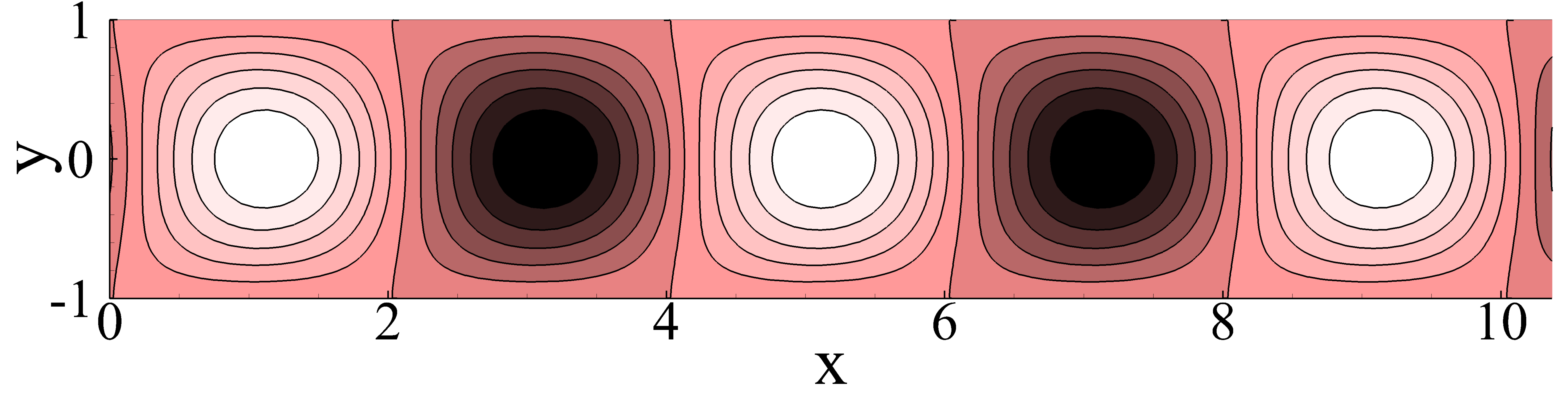}  \\
       \multicolumn{1}{l}{(\textit{c})~$\HH=5000$}\\
       \includegraphics[width=0.475\columnwidth]{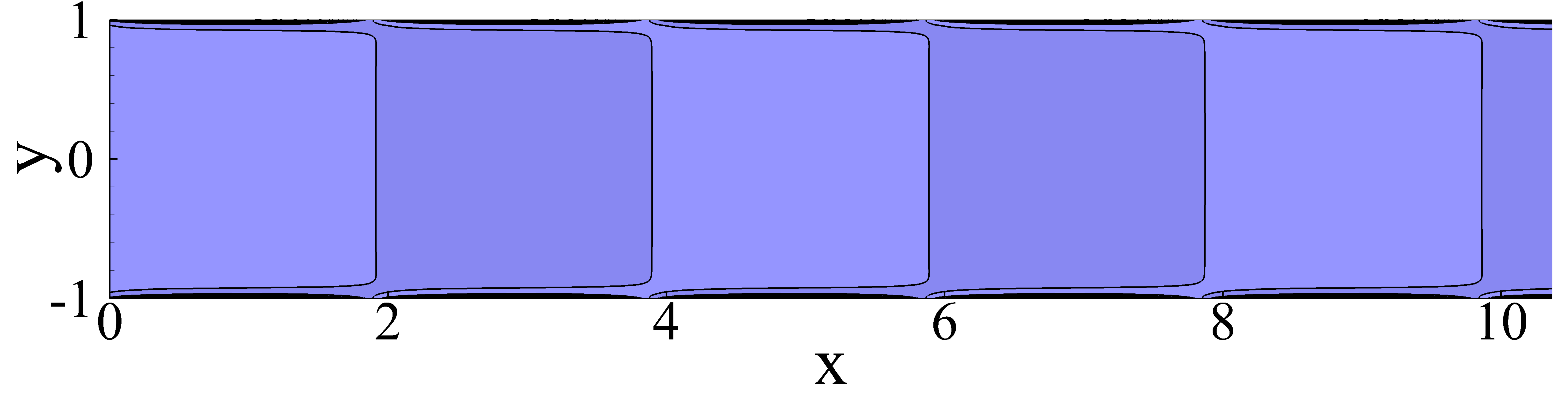}&
       \includegraphics[width=0.475\columnwidth]{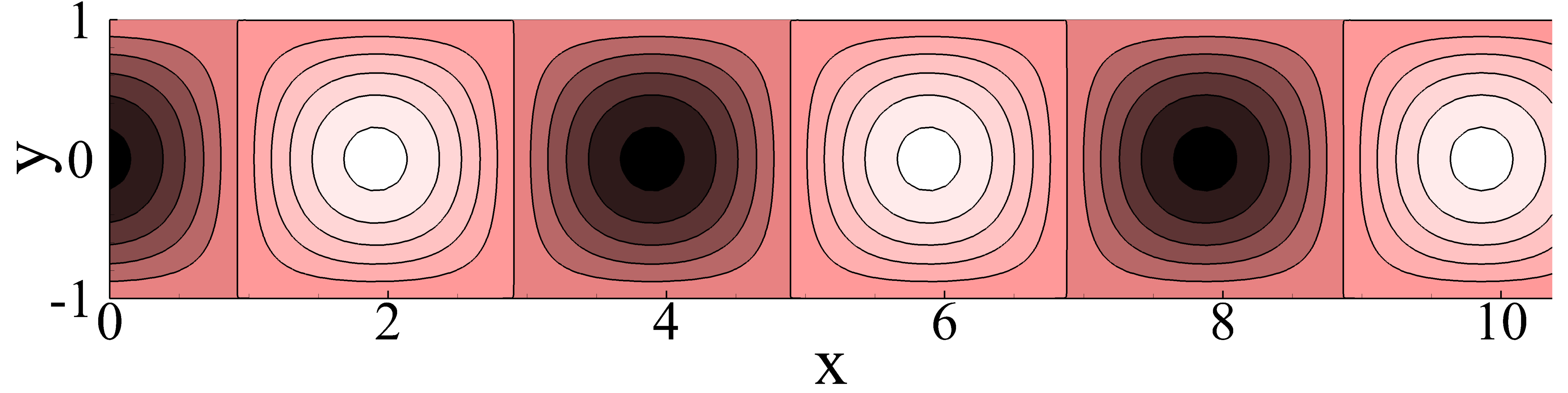}  \\
     \end{tabular}\\
     \begin{tabular}{cccc}
     \multicolumn{4}{l}{(\textit{d})}\\
     \multicolumn{1}{l}{(\textrm{i})~$\HH=0$}&
     \multicolumn{1}{l}{(\textrm{ii})~$\HH=100$}&
     \multicolumn{1}{l}{(\textrm{iii})~$\HH=1\times10^3$}&
     \multicolumn{1}{l}{(\textrm{iv})~$\HH=1\times10^4$}\\
       \includegraphics[width=0.24\columnwidth]{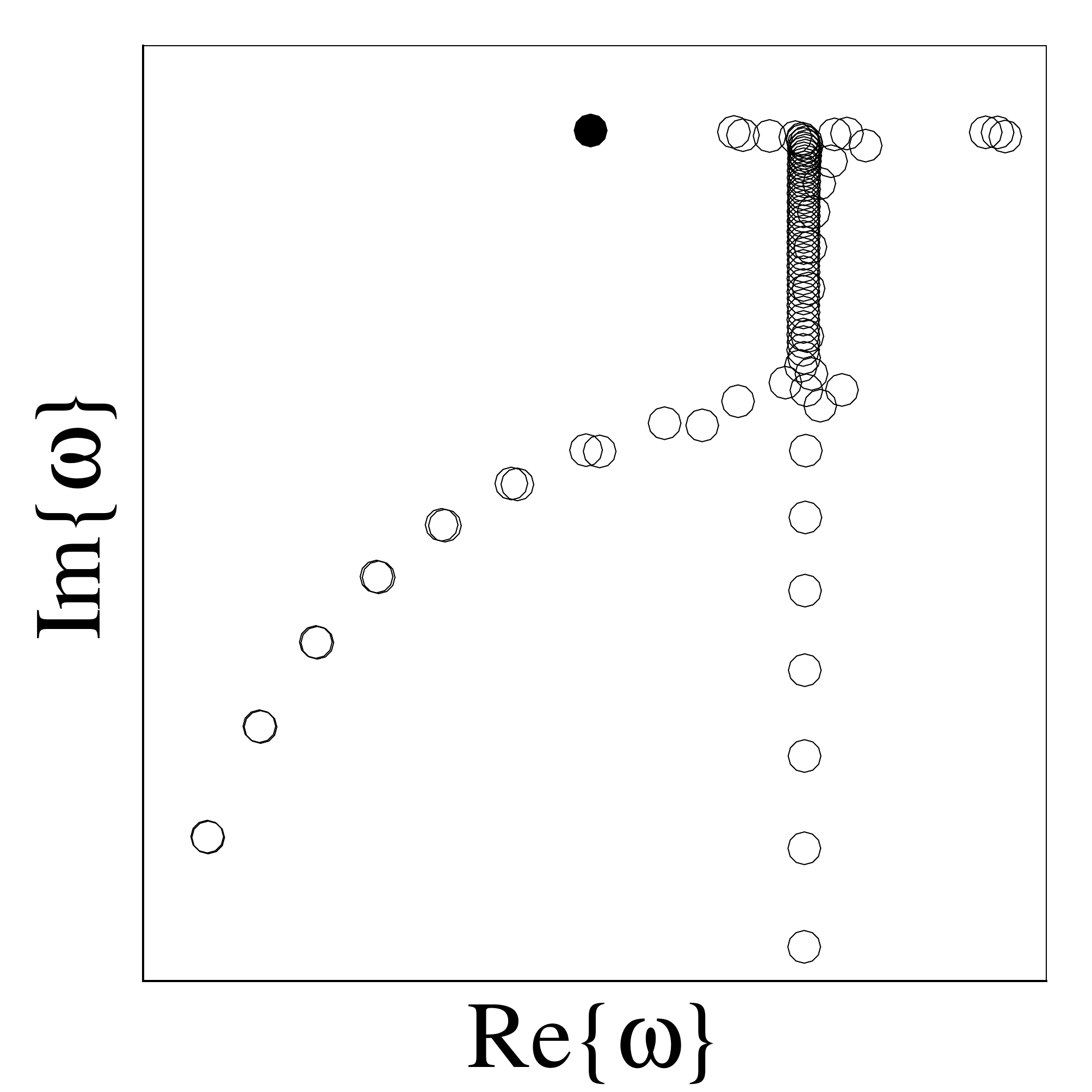}&
       \includegraphics[width=0.24\columnwidth]{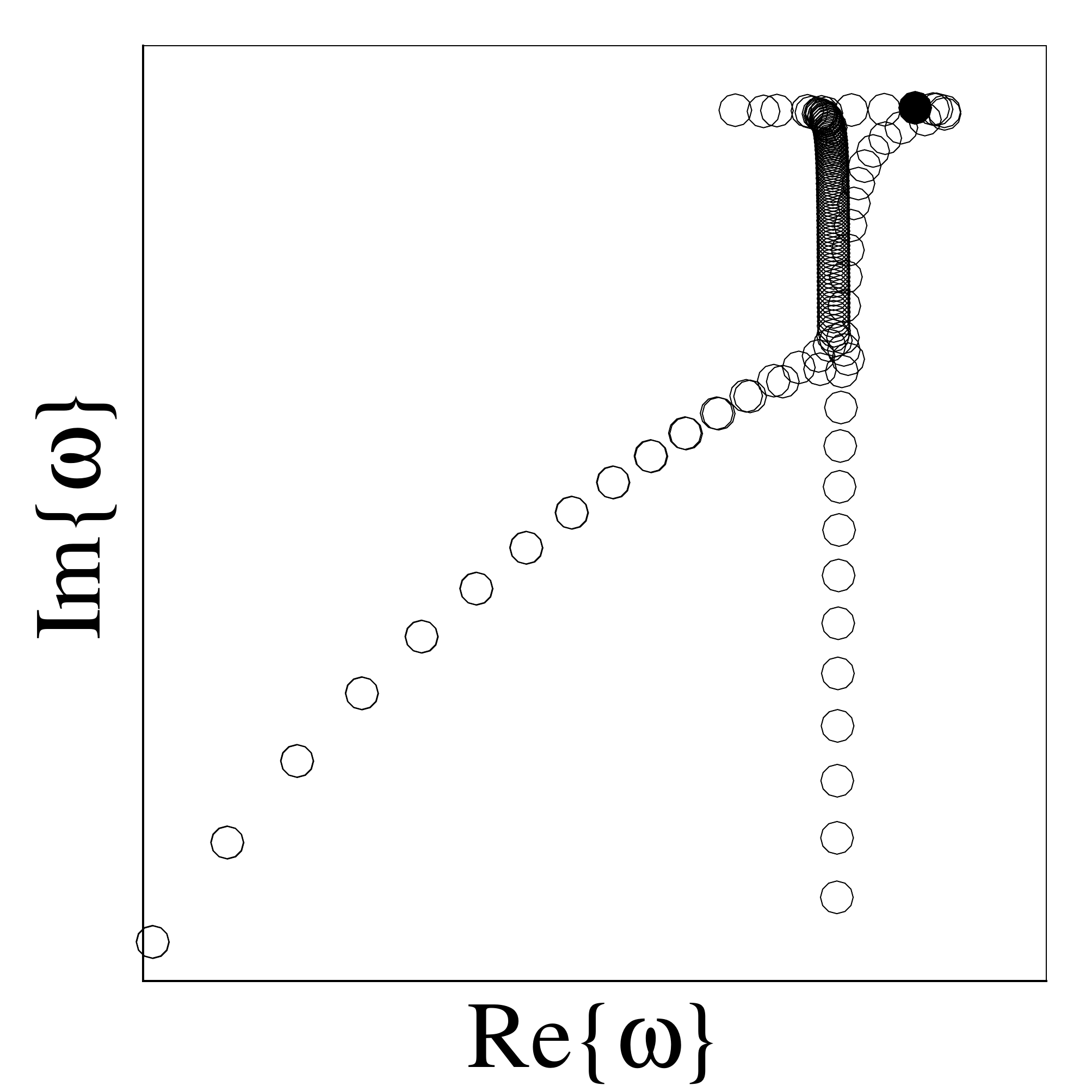}&
       \includegraphics[width=0.24\columnwidth]{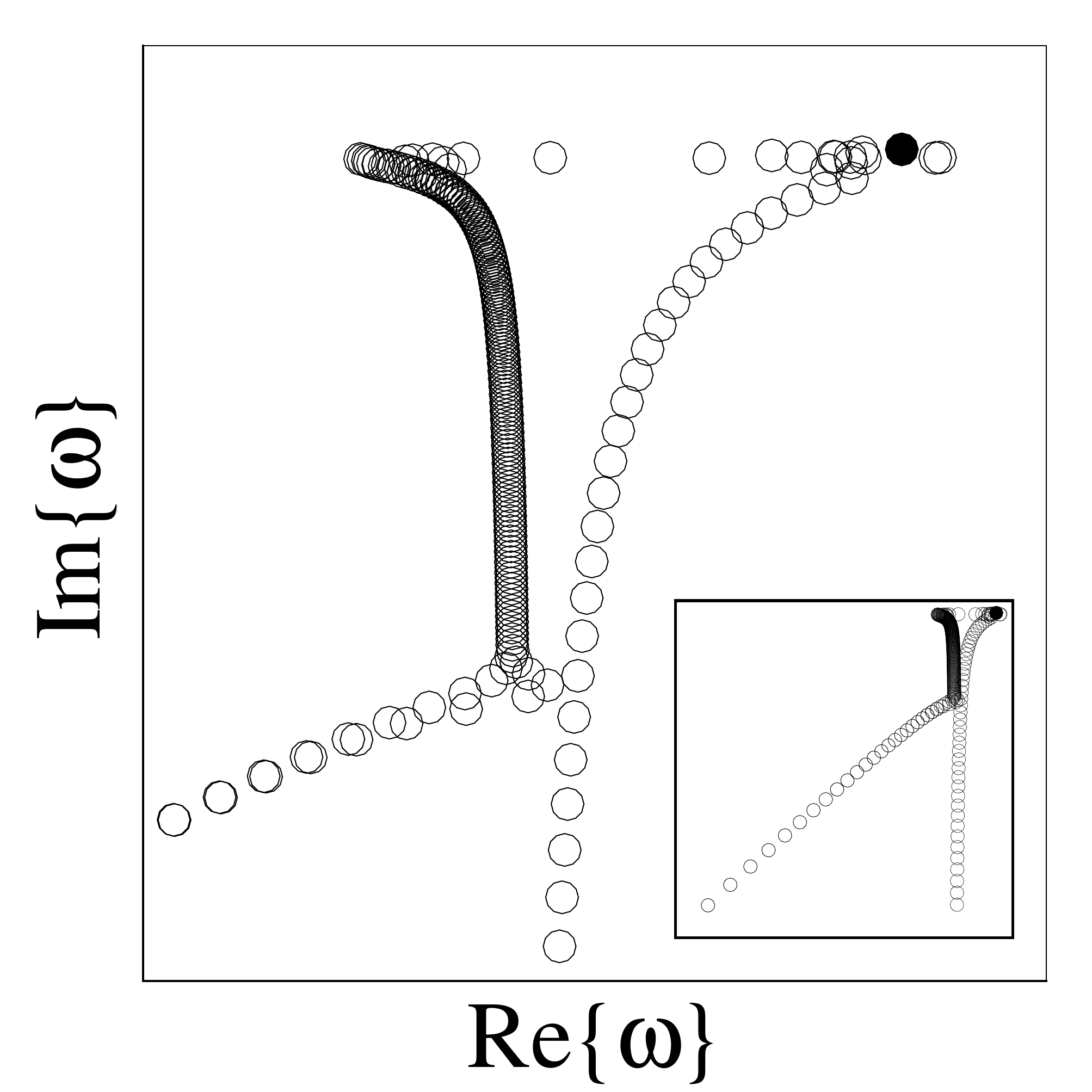}&
       \includegraphics[width=0.24\columnwidth]{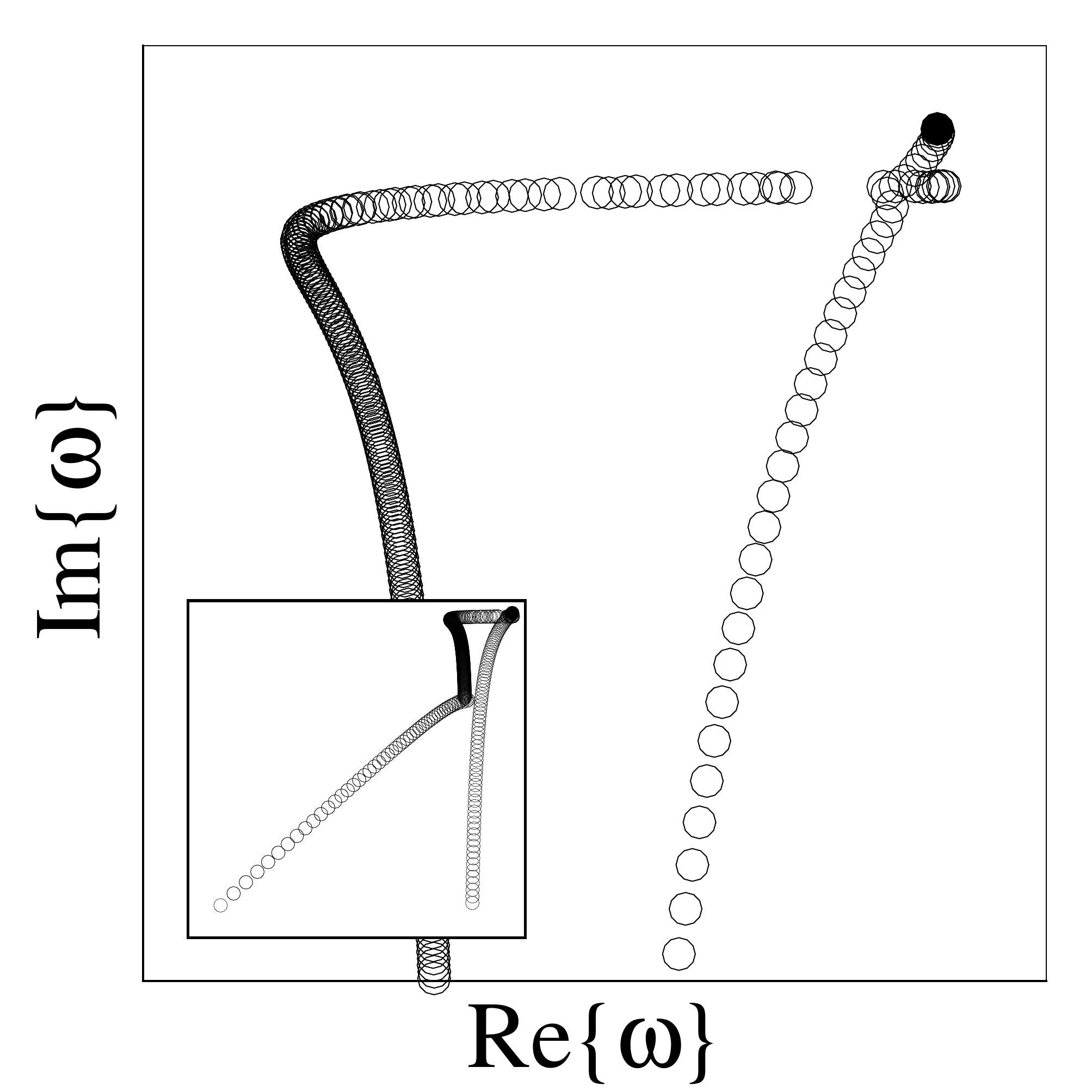}
     \end{tabular}
  \caption{(left)~Vorticity and (right)~temperature perturbation fields at the onset for $\Rey=300$ with (\textit{a})~$\HH=10$, (\textit{b})~$\HH=100$ and (\textit{c})~$\HH=5000$. Dark and light shading represent negative and positive values, respectively. (\textit{d})~Eigenvalue spectra of $k_c$ for $\Rey=300$ with (\textrm{i})~$\HH=0$, (\textrm{ii})~$\HH=100$, (\textrm{iii})~$\HH=1\times10^3$ and (\textrm{iv})~$\HH=1\times10^4$.}\label{fig:H1_Ra1065d336_Re300_vort_k1d2418}
  \end{center}
\end{figure}

Figure~\ref{fig:H1_Rac_Re0to300_growth_lsa}(\textit{b}) presents the eigenvalue spectra corresponding to the wavenumber of the peak growth rate for $\Rey=0$ and $\Rey=300$ at $\Ray_c=432.793$ and $\Ray_c=1933.058$, respectively, for $\HH=10$. A single vertical branch situated at $\textrm{Re}\{\omega\}=0$ is obtained for $\Rey=0$, while additional branches exist at finite $\Rey$ with the leading eigenvalue located on the leftmost branch. Unlike for plane Poiseuille flow, the location of the leading eigenvalue does not solely determine the type of instability mode observed and is instead strongly dependent on both $\Rey$ and $\HH$. However, the leading eigenvalue is mostly located to the left of the vertical branch at low $\HH$, and shifts to the right side as $\HH$ is increased. Correspondingly, low $\HH$ flows demonstrate a mixed mode with disturbances forming in the interior and along the horizontal walls while high $\HH$ flows demonstrate dominant wall modes. This is portrayed in figures~\ref{fig:H1_Ra1065d336_Re300_vort_k1d2418}(\textit{a,b,c}) with the change in eigenvalue spectrum shown in figure~\ref{fig:H1_Ra1065d336_Re300_vort_k1d2418}(\textit{d}). Finally, for all finite $\Rey$ considered in this paper, the $\textrm{Re}\{\omega\}$ of the leading eigenvalue is always non-zero. This seems like an intuitive result. However, \citet{aujogue2015onset} studied the case of magnetoconvection in an infinite plane geometry with rotation and found stationary modes as the most unstable. The only stationary modes found in this study were of pure \RB\ convection cells (\ie\ $\Rey=0$).

\subsubsection{Critical Rayleigh numbers for $\Rey\geq350$}\label{subsubsec:fixed_Re350_flows}

\begin{figure}
  \begin{center}
     \begin{tabular}{cc}
       \multicolumn{1}{l}{(\textit{a})}& \multicolumn{1}{l}{(\textit{b})~$\Ray=2470$}\\
       \includegraphics[width=0.475\columnwidth]{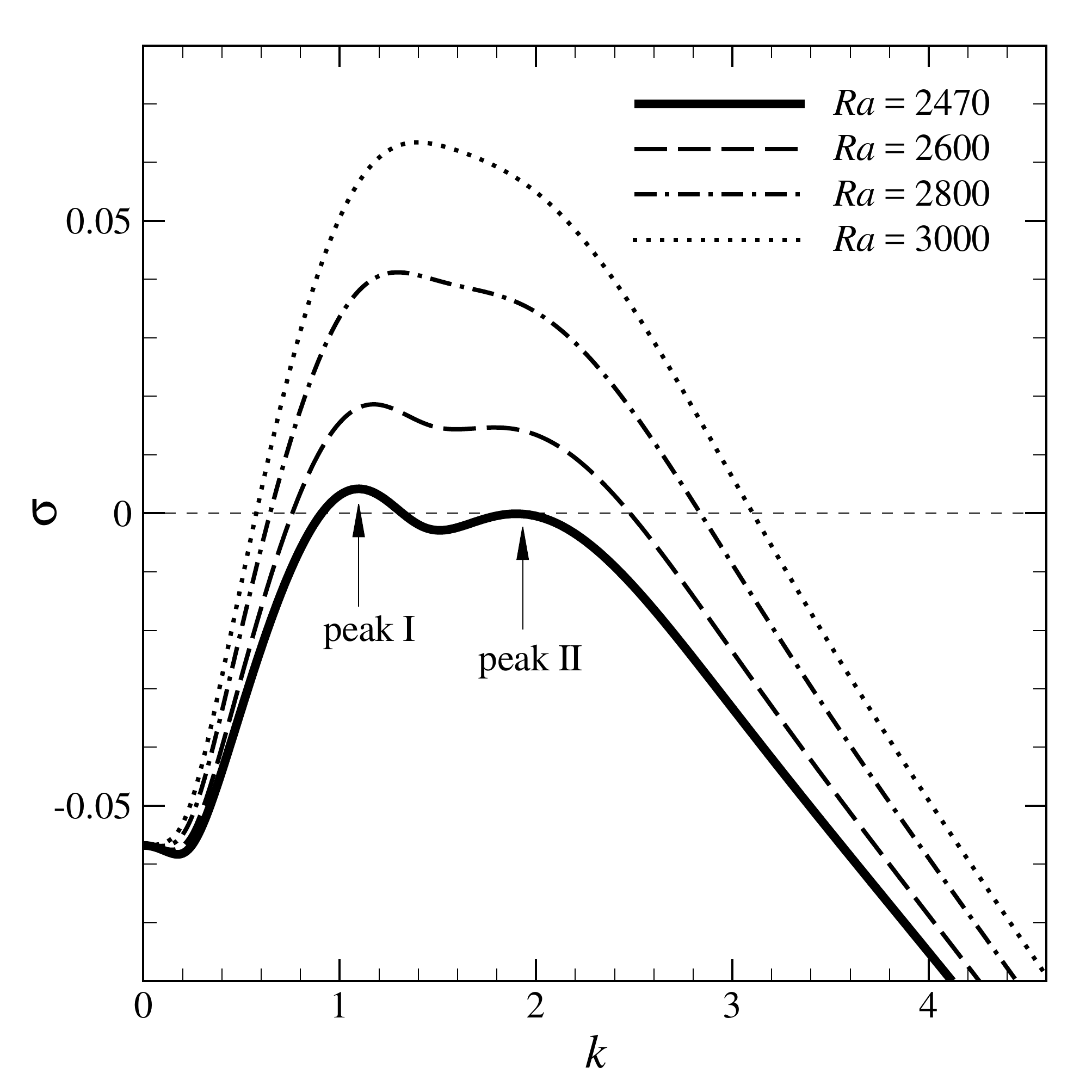}&
       \includegraphics[width=0.475\columnwidth]{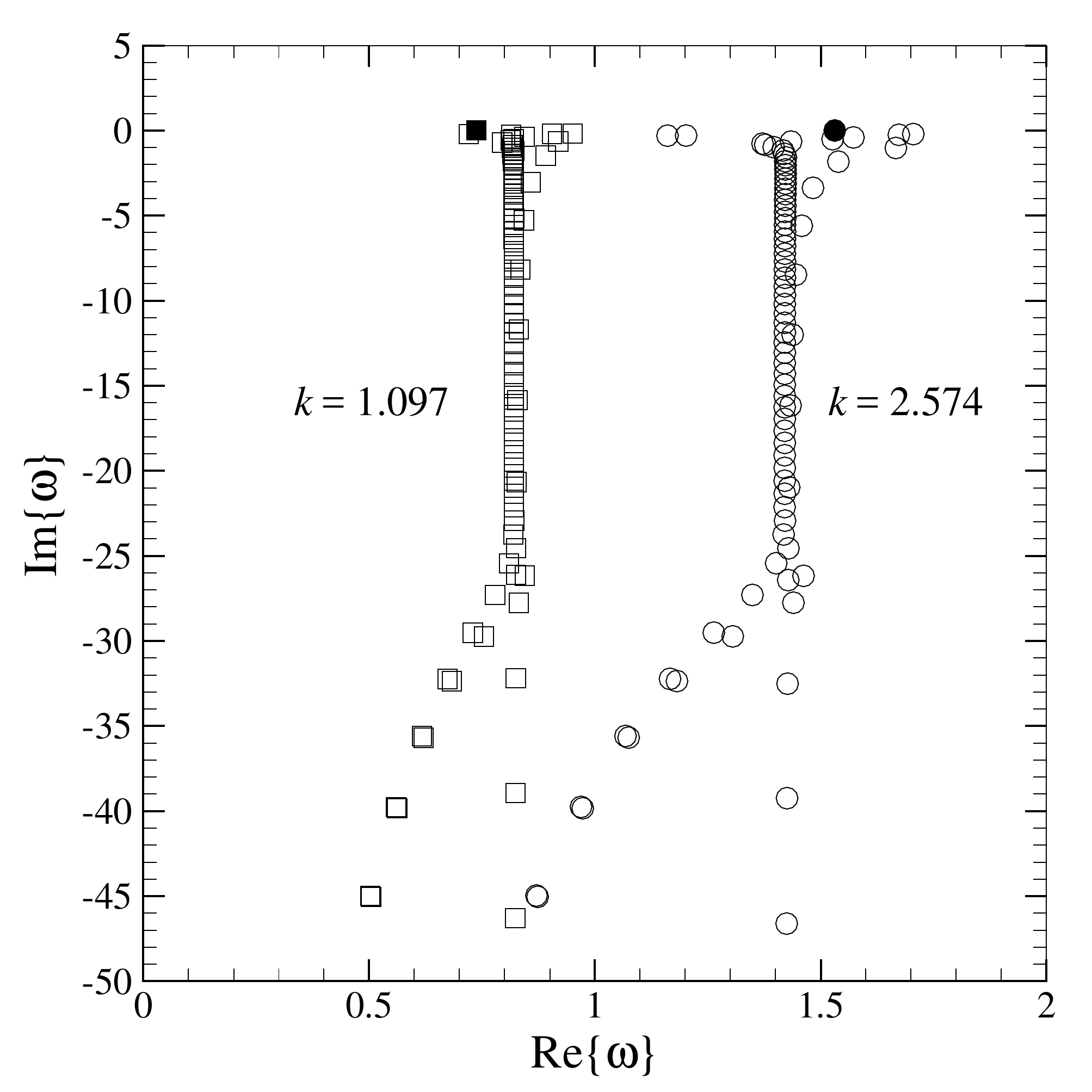}\\
       \multicolumn{1}{l}{(\textit{c})~$\Ray=2470$}\\
       \multicolumn{1}{l}{(\textrm{i})~$k=1.097$ (peak I)}\\
       \includegraphics[width=0.475\columnwidth]{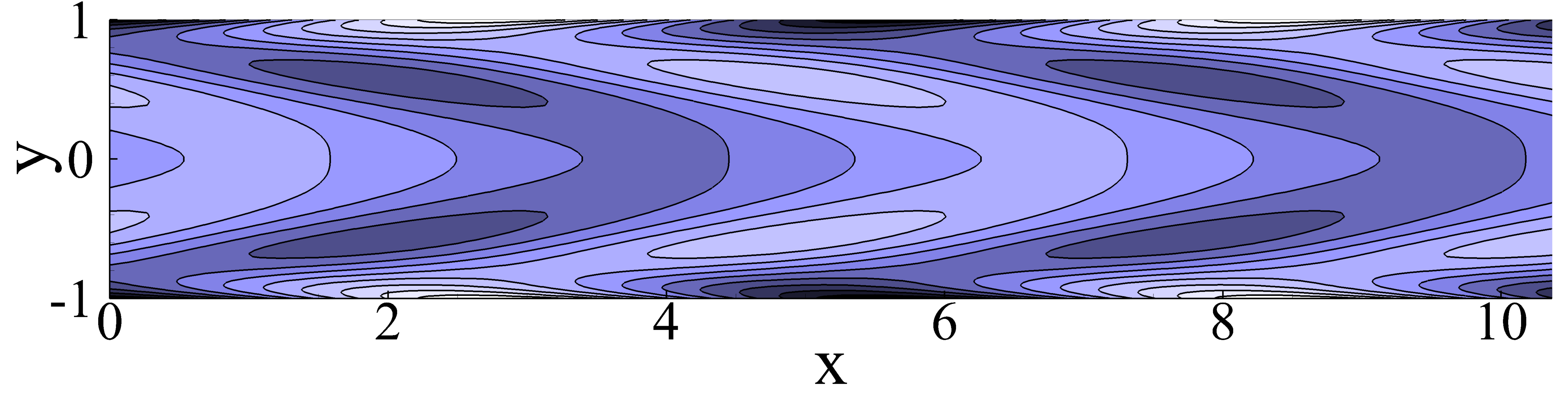} &
       \includegraphics[width=0.475\columnwidth]{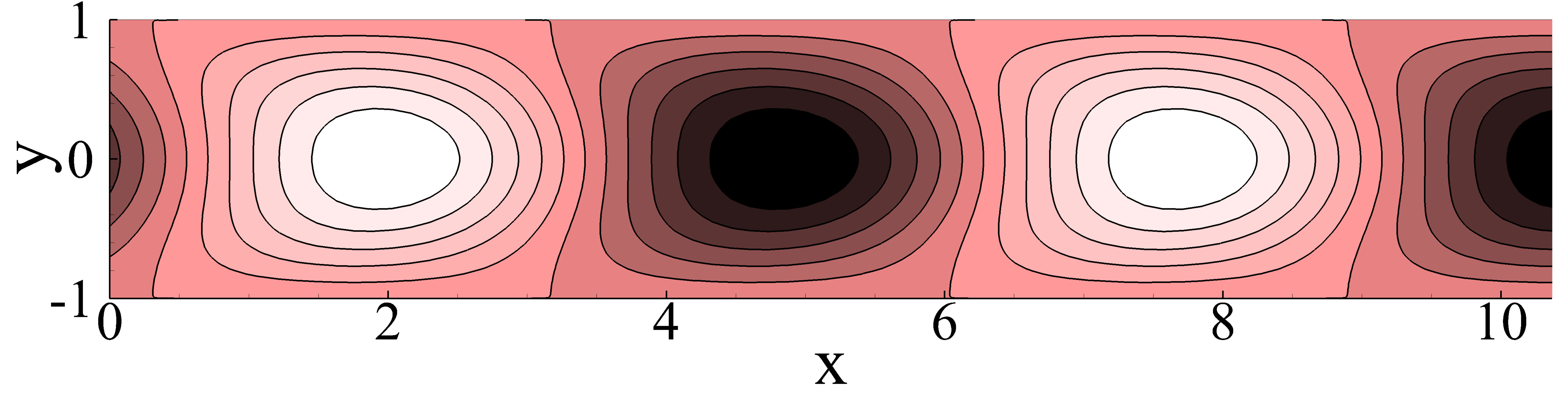}\\
       \multicolumn{1}{l}{(\textrm{ii})~$k=1.899$ (peak II)}\\
       \includegraphics[width=0.475\columnwidth]{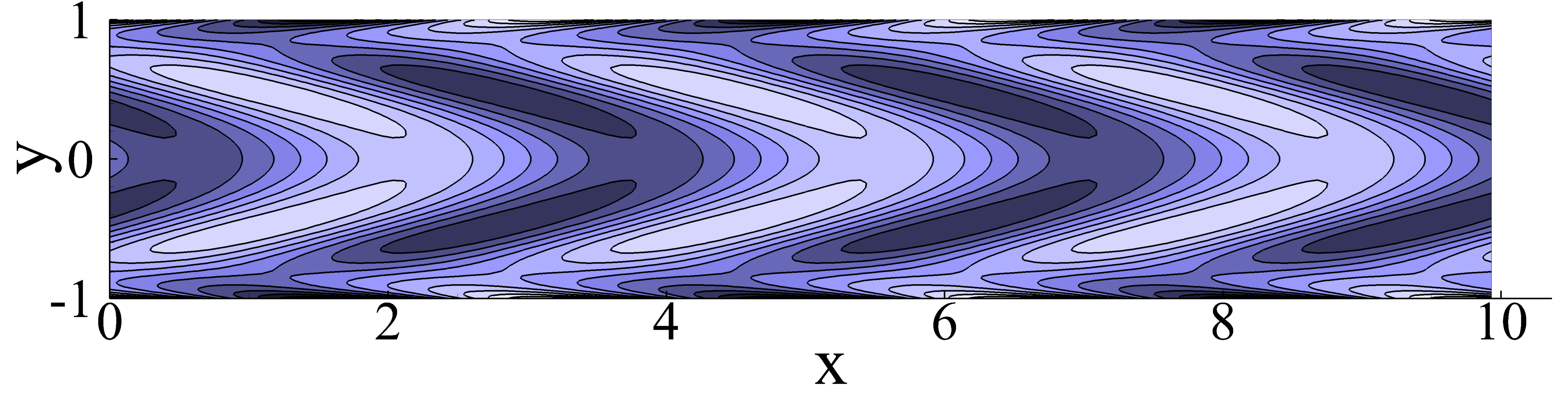} &
       \includegraphics[width=0.475\columnwidth]{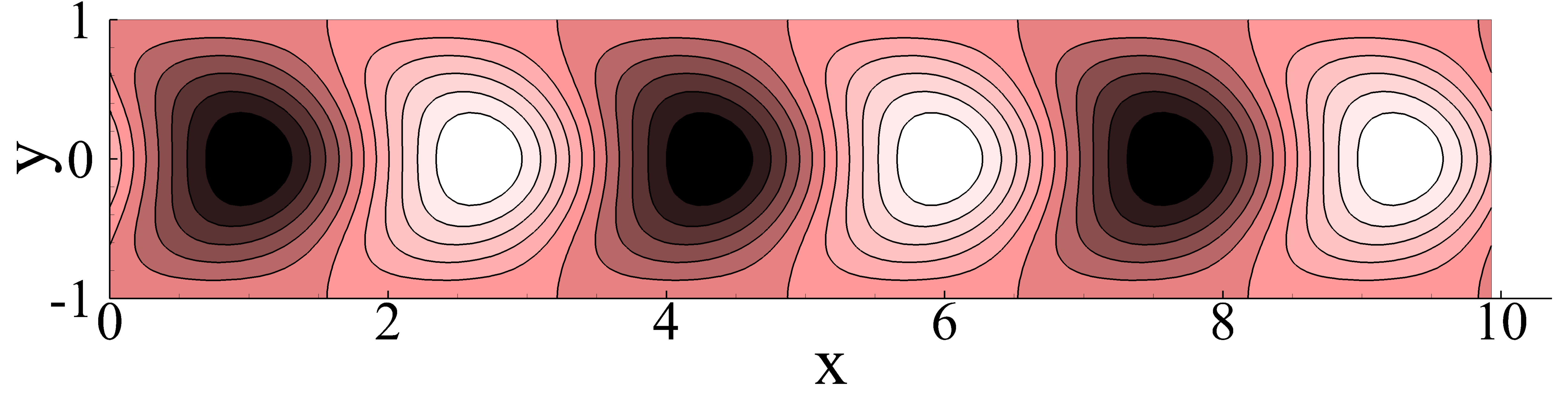}\\
       \end{tabular}
  \caption{(\textit{a})~Growth rate curves of various $\Ray=11405~(\square)$, $1\times10^4~(\triangle)$, $2\times10^4~(\diamond)$ and $3\times10^4~(\circ)$ for $\Rey=350$ and $\HH=10$. (\textit{b})~The eigenvalue spectra for $k=0.974$ (peak I) and $k=2.574$ (peak II) with $\Rey=350$, $\Ray=2470$ and $\HH=10$. (\textit{c})~The vorticity (left panels) and temperature (right panels) perturbation fields for (\textrm{i})~$k=1.097$ and (\textrm{ii})~$k=1.899$. Dark and light shading represent negative and positive values, respectively. }\label{fig:H0d1_Ra11405_Re1000_growth_lsa}
  \end{center}
\end{figure}

Increasing the Reynolds number above $\Rey=350$ sees the emergence of a second unstable mode (local maximum with $\sigma>0$) for $\Ray$ sufficiently larger than $\Ray_c$, in contrast with cases for $\Rey<350$. Examples of a second unstable local maximum for $\Rey=350$ are shown in figure~\ref{fig:H0d1_Ra11405_Re1000_growth_lsa}(\textit{a}). The instability modes associated with the low and high wavenumber peaks are labelled ``peak I'' and ``peak II'', respectively. The eigenvalue spectra for the wavenumbers corresponding to the maxima with $\Ray=2470$ are portrayed in figure~\ref{fig:H0d1_Ra11405_Re1000_growth_lsa}(\textit{b}). Hence, although A, P and S branches are clearly distinct in the eigenvalue spectra, their corresponding modes portray relatively similar structures exhibiting mixed mode features (\ie\ wall and interior). The typical structures of the instability modes are shown in figures~\ref{fig:H0d1_Ra11405_Re1000_growth_lsa}(\textit{c}).


When increasing $\Ray$ (\eg\ $\Ray=2600$ and higher), peak I recedes while peak II emerges (see figure~\ref{fig:H0d1_Ra11405_Re1000_growth_lsa}\textit{a}). The peak II wavenumbers eventually dominate, leading to a single peak at $\Ray=3\times10^3$. The most unstable wavenumber of $k=1.396$ at $\Ray=3\times10^3$ exhibits a mixed mode instability with its leading eigenvalue positioned on the right-side branch of the eigenvalue spectrum similar to the peak~II eigenvalue spectrum shown in figure~\ref{fig:H0d1_Ra11405_Re1000_growth_lsa}(\textit{b}). Further increasing the Rayleigh number causes the vorticity perturbations to only grow along the side wall, recovering a state similar to figure~\ref{fig:H1_Ra0_Re10033d15_vort}(\textit{d}).

\begin{figure}
  \begin{center}
     \begin{tabular}{cc}
     \multicolumn{1}{l}{(\textit{a})} & \multicolumn{1}{l}{(\textit{b})}\\
       \includegraphics[width=0.475\columnwidth]{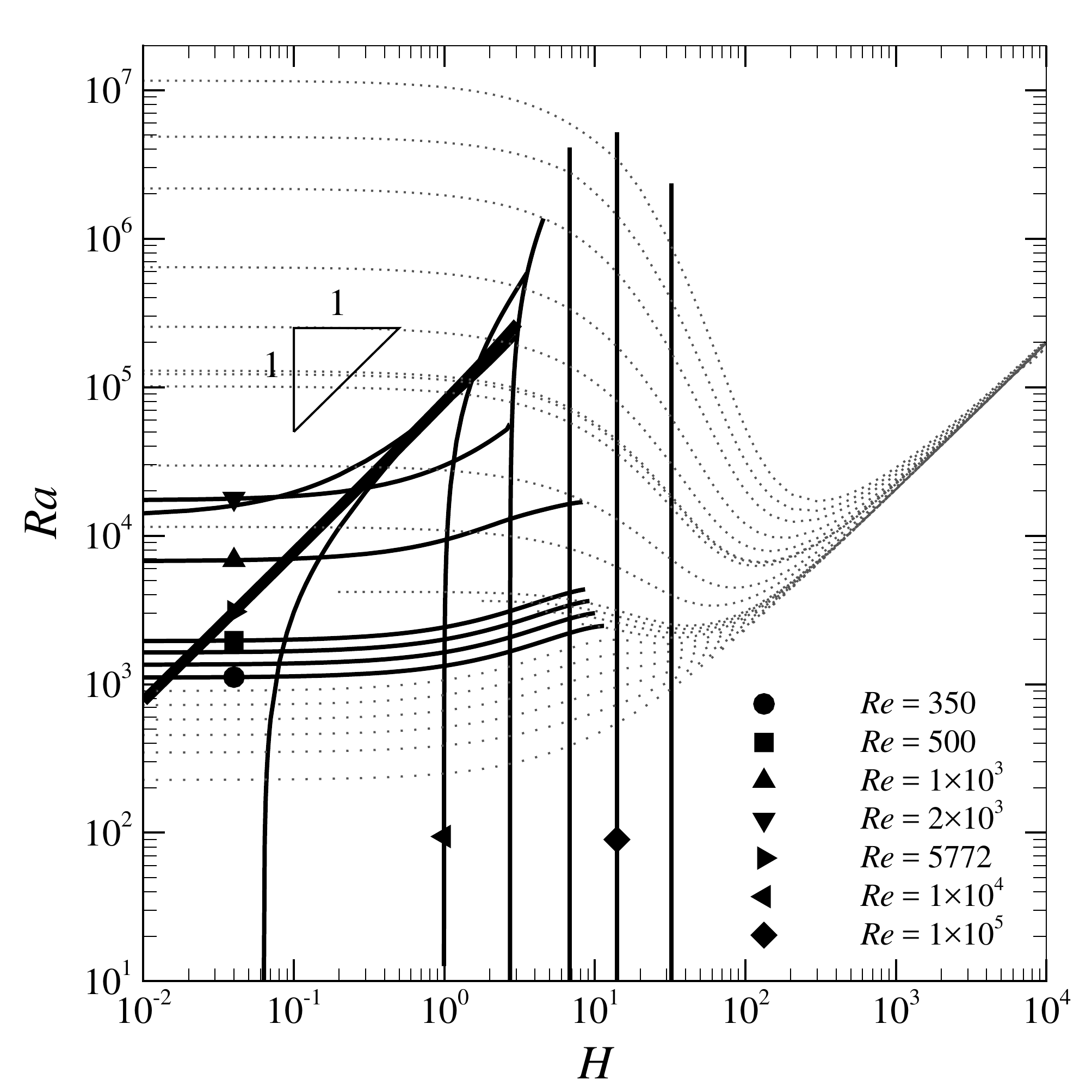}&
       \includegraphics[width=0.475\columnwidth]{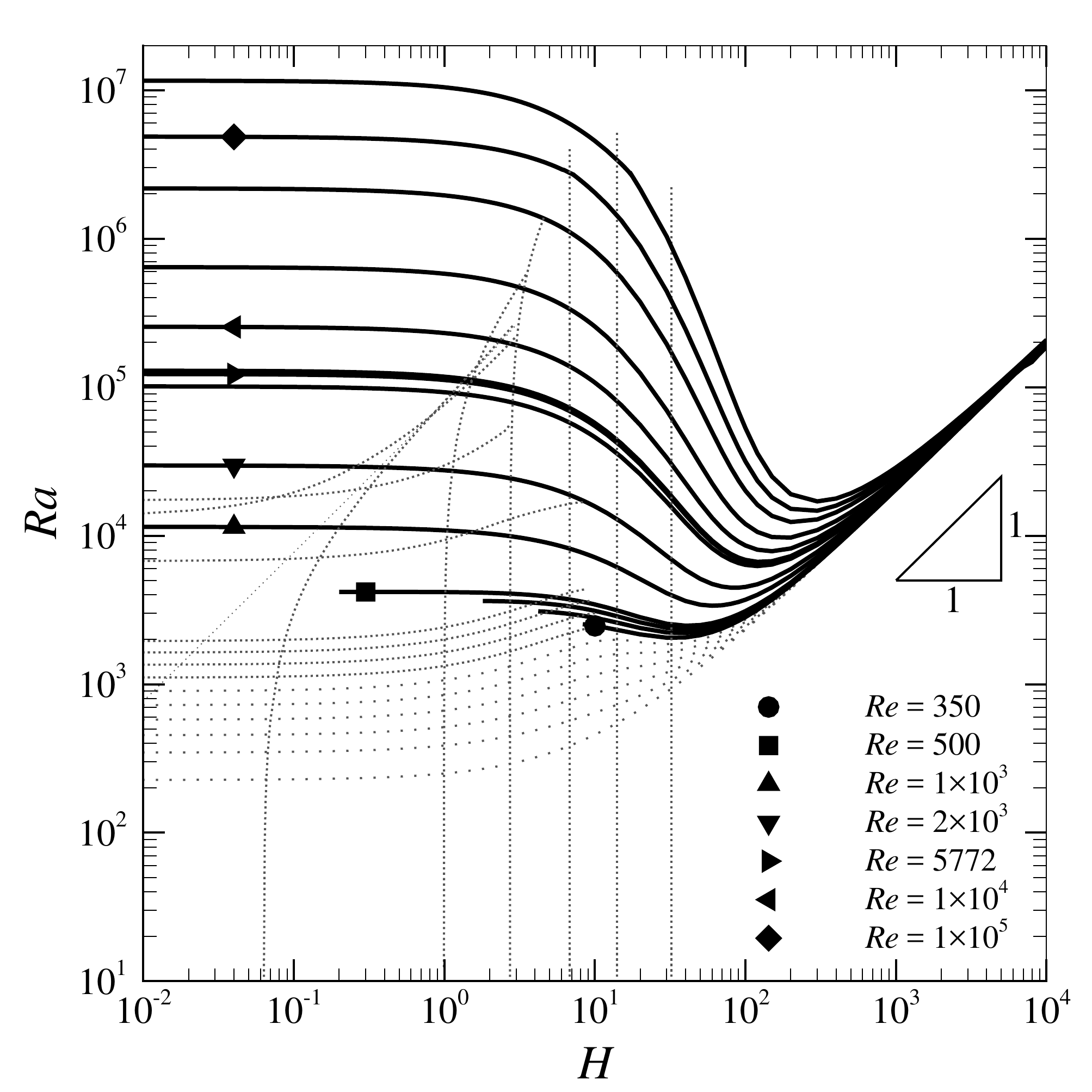}
       \end{tabular}

       \begin{tabular}{c}
     \multicolumn{1}{l}{(\textit{c})}\\
       \includegraphics[width=0.95\columnwidth]{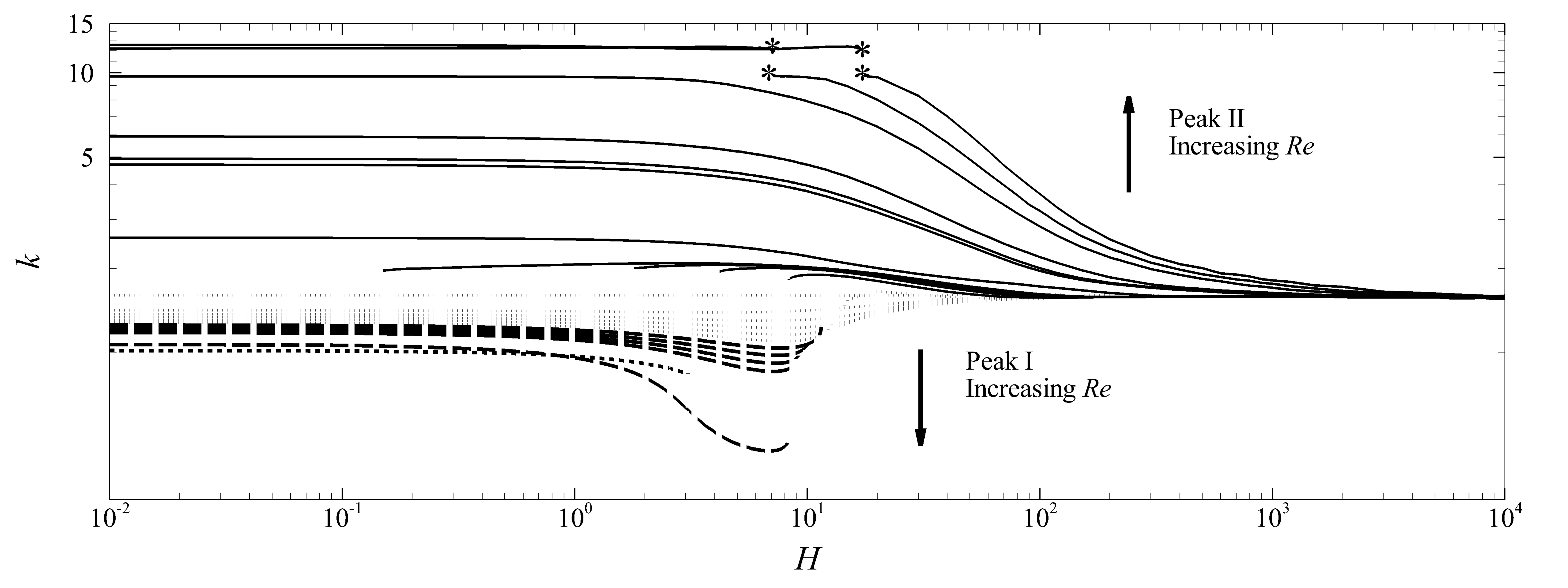}
     \end{tabular}
  \caption{Critical Rayleigh numbers for the onset of modes corresponding to (\textit{a})~peak I and (\textit{b})~peak II for $0\leq\Rey\leq2\times10^5$. The symbols identify the Reynolds numbers for each curve in panels (\textit{a,b}). The dotted lines are provided for guidance, representing the alternate instability (peak II in panel~\textit{a} and peak I in panel~\textit{b}) and the single instability for $0\leq\Rey\leq300$. The asterisks identify the higher-peak II wavenumber mode (lower $\HH$) and lower-peak II wavenumber mode (higher $\HH$). (\textit{c})~The corresponding critical wavenumbers $k_c$ with long-dashed and solid lines representing peak I and II instabilities, respectively. Data for $\Rey=350, 400, 450, 500, 1\times10^3, 2\times10^3, 5\times10^3, 5772, 6\times10^3, 1\times10^4, 2\times10^4, 5\times10^4, 1\times10^5, 2\times10^5$ are shown.}\label{fig:Rac_all}
  \end{center}
\end{figure}

\begin{figure}
  \begin{center}
     \begin{tabular}{cc}
     \multicolumn{1}{l}{(\textit{a})} & \multicolumn{1}{l}{(\textit{b})}\\
       \includegraphics[width=0.475\columnwidth]{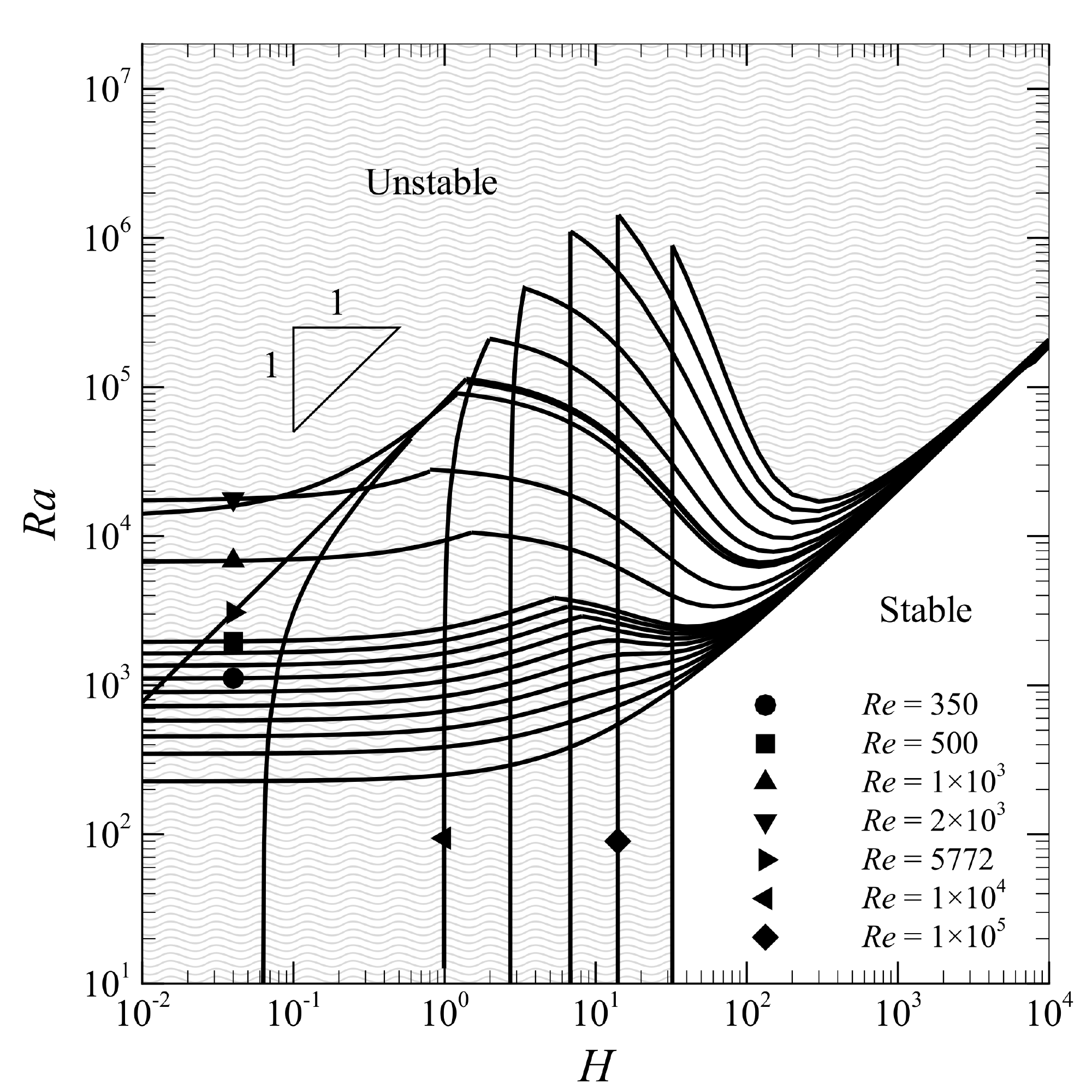}&
       \includegraphics[width=0.475\columnwidth]{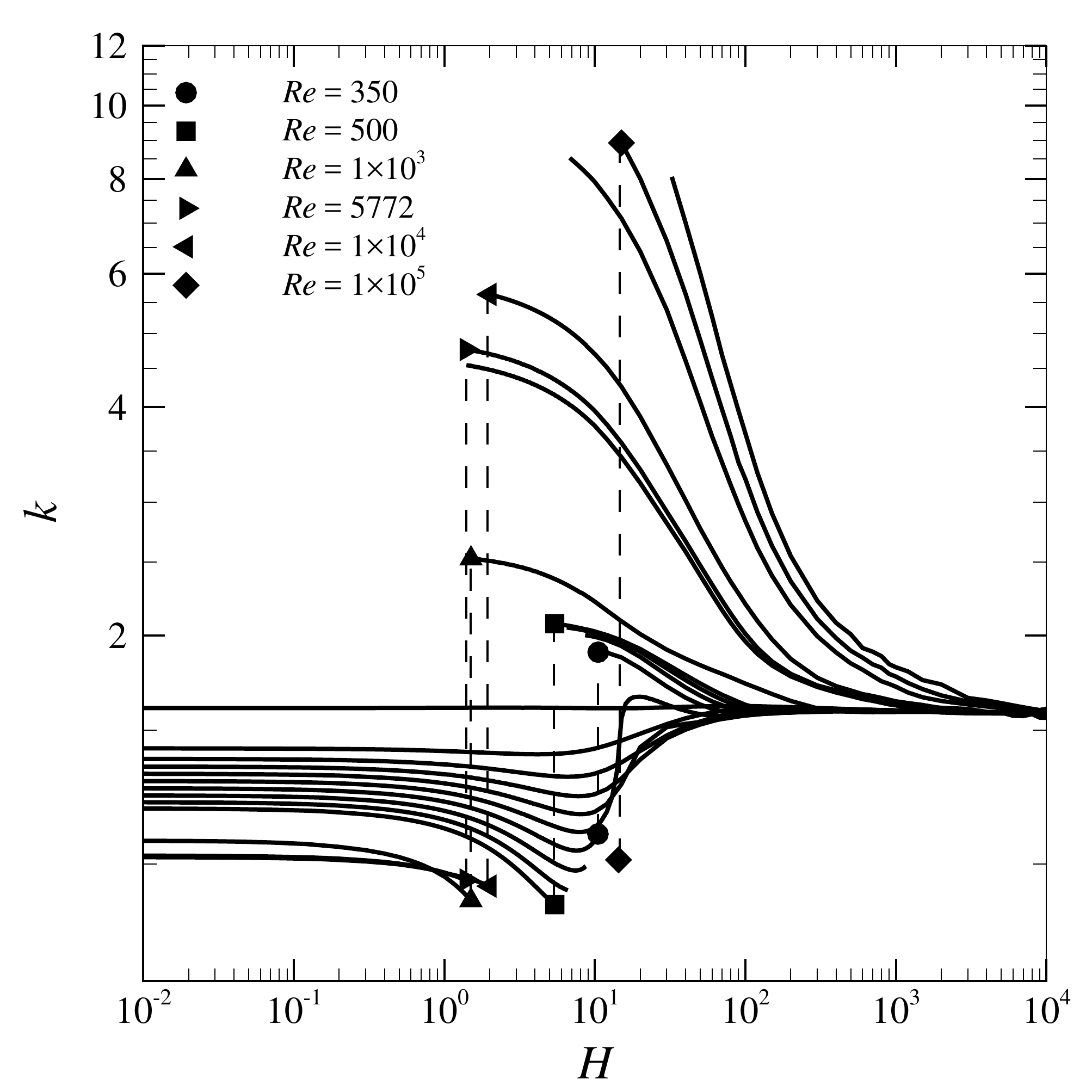}
       \end{tabular}
       \caption{(\textit{a})~Neutral stability of $\Ray_c$ as a function of $\HH$ for fixed Reynolds numbers $0\leq\Rey\leq2\times10^5$ and (\textit{b})~corresponding $k_c$ curves are presented. Unstable conditions are represented by flow conditions above and to the left of all lines for panel~(\textit{a}). The symbols identify the Reynolds numbers for each curve. The dashed lines mark the discontinuous jump in wavenumber for curves highlighted with symbols. Shaded region represents the unstable flow conditions for $\Rey=2\times10^5$.}\label{fig:Rac_all_neutral}
  \end{center}
\end{figure}

The onset of peak I and II instabilities for $\Rey\geq350$ are highlighted separately in figures~\ref{fig:Rac_all}(\textit{a,b}), respectively. It is important to note that these curves describe the onset of individual instability modes rather than the neutral stability of a fixed $\Rey$ flow. Figure~\ref{fig:Rac_all_neutral}(\textit{a}) illustrates this for all fixed Reynolds numbers considered in this study with the corresponding critical wavenumbers shown in panel~(\textit{b}). Consequently, there are discontinuous jumps (marked by dashed lines) in the critical wavenumber curves when the most unstable mode switches from one mode to the other. Similar jumps in unstable critical modes were observed experimentally \citep{nakagawa1957experiments,aujogue2016little} and theoretically predicted \citep{chandrasekhar1961hydrodynamics,aujogue2015onset} for the onset of \RB\ convection subjected to the influence of an external magnetic field and to background rotation. In such systems, a magnetic mode with long wavelength competes with a short wavelength viscous mode. The dominant mode is determined by the dominant force, with a jump in critical wavelength when the balance between these forces reverses.

Referring back to figure~\ref{fig:Rac_all}(\textit{a,b}), at $\Rey=350$ the curve for the onset of peak I only spans the low $\HH$ regime while peak II is present only in the intermediate and high $\HH$ regimes. With increasing Reynolds number up to $\Rey=10^3$, peak II extends its presence into lower $\HH$ and eventually covers the entire range of $\HH$ investigated. In contrast, the onset of peak I remains limited to $\HH\lesssim1$. This is because the increase in $\HH$ leads to an increase in $\Ray_c$ which results in a wider range of wavenumbers becoming associated with peak II. The transition from peak I dominant wavenumbers to peak II dominance is also visible in the growth rate curves in figure~\ref{fig:H0d1_Ra11405_Re1000_growth_lsa}(\textit{a}). For the higher range of fixed $\Rey$ investigated in this paper, the curve for the onset of peak II continues to maintain its profile with increasing $\Rey$ throughout the entire $\HH$ regime. It turns out that the minima in the curves for the onset of peak II represent an important transition point. Peak II instabilities exhibit mixed modes for $\HH$ below this transition point and wall modes for $\HH$ above it. Also, at high $\HH$ values where $\Ray_c$ becomes independent of $\Rey$, the phase speed of the peak II instability exhibits a linear dependence on the wavenumber (\ie\ $\textrm{Re}\{\omega\}\sim k$).

Since the critical curves for peak II do not change significantly with $\Rey$, we shall now focus on the peak I instability. A growing dependence on $\HH$ is observed as the Reynolds number is increased between $5\times10^3\leq\Rey<10^4$. This is demonstrated by the transition from a near horizontal curve at $\Rey=5\times10^3$ to a vertical curve at $\Rey=10^4$ (figure~\ref{fig:Rac_all}\textit{a}). This transition is a result of shear becoming more significant, thereby disrupting the balance between buoyancy and dissipation. Ultimately, at sufficiently high $\Rey$, the onset of the peak I instability is governed by the balance between shear and Hartmann friction, and becomes independent of $\Ray$. The critical value of $\HH$ and the independence of $\Ray$ agree with \citet{potherat2007quasi} (see figure~\ref{fig:Rec_Ra0}\textit{a}). The onset of peak I still matches that of $\Ray=0$ as the Reynolds number is further increased beyond $\Rey=10^4$. The dominance of shear above $\Rey=10^3$ is supported by the change in eigenvalue spectrum structure of the critical wavenumber. In the shear-dominated state, the eigenvalue spectrum portrays the distinct A, P and S branches that are exhibited by plane \Poi\ flow.

Interestingly, the critical Rayleigh number for the onset of peak I instability develops a linear dependence with $\HH$ at $\Rey\approx5772$, which is the critical $\Rey$ value for linear instability of plane \Poi\ flow. For $\Rey\lesssim5772$ the flow becomes unstable to the peak I instability through the onset of \RB-type modes above the critical Rayleigh number at low $\HH$. However, for $\Rey\gtrsim5772$, the flow becomes unstable to the peak I instability via plane \Poi\ instability below $\HH_c$ for all $\Ray>0$. As $\Rey\rightarrow5772$, the critical $\Ray$ and $\HH$ values become smaller to approach the point ($\Ray=0$, $\HH=0$) in the $\Ray$--$\HH$ parameter space. The linear scaling describing this neutral stability reflects the balance between the destabilising thermal forcing ($\Ray$) and stabilising Hartmann friction ($\HH$) (see Sec.~\ref{subsec:Re0_Ra0_flows}).

The critical wavenumbers for fixed $0\leq\Rey\leq2\times10^5$ are shown in figure~\ref{fig:Rac_all}(\textit{c}) for both peak I and peak II instabilities. All peak I wavenumbers were found to be smaller than the baseline value of $k\approx1.5857$ while peak II wavenumbers were larger. As $\HH\rightarrow\infty$, the peak II wavenumbers asymptote towards the Reynolds-number-independence (\ie\ $\Rey=0$). Increasing the Reynolds number leads to a decrease and increase in wavenumber for peaks I and II, respectively. However, the peak I wavenumbers eventually become constant as the system develops a strong and exclusive dependence on $\HH$. This figure explains the discontinuous jumps in preferred wavenumber in the neutral stability curves shown in figure~\ref{fig:Rac_all_neutral}(\textit{b}).

\begin{figure}
  \begin{center}
     \begin{tabular}{ccc}
       \multicolumn{1}{l}{(\textit{a})} & \multicolumn{1}{l}{(\textit{b})}  & \multicolumn{1}{l}{(\textit{c})}\\
       \includegraphics[width=0.32\columnwidth]{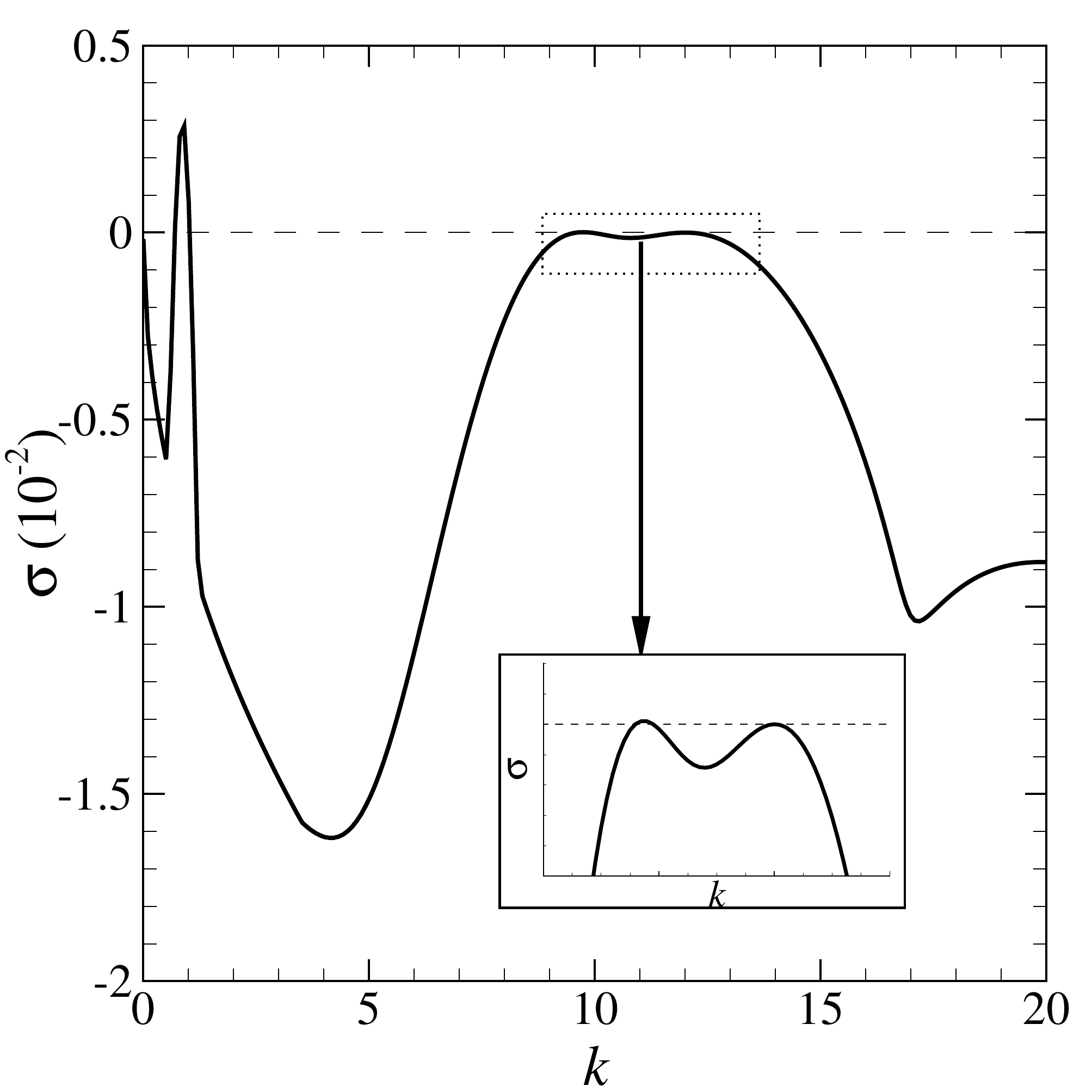}&
       \includegraphics[width=0.32\columnwidth]{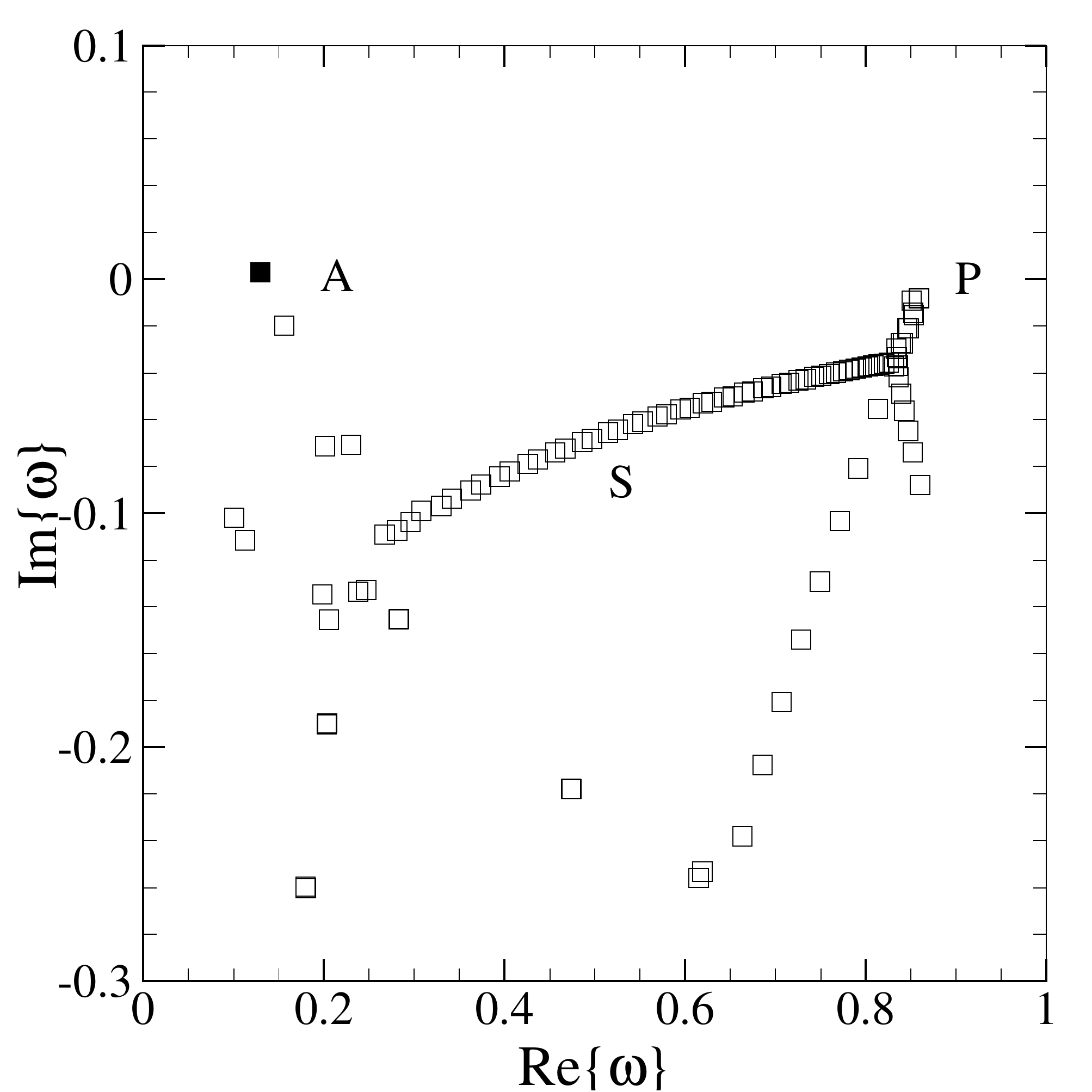}&
       \includegraphics[width=0.32\columnwidth]{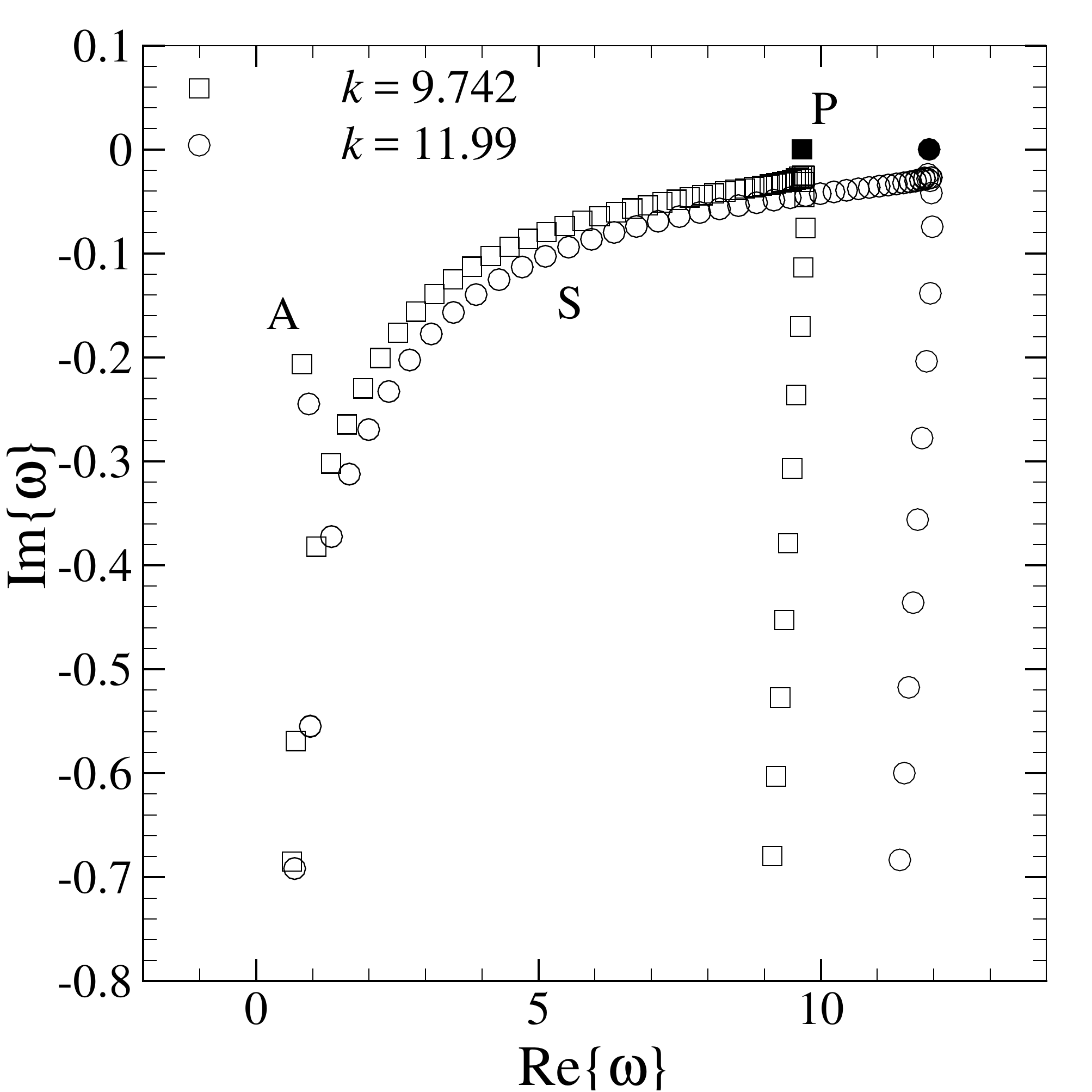}
     \end{tabular}
  \caption{(\textit{a})~Growth rate curve for $\HH=7.2$, $\Rey=10^5$ and $\Ray=2.6934\times10^6$. The eigenvalue spectra of (\textit{b})~$k_c=0.867$, and (\textit{c})~$k_c=9.742$~($\square$) and $k_c=11.99$~($\circ$). The solid symbols represent the leading eigenvalue.}\label{fig:H7d2_Ra2d6933e6_Re10000_growth_lsa}
  \end{center}
\end{figure}

Further increasing the Reynolds number beyond $\Rey=1\times10^5$ results in the peak II wavenumbers developing two maxima in the growth rate curve as shown in figure~\ref{fig:H7d2_Ra2d6933e6_Re10000_growth_lsa}(\textit{a}). The two newly developed growth rate peaks demonstrate characteristics that are consistent with what has been described for peak II previously. The eigenvalue spectra for each of the three growth rate peaks ($k=0.867, 9.742, 11.99$) are illustrated in panels~(\textit{b,c}). At this Reynolds number, the peak I instability is significantly affected by the shear. In fact, the Rayleigh number no longer has an influence on the onset of the peak I instability (\ie\ the curve is vertical). Thus, the peak I instability now represents a wall mode rather than the mixed mode observed at lower $\Rey$. The corresponding critical wavenumbers curves for this double peak II mode are shown in figure~\ref{fig:Rac_all}(\textit{c}). The discontinuity in $k_c$ over the range of $7\lesssim\HH \lesssim11$ for $\Rey=10^5$ and $2\times10^5$ is caused by a switch in dominance from the higher wavenumber peak associated with peak II to the lower one as $\HH$ increases. However, when considering the full neutral stability, the higher-wavenumber peak II instability is never the most unstable (see figure~\ref{fig:Rac_all_neutral}\textit{b}).

Investigating the stability of fixed Reynolds number flows has revealed a transition from the neutral stability being described by a single instability mechanism for $\Rey\leq300$ to the onset of multiple instability mechanisms for $\Rey\geq350$. The stability of the system is complex and has shown a preference to wall, center and mixed modes along the neutral stability curve of a fixed $\Rey$ flow. The type of mode observed is dependent on both $\Rey$ and $\HH$. A different progression of instability onset is observed with increasing Rayleigh numbers. Hence, the study of fixed $\Ray$ flows exposes an additional transition from a thermal-dominant instability to a shear-dominant instability which is discussed in the next section.

\subsection{Critical Reynolds number at finite Rayleigh numbers}\label{subsec:fixed_Ra_flows}
\begin{figure}
  \begin{center}
     \begin{tabular}{cc}
     \multicolumn{1}{l}{(\textit{a})}  &  \multicolumn{1}{l}{(\textit{b})}\\
     \includegraphics[width=0.475\columnwidth]{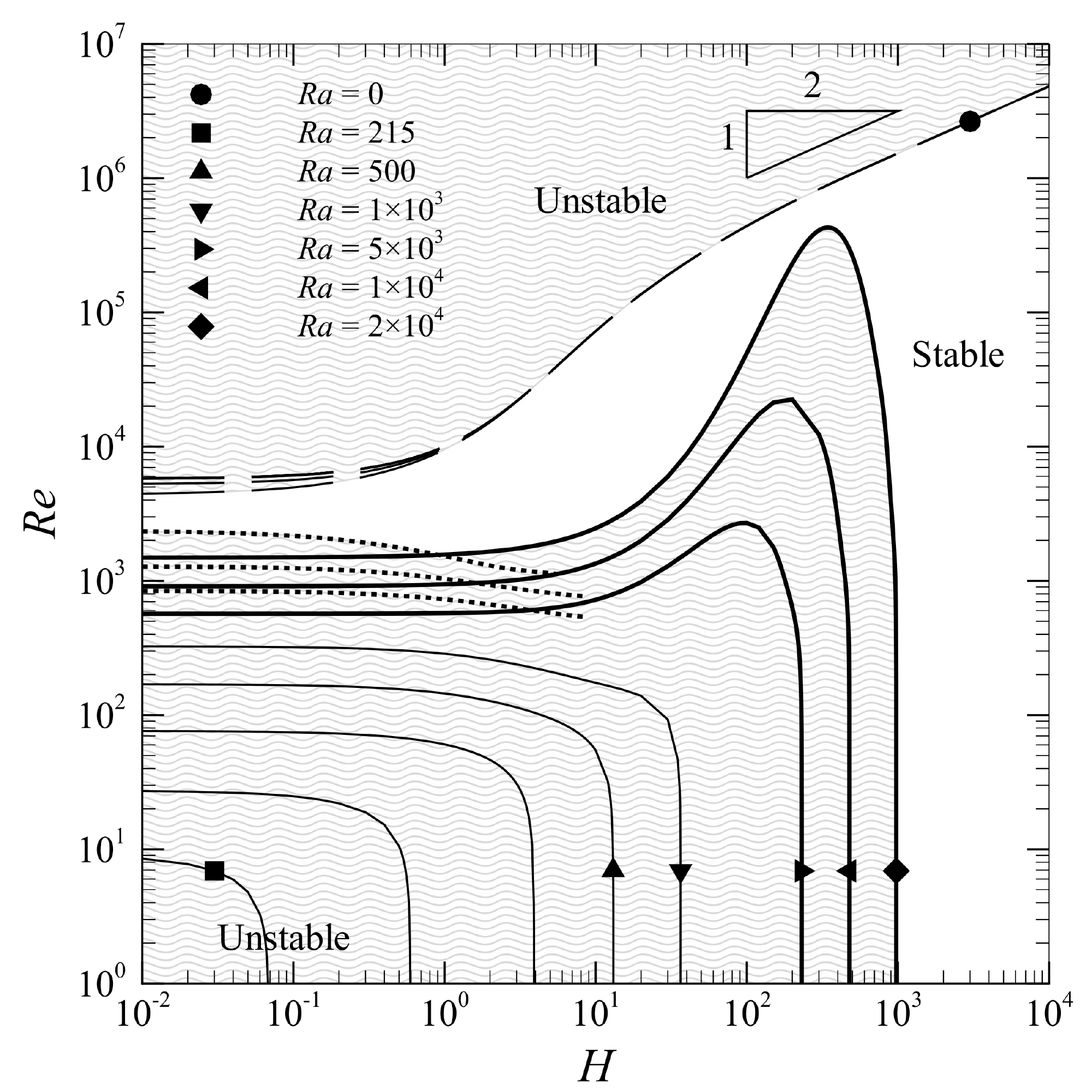}&
     \includegraphics[width=0.475\columnwidth]{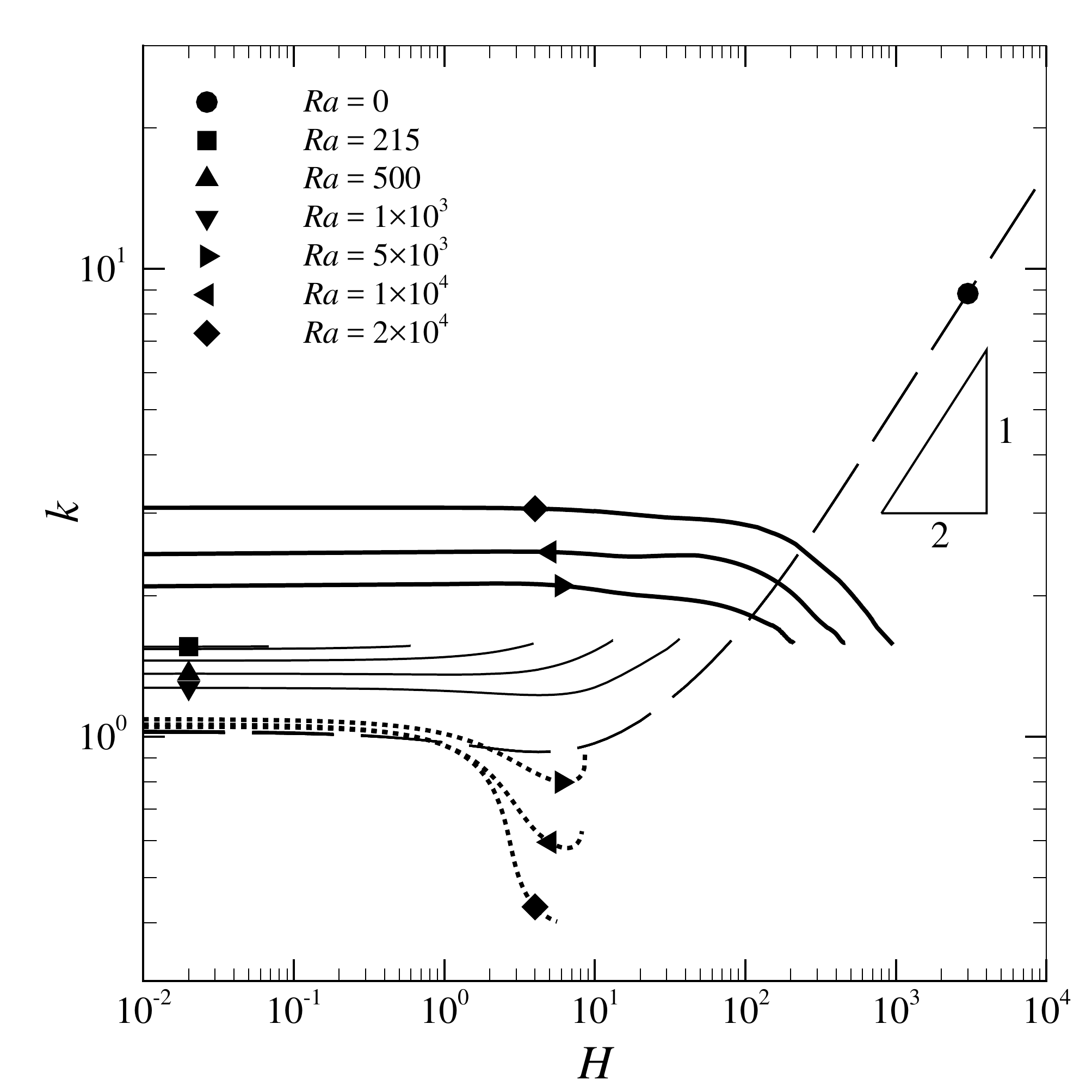}
     \end{tabular}
  \caption{Stability diagrams of (\textit{a})~$\Rey_c$ and (\textit{b})~$k_c$ as a function of $\HH$ for fixed Raleigh numbers $0\leq\Ray\leq2\times10^4$. The onset/suppression curves with long-dashed lines represent shear-dominant instability, thin-solid lines representing thermal-dominant instability, dotted lines representing a low-wavenumber instability (peak I) and thick-solid lines representing a high-wavenumber instability (peak II). The symbols identify the Rayleigh number for each curve. The curves for the onset of shear-dominant instability are only marked for $\Ray=0$ since the other curves are weakly sensitive over the range of $\Ray$ studied. Unstable conditions are represented by flow conditions above the dashed lines, below the dotted lines and below all solid lines. Data for $\Ray=0, 215, 226.67, 300, 500, 1\times10^3, 5\times10^3, 1\times10^4, 2\times10^4$ are shown here. Shaded regions represent the unstable flow conditions for $\Ray=2\times10^4$.}\label{fig:Rec_all}
  \end{center}
\end{figure}

In the range of $0\leq\Ray\leq1\times10^3$, a single thermal-dominant instability mechanism is observed while the higher $\Ray$ range ($\Ray\geq2\times10^4$) exhibits two instability mechanisms corresponding to peaks I and II described previously. These instability modes are in addition to the shear-dominant instability that is observed for all $\Ray$. The critical curves and critical wavenumbers as a function of $\HH$ are presented in figure~\ref{fig:Rec_all}.

\subsubsection{Critical Reynolds numbers for $0<\Ray\leq1\times10^3$}\label{subsubsec:fixed_Ra1e3_flows}
The $\Rey_c$ curve associated with the onset of shear-dominant instability at $\Ray=0$ (see figure~\ref{fig:Rec_Ra0}\textit{a}) is strongly insensitive to all $\Ray\leq2\times10^4$ (see dashed lines in figure~\ref{fig:Rec_all}\textit{a}). The instability is a \TollSchl\ wave whose phase speed is linearly dependent on the critical wavenumber (\ie\ $\textrm{Re}\{\omega\}\sim k$, not shown here). Since the shear-dominant instability is quite insensitive to $\Ray$ throughout the investigated $\Ray$ range, the remaining sections focus on the thermal-dominant instabilities.

The only Rayleigh number which demonstrates a single instability through the entire range of $\Rey$ investigated is $\Ray=0$. A second, thermally-dominant instability develops for $213.47\leq\Ray\leq1\times10^3$ provided that the through-flow is sufficiently weak (this is exemplified by the rapid change in direction of the neutral curves from nearly horizontal at lower $\HH$ towards the vertical at the critical $\HH$ values). As $\Rey\rightarrow0$ the flow becomes more prone to thermal instability whereas the flow is more susceptible to shear instability as $\Rey\rightarrow\infty$.  Hence, there are two unstable regions marked with shading in figure~\ref{fig:Rec_all}(\textit{a}). This result confirms that increasing the Reynolds number acts to suppress the thermally-dominant transverse rolls. However, this approach also demonstrates a progression from an unstable flow to a stable flow and to an unstable flow again with increasing $\Rey$, which was not observable in previous stability diagrams.

The $\Rey$--$\HH$ stability diagram shows that thermal disturbances are weakly sensitive to the magnetic damping at low $\HH$ but are abruptly suppressed when $\HH$ is sufficiently large. Indeed the $\HH$ values corresponding to the vertical neutral curves are precisely the critical values found in the $\Ray$--$\HH$ regime for $\Rey=0$ (see figure~\ref{fig:Rec_Ra0}\textit{b}). Additionally, the diagram shows that for an appropriate $\Rey$, increasing $\HH$ can cause the flow to transition through both stable and unstable states, as has been observed in previous studies \citep{genin2011temperature,zikanov2013natural,belyaev2015temperature}. The corresponding $k_c$ curves are shown in figure~\ref{fig:Rec_all}(\textit{b}). The critical wavenumbers associated with the shear-dominant instability are weakly sensitive to $\Ray$ in the range investigated here. The thermal-dominant instability adopts a larger wavenumber structure for $0\leq\Ray\leq1\times10^3$, which decreases with increasing $\Ray$. 

\subsubsection{Critical Reynolds numbers for $\Ray\geq5\times10^3$}\label{subsubsec:fixed_Ra5e3_flows}
\begin{figure}
  \begin{center}
     \begin{tabular}{cc}
     \multicolumn{1}{l}{(\textit{a})}  &  \multicolumn{1}{l}{(\textit{b})}\\
     \includegraphics[width=0.475\columnwidth]{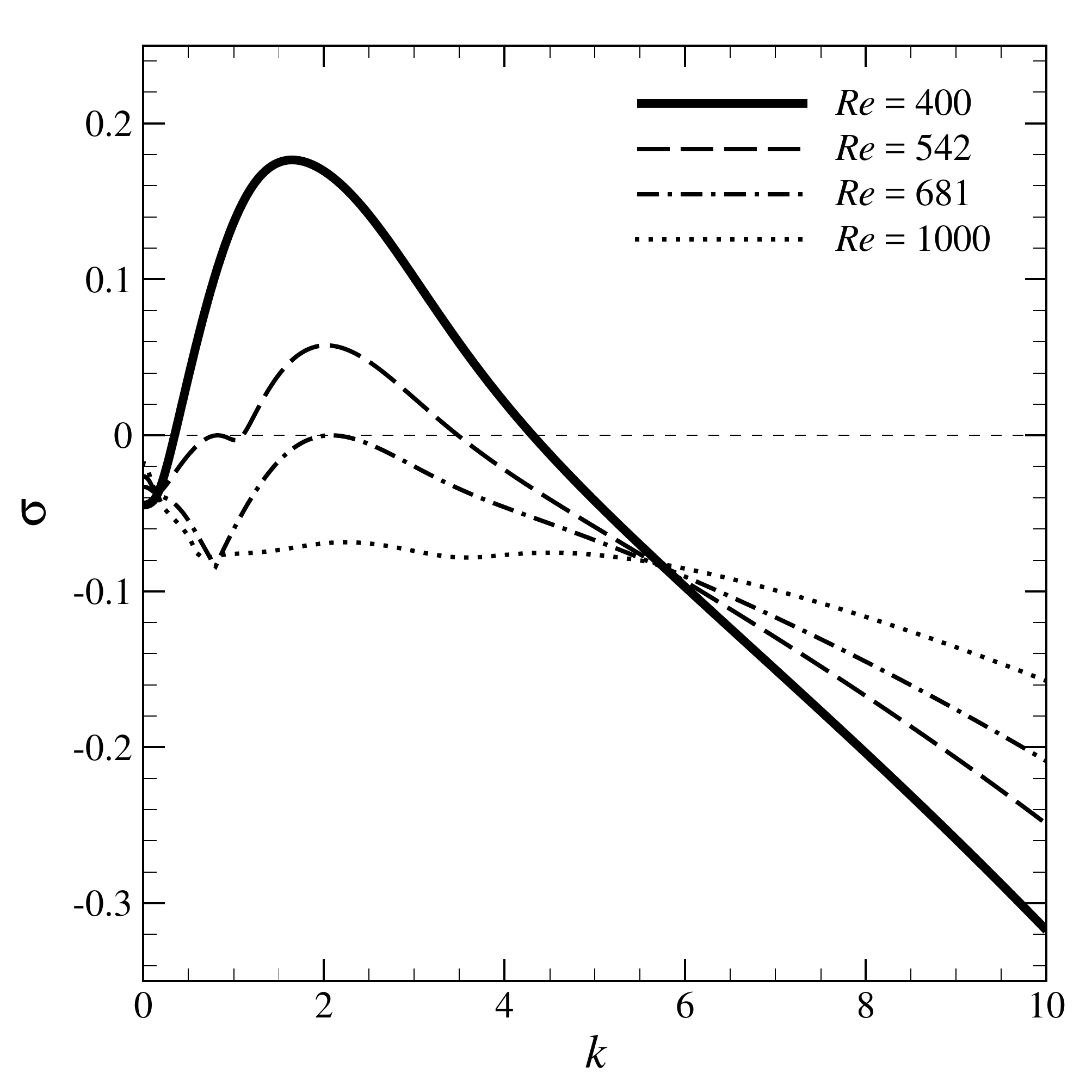}&
     \includegraphics[width=0.475\columnwidth]{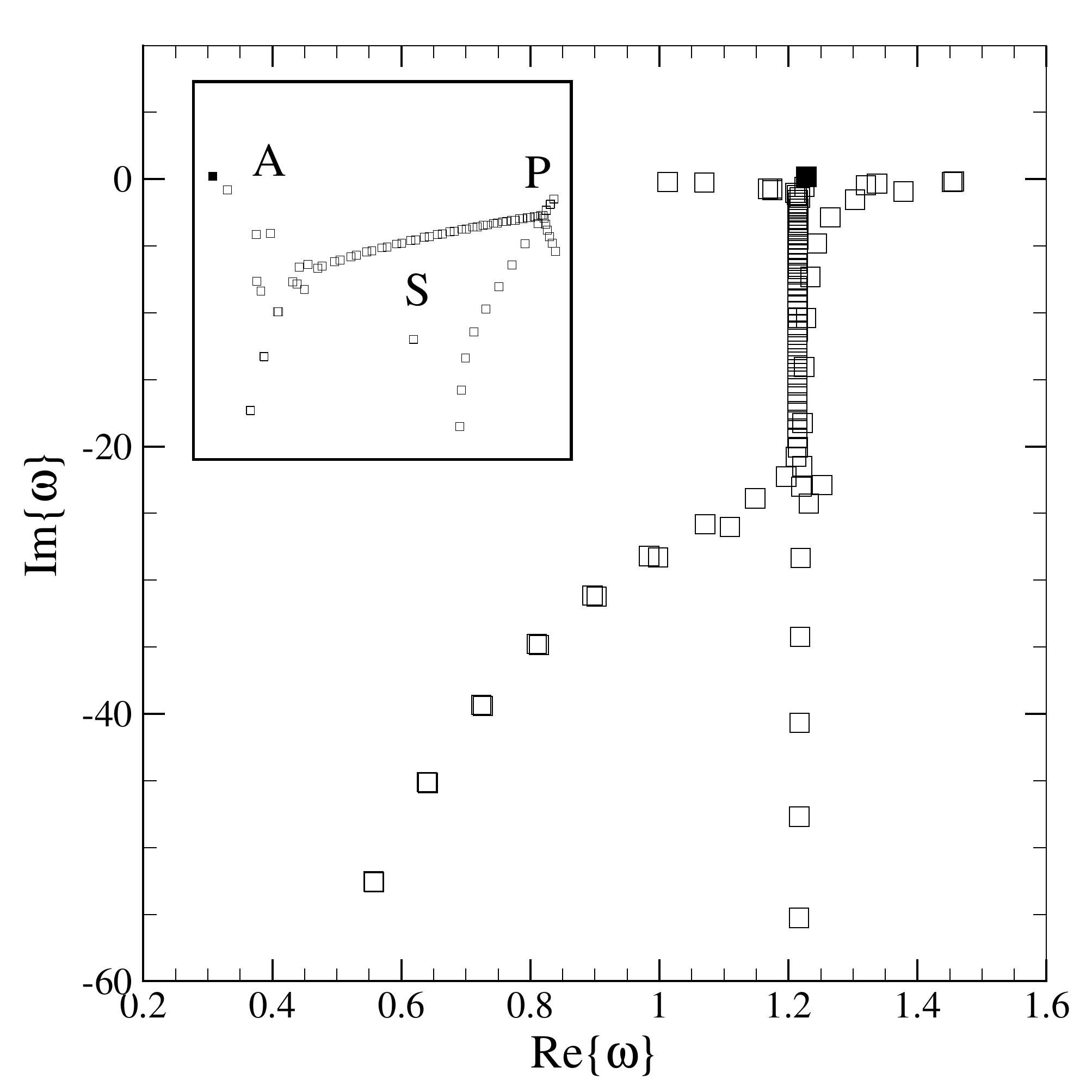}
     \end{tabular}
  \caption{(\textit{a})~Growth rate curves for $\HH=1$ and $\Ray=5\times10^3$ for various $\Rey$. The Reynolds numbers are $\Rey=400$~($\square$), $542$~($\triangle$), $681$~($\diamond$) and $\times10^3$~($\circ$). (\textit{b})~The eigenvalue spectra corresponding to the wavenumber of peak growth rate for $\Rey=400$ and (inset) $\Rey=57975$. The solid symbols denote the leading eigenvalue.}\label{fig:H1_Ra5000_varyingRe_growth_lsa}
  \end{center}
\end{figure}

Increasing the Rayleigh number up to $\Ray=5\times10^3$ spawns the onset of multiple instabilities evidenced by the multiple peaks appearing in the growth rate curves. These peaks are respectively the low and high wavenumber peaks I and II instabilities observed for fixed Reynolds number flows with $\Rey>350$ (section~\ref{subsubsec:fixed_Re350_flows}). Figure~\ref{fig:H1_Ra5000_varyingRe_growth_lsa}(\textit{a}) illustrates peaks I and II in the growth rate curves for $\Ray=5\times10^3$ with various $\Rey$. The stabilising effect of the Reynolds number is observed through the decrease in their maximum growth rate with increasing $\Rey$. The single thermal-dominant instability at $\Rey=400$ divides into two instability modes at $\Rey=542$. Further increases to $\Rey$ causes the flow to become stable. The eigenvalue spectrum for $\Rey=400$ is shown in figure~\ref{fig:H1_Ra5000_varyingRe_growth_lsa}(\textit{b}) and is representative of all $\Rey$ presented in panel~(\textit{a}). Although not shown in the figure, for $\Rey\geq\Rey_c=57974$ the flow becomes unstable again to the shear-dominant instability (mode A).

The corresponding critical value of $\Rey$ are approximately constant at low $\HH$ and increases with increasing $\HH$ before suddenly decreasing at higher $\HH$. The maximum point on the curve describing the onset of peak II is significant as it denotes the transition in mode type (\ie\ wall, interior or mixed). Typically, the flow exhibits a mixed instability mode for $\HH$ values to the left of the turning point and a wall mode for $\HH$ values to the right of the turning point. This reflects behaviours of fixed-$\Rey$ flows in that the type of instability changes with increasing $\HH$ along the peak II curve. The of emergence peaks I and II at $\Ray=5\times10^3$ introduce one smaller and one larger wavenumber than the shear-dominant instabilities. The critical wavenumbers of these modes are shown in figure~\ref{fig:Rec_all}(\textit{b}) and demonstrate a greater insensitivity to $\HH$ and $\Ray$ than shear-dominant instabilities. The critical wavenumber for the peak I instability decreases with increasing $\Ray$ while it increases with increasing $\Ray$ for peak II instability.

\begin{figure}
  \begin{center}
     \begin{tabular}{c}
  \includegraphics[width=0.6\columnwidth]{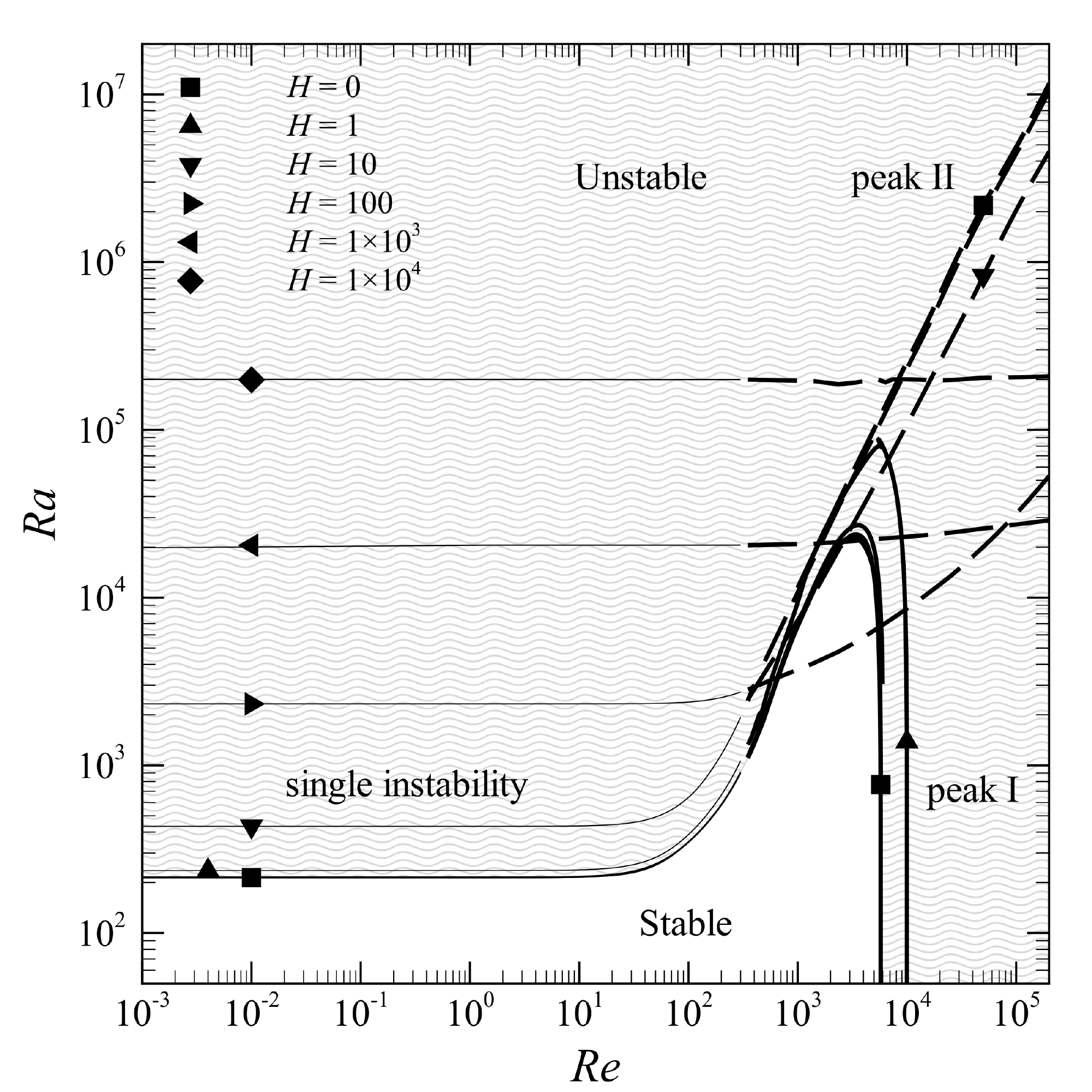}  \\
     \end{tabular}
  \caption{Stability diagram for $\Ray_c$ against $\Rey$ for fixed $\HH$. The symbols identify the modified Hartmann number for each curve. The curves for the onset of single instability are represented by thin solid lines, peak I by thick solid lines, and peak II by dashed lines. Unstable conditions are represented by flow conditions above all dashed and solid lines. Data for $\HH=0.01, 1, 10, 100, 1\times10^3, 1\times10^4, 1\times10^5$ are shown here. Shaded region represents the unstable flow conditions for $\HH=0$.}\label{fig:Rac_vs_Re_fnH}
  \end{center}
\end{figure}

A stability diagram of $\Ray_c$ against $\Rey$ for fixed $\HH$, for the single, peak I and peak II instabilities is presented in figure~\ref{fig:Rac_vs_Re_fnH}. The diagram depicts the critical Rayleigh number demonstrating weak sensitivity to $\HH$ for any given $\Rey\leq10$. This is in agreement with the results of \citet{nicolas2000linear} for an infinite duct aspect ratio and $\Prn=1\times10^{-6}$. Increasing $\Rey$ beyond $\Rey=350$ incurs a deviation from the constant $\Ray_c$ value, into a regime where two instability modes are observed. Peak I instability is characterised by an initial increase and then decrease in $\Ray_c$ with increasing $\Rey$. In contrast, $\Ray_c$ for peak II instability increases monotonically with increasing $\Rey$. Note that the critical $\Ray$ for the onset of peak II mode initially decreases with increasing $\HH$ but begins to increase again beyond $\HH=100$. It is worth mentioning that qualitatively similar stability diagrams have been produced by \citet{fakhfakh2010selective} for different Prandtl number liquids in a flow heated from below with vertical and horizontal magnetic fields. However, that study investigated an infinite domain where friction from the Hartmann walls are absent and therefore direct comparisons with the present study cannot be performed.

Several critical relationships can be established for the case of $\Rey>0$, $\Ray>0$ and $\HH=0$. The single instability takes place at $\Ray_c\approx213$ and $k_c\approx1.55$ for $\Rey\leq10$. The onset of the peak I instability data does not easily lend itself to being described by a simple function. However, the onset of the peak II instability follows $\Ray_c=(1.4693\pm0.3171)\Rey^{(1.307\pm0.0206)}$ and $k_c=0.3383\Rey^{0.3072}$ for $\Rey\geq350$. As $\HH\rightarrow\infty$, the stability diagram show that the single and peak II instabilities become independent of $\Rey$. These results along with all other key relationships established in this study are highlighted in table~\ref{table:results_summary}.

\begin{table}
  \begin{center}
\def~{\hphantom{0}}
  \begin{tabular}{ll}
      \hline\hline\\[-0.25cm]
       Parameter values\hspace{1.5cm}               & Flow ($\Prn=0.022$)\\[2pt]
      \hline\\[-0.25cm]
      $\Rey>0$, $\Ray=0$, $\HH=0$ & Plane \Poi\ flow\\
      {}                          & $\Rey_c=5772.22$, $k_c=1.02$\\[0.15cm]
      $\Rey=0$, $\Ray>0$, $\HH=0$ & \RB\ convection\\
      {}                          & $\Ray_c=213.47$, $k_c=1.5582$\\[0.15cm]
      $\Rey=0$, $\Ray=0$, $\HH>0$ & Magnetically damped, motionless flow\\
      {}                          & Stable\\[0.15cm]
      $\Rey>0$, $\Ray>0$, $\HH=0$ & \PRB\ flow (transverse roll instability)\\
      {}                          & $\Rey\leq10$ (single): $\Ray_c\approx213$, $k_c\approx1.5572$\\
      {}        & $\Rey\geq1000$ (peak II): $\Ray_c=(1.4693\pm0.3171)\Rey^{(1.307\pm0.0206)}$, $k_c=0.3383\Rey^{0.3072}$\\[0.15cm]
      $\Rey>0$, $\Ray=0$, $\HH>0$ & Fully developed SM82 duct flow\\
      {}                          & As $\HH\rightarrow\infty$, $\Rey_c=48347\HH^{\,1/2}$, $k_c=0.1615\HH^{\,1/2}$\\[0.15cm]
      $\Rey=0$, $\Ray>0$, $\HH>0$ & SM82 flow between horizontal plates heated from below\\
      {}                          & As $\HH\rightarrow\infty$, $\Ray_c=(21.672\pm0.212)\HH^{\,(0.991\pm0.001)}$, $k_c=\pi/2$\\[0.15cm]
      $\Rey>0$, $\Ray>0$, $\HH>0$ & \PRB\ with SM82 implementation\\
      {}                          & As $\HH\rightarrow\infty$, $\Rey_c=48347\HH^{\,1/2}$, $k_c=0.1615\HH^{\,1/2}$\\
      {}                          & As $\HH\rightarrow\infty$, $\Ray_c=(21.672\pm0.212)\HH^{\,(0.991\pm0.001)}$, $k_c=\pi/2$\\[0.15cm]
      \hline\hline
  \end{tabular}
  \caption{Key relationships obtained in this study for a variety of parameter values.}
  \label{table:results_summary}
  \end{center}
\end{table}

\section{Conclusions}\label{sec:conclusions}
This paper has systematically investigated the linear stability of \PRB\ flows under the effect of a transverse magnetic field. This study has extended the investigation of \citet{potherat2007quasi} by introducing vertical thermal stratification into the system. A \qtwod\ model following \citet{sommeria1982mhd} was used to describe the duct flow which included the modelling of the friction induced by the Hartmann layers. Since the system is governed by three non-dimensional parameters, two primary approaches were undertaken to understand the linear stability of the flow, fixing either $\Rey$ or $\Ray$ and then determining the corresponding $\Ray_c$ or $\Rey_c$.

The onset and suppression of multiple instabilities were determined and mapped onto $\Rey$--$\HH$ and $\Ray$--$\HH$ stability diagrams over $0\leq\HH\leq10^4$. A remarkable consequence of the competition between several instability mechanisms is the existence of a sharp discontinuity in critical wavenumber when increasing $\HH$ through the changeover point through between the two dominant modes. The discontinuity takes place with increasing $\HH$ and strongly resembles that observed when switching between magnetic and viscous modes both in plane and confined rotating magnetoconvection \citep{nakagawa1955experiment,aujogue2015onset,aujogue2016little}. A second discontinuous drop in wavenumber with increasing $\HH$ was also observed at very high $\Rey$, though these two modes exhibited peak II characteristics. Asymptotic relationships described by $\Rey_c\propto\HH^{\,1/2}$ and $\Ray_c\propto\HH$ for $\HH\rightarrow\infty$ were obtained. In the former case, the stability of the flow is governed by the individual Shercliff layers while in the latter case, the stability is dictated by the balance between buoyancy and Hartmann friction.

The disturbance fields for vorticity and temperature depicted two distinct regions for growth, namely within the Shercliff layers and the interior region. Generally, it was found that instabilities manifesting along the side walls and in the interior flow (\ie\ mixed modes) only existed at low $\HH$. A strong magnetic field was found to suppress the interior structures leaving only wall modes. The inclusion of thermal stratification has been shown to be able to encourage mixing within the interior of the duct which would otherwise be limited to the boundary layers.

\appendix
\section{Governing equations for cases of $\Ray=0$ and $\Rey=0$}\label{appsec:governing_eqs}

The SM82 equations are coupled with a thermal transport equation through a Boussinesq approximation to describe the \mhd\ duct flow with vertical thermal stratification (\ie\ $\Rey>0$, $\Ray>0$). These equations are given in dimensional form as
\begin{subequations}\label{eq:dimensional_SM82}
\begin{align}
\dfrac{\partial \mathbf{u}}{\partial t} + (\mathbf{u} \cdot \nabla)\mathbf{u} & = -\dfrac{1}{\rho}\nabla p + \nu{\nabla}^2\mathbf{u} - \dfrac{n}{t_H}\mathbf{u}-\alpha\mathbf{g}\mathbf{\theta},\\
\dfrac{\partial \mathbf{\theta}}{\partial t} +(\mathbf{u} \cdot \nabla)\mathbf{\theta} & = \kappa{\nabla}^2\mathbf{\theta}\\
\nabla \cdot \mathbf{u} & = 0,
\end{align}
\end{subequations}
where $\mathbf{u}$ is the velocity vector, $t$ is time, $p$ is pressure, $\mathbf{g}$ is the gravitational acceleration acting in the negative $y$ direction, and $t_H=(a/B)\sqrt{\rho/(\xi\nu)}$ as the Hartmann damping time.

The non-dimensional equations for the case of no heating ($\Ray=0$)
\begin{subequations}
\begin{align}
\dfrac{\partial \mathbf{u}}{\partial t} + (\mathbf{u} \cdot \nabla)\mathbf{u} & = -\nabla p + \dfrac{1}{\Rey}{\nabla}^2 \mathbf{u} - \dfrac{\HH}{\Rey}\mathbf{u},\\
\nabla \cdot \mathbf{u} & = 0,
\end{align}
\end{subequations}
are obtained by normalising lengths by $L$, velocity by $U_0$, time by $L/U_0$ and pressure by $\rho U_0^2$.

For no through-flow ($\Rey=0$), the non-dimensional equations
\begin{subequations}\label{eq:PRB_Re0}
\begin{align}
\dfrac{\partial \mathbf{u}}{\partial t} + (\mathbf{u} \cdot \nabla)\mathbf{u} & = -\nabla p + \Prn{\nabla}^2 \mathbf{u}-\Prn \HH\mathbf{u} + \Prn \Ray\theta\mathbf{\hat{e}}_y,\\
\dfrac{\partial \mathbf{\theta}}{\partial t} +(\mathbf{u} \cdot \nabla)\mathbf{\theta} & = {\nabla}^2\mathbf{\theta},\\
\nabla \cdot \mathbf{u} & = 0,
\end{align}
\end{subequations}
are obtained by normalising lengths by $L$, velocity by $\kappa/L$, time by $L^2/\kappa$, pressure by $\rho(\kappa/L)^2$ and temperature by $\Delta\theta$.

The linearised governing equations for the $\Rey=0$ and $\Ray=0$ cases can be obtained using the same derivation process described in Sec.~\ref{subsec:linear}. The corresponding eigenvalue equations for the $\Ray=0$ and $\Rey=0$ cases are respectively given by
\begin{subequations}\label{eq:Ray0_linear}
\begin{align}
  \begin{split}
    \dfrac{1}{\Rey}\left(\mathrm{D}^2 - k^2\right)^2\vp +\ci k\ub^{\prime\prime}\vp - \ci k\ub\left(\mathrm{D}^2-k^2\right)\vp \\
    -\dfrac{\HH}{\Rey}\left(\mathrm{D}^2-k^2\right)\vp &= -\ci\omega\left(\mathrm{D}^2 - k^2\right)\vp,
    \end{split}
\end{align}
\end{subequations}
and
\begin{subequations} \label{eq:Rey0_linear}
\begin{align}
  \begin{split}
    \Prn\left(\mathrm{D}^2 - k^2\right)^2\vp -\HH\left(\mathrm{D}^2-k^2\right)\vp \\
    - \Prn\Ray k^2\tp &= -\ci\omega\left(\mathrm{D}^2 - k^2\right)\vp,
  \end{split}\\
      -\tb^\prime\vp + \left(\mathrm{D}^2 - k^2\right)\tp & = -\ci\omega\tp.
\end{align}
\end{subequations}

\begin{acknowledgments}
The authors thank TzeKih Tsai and Wisam K. Hussam from Monash University for their help in validating the \oned\ linear stability solver used in the present study. This research was supported by ARC Discovery Grants DP120100153 and DP150102920. A. P. acknowledges support from the Royal Society under the Wolfson Research Merit Award Scheme (grant WM140032). Additional assistance was provided through high-performance computing time allocations from the National Computational Infrastructure (NCI), the Victorian Life Sciences Computation Initiative (VLSCI), and the Monash SunGRID.
\end{acknowledgments}


\begin{thebibliography}{54}%
\makeatletter
\providecommand \@ifxundefined [1]{%
 \@ifx{#1\undefined}
}%
\providecommand \@ifnum [1]{%
 \ifnum #1\expandafter \@firstoftwo
 \else \expandafter \@secondoftwo
 \fi
}%
\providecommand \@ifx [1]{%
 \ifx #1\expandafter \@firstoftwo
 \else \expandafter \@secondoftwo
 \fi
}%
\providecommand \natexlab [1]{#1}%
\providecommand \enquote  [1]{``#1''}%
\providecommand \bibnamefont  [1]{#1}%
\providecommand \bibfnamefont [1]{#1}%
\providecommand \citenamefont [1]{#1}%
\providecommand \href@noop [0]{\@secondoftwo}%
\providecommand \href [0]{\begingroup \@sanitize@url \@href}%
\providecommand \@href[1]{\@@startlink{#1}\@@href}%
\providecommand \@@href[1]{\endgroup#1\@@endlink}%
\providecommand \@sanitize@url [0]{\catcode `\\12\catcode `\$12\catcode
  `\&12\catcode `\#12\catcode `\^12\catcode `\_12\catcode `\%12\relax}%
\providecommand \@@startlink[1]{}%
\providecommand \@@endlink[0]{}%
\providecommand \url  [0]{\begingroup\@sanitize@url \@url }%
\providecommand \@url [1]{\endgroup\@href {#1}{\urlprefix }}%
\providecommand \urlprefix  [0]{URL }%
\providecommand \Eprint [0]{\href }%
\providecommand \doibase [0]{http://dx.doi.org/}%
\providecommand \selectlanguage [0]{\@gobble}%
\providecommand \bibinfo  [0]{\@secondoftwo}%
\providecommand \bibfield  [0]{\@secondoftwo}%
\providecommand \translation [1]{[#1]}%
\providecommand \BibitemOpen [0]{}%
\providecommand \bibitemStop [0]{}%
\providecommand \bibitemNoStop [0]{.\EOS\space}%
\providecommand \EOS [0]{\spacefactor3000\relax}%
\providecommand \BibitemShut  [1]{\csname bibitem#1\endcsname}%
\let\auto@bib@innerbib\@empty
\bibitem [{\citenamefont {Sommeria}\ and\ \citenamefont
  {Moreau}(1982)}]{sommeria1982mhd}%
  \BibitemOpen
  \bibfield  {author} {\bibinfo {author} {\bibfnamefont {J.}~\bibnamefont
  {Sommeria}}\ and\ \bibinfo {author} {\bibfnamefont {R.}~\bibnamefont
  {Moreau}},\ }\bibfield  {title} {\enquote {\bibinfo {title} {Why, how, and
  when, {MHD} turbulence becomes two-dimensional},}\ }\href@noop {} {\bibfield
  {journal} {\bibinfo  {journal} {J.\,Fluid Mech.}\ }\textbf {\bibinfo {volume}
  {118}},\ \bibinfo {pages} {507--518} (\bibinfo {year} {1982})}\BibitemShut
  {NoStop}%
\bibitem [{\citenamefont {Abdou}\ \emph {et~al.}(2015)\citenamefont {Abdou},
  \citenamefont {Morley}, \citenamefont {Smolentsev}, \citenamefont {Ying},
  \citenamefont {Malang}, \citenamefont {Rowcliffe},\ and\ \citenamefont
  {Ulrickson}}]{abdou2015blanket}%
  \BibitemOpen
  \bibfield  {author} {\bibinfo {author} {\bibfnamefont {M.}~\bibnamefont
  {Abdou}}, \bibinfo {author} {\bibfnamefont {N.~B.}\ \bibnamefont {Morley}},
  \bibinfo {author} {\bibfnamefont {S.}~\bibnamefont {Smolentsev}}, \bibinfo
  {author} {\bibfnamefont {A.}~\bibnamefont {Ying}}, \bibinfo {author}
  {\bibfnamefont {S.}~\bibnamefont {Malang}}, \bibinfo {author} {\bibfnamefont
  {A.}~\bibnamefont {Rowcliffe}}, \ and\ \bibinfo {author} {\bibfnamefont
  {M.}~\bibnamefont {Ulrickson}},\ }\bibfield  {title} {\enquote {\bibinfo
  {title} {Blanket/first wall challenges and required {R\&D} on the pathway to
  {DEMO}},}\ }\href@noop {} {\bibfield  {journal} {\bibinfo  {journal} {Fusion
  Eng.\ Des.}\ }\textbf {\bibinfo {volume} {100}},\ \bibinfo {pages} {2--43}
  (\bibinfo {year} {2015})}\BibitemShut {NoStop}%
\bibitem [{\citenamefont {Drazin}\ and\ \citenamefont
  {Reid}(2004)}]{drazin2004hydrodynamic}%
  \BibitemOpen
  \bibfield  {author} {\bibinfo {author} {\bibfnamefont {P.~G.}\ \bibnamefont
  {Drazin}}\ and\ \bibinfo {author} {\bibfnamefont {W.~H.}\ \bibnamefont
  {Reid}},\ }\href@noop {} {\emph {\bibinfo {title} {Hydrodynamic stability}}}\
  (\bibinfo  {publisher} {Cambridge University Press},\ \bibinfo {year}
  {2004})\BibitemShut {NoStop}%
\bibitem [{\citenamefont {Gage}\ and\ \citenamefont
  {Reid}(1968)}]{gage1968stability}%
  \BibitemOpen
  \bibfield  {author} {\bibinfo {author} {\bibfnamefont {K.~S.}\ \bibnamefont
  {Gage}}\ and\ \bibinfo {author} {\bibfnamefont {W.~H.}\ \bibnamefont
  {Reid}},\ }\bibfield  {title} {\enquote {\bibinfo {title} {The stability of
  thermally stratified plane {P}oiseuille flow},}\ }\href@noop {} {\bibfield
  {journal} {\bibinfo  {journal} {J.\,Fluid Mech.}\ }\textbf {\bibinfo {volume}
  {33}},\ \bibinfo {pages} {21--32} (\bibinfo {year} {1968})}\BibitemShut
  {NoStop}%
\bibitem [{\citenamefont {Yasuo}\ and\ \citenamefont
  {Yutaka}(1966)}]{yasuo1966forced}%
  \BibitemOpen
  \bibfield  {author} {\bibinfo {author} {\bibfnamefont {M.}~\bibnamefont
  {Yasuo}}\ and\ \bibinfo {author} {\bibfnamefont {U.}~\bibnamefont {Yutaka}},\
  }\bibfield  {title} {\enquote {\bibinfo {title} {Forced convective heat
  transfer between horizontal flat plates},}\ }\href@noop {} {\bibfield
  {journal} {\bibinfo  {journal} {Int.\ J.\ Heat Mass Trans.}\ }\textbf
  {\bibinfo {volume} {9}},\ \bibinfo {pages} {803--817} (\bibinfo {year}
  {1966})}\BibitemShut {NoStop}%
\bibitem [{\citenamefont {Akiyama}\ \emph {et~al.}(1971)\citenamefont
  {Akiyama}, \citenamefont {Hwang},\ and\ \citenamefont
  {Cheng}}]{akiyama1971experiments}%
  \BibitemOpen
  \bibfield  {author} {\bibinfo {author} {\bibfnamefont {M.}~\bibnamefont
  {Akiyama}}, \bibinfo {author} {\bibfnamefont {G.-J.}\ \bibnamefont {Hwang}},
  \ and\ \bibinfo {author} {\bibfnamefont {K.~C.}\ \bibnamefont {Cheng}},\
  }\bibfield  {title} {\enquote {\bibinfo {title} {Experiments on the onset of
  longitudinal vortices in laminar forced convection between horizontal
  plates},}\ }\href@noop {} {\bibfield  {journal} {\bibinfo  {journal}
  {J.\,Heat Trans.}\ }\textbf {\bibinfo {volume} {93}},\ \bibinfo {pages}
  {335--341} (\bibinfo {year} {1971})}\BibitemShut {NoStop}%
\bibitem [{\citenamefont {Fukui}\ \emph {et~al.}(1983)\citenamefont {Fukui},
  \citenamefont {Nakajima},\ and\ \citenamefont
  {Ueda}}]{fukui1983longitudinal}%
  \BibitemOpen
  \bibfield  {author} {\bibinfo {author} {\bibfnamefont {K.}~\bibnamefont
  {Fukui}}, \bibinfo {author} {\bibfnamefont {M.}~\bibnamefont {Nakajima}}, \
  and\ \bibinfo {author} {\bibfnamefont {H.}~\bibnamefont {Ueda}},\ }\bibfield
  {title} {\enquote {\bibinfo {title} {The longitudinal vortex and its effects
  on the transport processes in combined free and forced laminar convection
  between horizontal and inclined parallel plates},}\ }\href@noop {} {\bibfield
   {journal} {\bibinfo  {journal} {Int.\ J.\ Heat Mass Trans.}\ }\textbf
  {\bibinfo {volume} {26}},\ \bibinfo {pages} {109--120} (\bibinfo {year}
  {1983})}\BibitemShut {NoStop}%
\bibitem [{\citenamefont {M{\"u}ller}\ \emph {et~al.}(1992)\citenamefont
  {M{\"u}ller}, \citenamefont {L{\"u}cke},\ and\ \citenamefont
  {Kamps}}]{muller1992effect}%
  \BibitemOpen
  \bibfield  {author} {\bibinfo {author} {\bibfnamefont {H.~W.}\ \bibnamefont
  {M{\"u}ller}}, \bibinfo {author} {\bibfnamefont {M.}~\bibnamefont
  {L{\"u}cke}}, \ and\ \bibinfo {author} {\bibfnamefont {M.}~\bibnamefont
  {Kamps}},\ }\bibfield  {title} {\enquote {\bibinfo {title} {The effect of
  throughflow on {R}ayleigh {B}ernard convective rolls},}\ }in\ \href@noop {}
  {\emph {\bibinfo {booktitle} {Ordered and Turbulent Patterns in
  {T}aylor--{C}ouette Flow}}}\ (\bibinfo  {publisher} {Springer},\ \bibinfo
  {year} {1992})\ pp.\ \bibinfo {pages} {187--196}\BibitemShut {NoStop}%
\bibitem [{\citenamefont {Carriere}\ and\ \citenamefont
  {Monkewitz}(1999)}]{carriere1999convective}%
  \BibitemOpen
  \bibfield  {author} {\bibinfo {author} {\bibfnamefont {P.}~\bibnamefont
  {Carriere}}\ and\ \bibinfo {author} {\bibfnamefont {P.~A.}\ \bibnamefont
  {Monkewitz}},\ }\bibfield  {title} {\enquote {\bibinfo {title} {Convective
  versus absolute instability in mixed {R}ayleigh--{B}{\'e}nard--{P}oiseuille
  convection},}\ }\href@noop {} {\bibfield  {journal} {\bibinfo  {journal}
  {J.\,Fluid Mech.}\ }\textbf {\bibinfo {volume} {384}},\ \bibinfo {pages}
  {243--262} (\bibinfo {year} {1999})}\BibitemShut {NoStop}%
\bibitem [{\citenamefont {Nicolas}\ \emph {et~al.}(2000)\citenamefont
  {Nicolas}, \citenamefont {Luijkx},\ and\ \citenamefont
  {Platten}}]{nicolas2000linear}%
  \BibitemOpen
  \bibfield  {author} {\bibinfo {author} {\bibfnamefont {X.}~\bibnamefont
  {Nicolas}}, \bibinfo {author} {\bibfnamefont {J.-M.}\ \bibnamefont {Luijkx}},
  \ and\ \bibinfo {author} {\bibfnamefont {J.-K.}\ \bibnamefont {Platten}},\
  }\bibfield  {title} {\enquote {\bibinfo {title} {Linear stability of mixed
  convection flows in horizontal rectangular channels of finite transversal
  extension heated from below},}\ }\href@noop {} {\bibfield  {journal}
  {\bibinfo  {journal} {Int.\ J.\ Heat Mass Trans.}\ }\textbf {\bibinfo
  {volume} {43}},\ \bibinfo {pages} {589--610} (\bibinfo {year}
  {2000})}\BibitemShut {NoStop}%
\bibitem [{\citenamefont {Grandjean}\ and\ \citenamefont
  {Monkewitz}(2009)}]{grandjean2009experimental}%
  \BibitemOpen
  \bibfield  {author} {\bibinfo {author} {\bibfnamefont {E.}~\bibnamefont
  {Grandjean}}\ and\ \bibinfo {author} {\bibfnamefont {P.~A.}\ \bibnamefont
  {Monkewitz}},\ }\bibfield  {title} {\enquote {\bibinfo {title} {Experimental
  investigation into localized instabilities of mixed
  {R}ayleigh--{B}{\'e}nard--{P}oiseuille convection},}\ }\href@noop {}
  {\bibfield  {journal} {\bibinfo  {journal} {J.\,Fluid Mech.}\ }\textbf
  {\bibinfo {volume} {640}},\ \bibinfo {pages} {401--419} (\bibinfo {year}
  {2009})}\BibitemShut {NoStop}%
\bibitem [{\citenamefont {Mergui}\ \emph {et~al.}(2011)\citenamefont {Mergui},
  \citenamefont {Nicolas},\ and\ \citenamefont {Hirata}}]{mergui2011sidewall}%
  \BibitemOpen
  \bibfield  {author} {\bibinfo {author} {\bibfnamefont {S.}~\bibnamefont
  {Mergui}}, \bibinfo {author} {\bibfnamefont {X.}~\bibnamefont {Nicolas}}, \
  and\ \bibinfo {author} {\bibfnamefont {S.}~\bibnamefont {Hirata}},\
  }\bibfield  {title} {\enquote {\bibinfo {title} {Sidewall and thermal
  boundary condition effects on the evolution of longitudinal rolls in
  {R}ayleigh-{B}{\'e}nard-{P}oiseuille convection},}\ }\href@noop {} {\ \textbf
  {\bibinfo {volume} {23}},\ \bibinfo {pages} {084101} (\bibinfo {year}
  {2011})}\BibitemShut {NoStop}%
\bibitem [{\citenamefont {Nicolas}\ \emph {et~al.}(2012)\citenamefont
  {Nicolas}, \citenamefont {Zoueidi},\ and\ \citenamefont
  {Xin}}]{nicolas2012influence}%
  \BibitemOpen
  \bibfield  {author} {\bibinfo {author} {\bibfnamefont {X.}~\bibnamefont
  {Nicolas}}, \bibinfo {author} {\bibfnamefont {N.}~\bibnamefont {Zoueidi}}, \
  and\ \bibinfo {author} {\bibfnamefont {S.}~\bibnamefont {Xin}},\ }\bibfield
  {title} {\enquote {\bibinfo {title} {Influence of a white noise at channel
  inlet on the parallel and wavy convective instabilities of
  {P}oiseuille-{R}ayleigh-{B}{\'e}nard flows},}\ }\href@noop {} {\ \textbf
  {\bibinfo {volume} {24}},\ \bibinfo {pages} {084101} (\bibinfo {year}
  {2012})}\BibitemShut {NoStop}%
\bibitem [{\citenamefont {Luijkx}\ \emph {et~al.}(1981)\citenamefont {Luijkx},
  \citenamefont {Platten},\ and\ \citenamefont {Legros}}]{luijkx1981existence}%
  \BibitemOpen
  \bibfield  {author} {\bibinfo {author} {\bibfnamefont {J.-M.}\ \bibnamefont
  {Luijkx}}, \bibinfo {author} {\bibfnamefont {J.~K.}\ \bibnamefont {Platten}},
  \ and\ \bibinfo {author} {\bibfnamefont {J.~Cl.}\ \bibnamefont {Legros}},\
  }\bibfield  {title} {\enquote {\bibinfo {title} {On the existence of
  thermoconvective rolls, transverse to a superimposed mean {P}oiseuille
  flow},}\ }\href@noop {} {\bibfield  {journal} {\bibinfo  {journal} {Int.\ J.\
  Heat Mass Trans.}\ }\textbf {\bibinfo {volume} {24}},\ \bibinfo {pages}
  {1287--1291} (\bibinfo {year} {1981})}\BibitemShut {NoStop}%
\bibitem [{\citenamefont {Poth{\'e}rat}(2007)}]{potherat2007quasi}%
  \BibitemOpen
  \bibfield  {author} {\bibinfo {author} {\bibfnamefont {A.}~\bibnamefont
  {Poth{\'e}rat}},\ }\bibfield  {title} {\enquote {\bibinfo {title}
  {Quasi-two-dimensional perturbations in duct flows under transverse magnetic
  field},}\ }\href@noop {} {\bibfield  {journal} {\bibinfo  {journal} {Phys.\
  Fluids}\ }\textbf {\bibinfo {volume} {19}},\ \bibinfo {pages} {074104}
  (\bibinfo {year} {2007})}\BibitemShut {NoStop}%
\bibitem [{\citenamefont {Fakhfakh}\ \emph {et~al.}(2010)\citenamefont
  {Fakhfakh}, \citenamefont {Kaddeche}, \citenamefont {Henry},\ and\
  \citenamefont {Hadid}}]{fakhfakh2010selective}%
  \BibitemOpen
  \bibfield  {author} {\bibinfo {author} {\bibfnamefont {W.}~\bibnamefont
  {Fakhfakh}}, \bibinfo {author} {\bibfnamefont {S.}~\bibnamefont {Kaddeche}},
  \bibinfo {author} {\bibfnamefont {D.}~\bibnamefont {Henry}}, \ and\ \bibinfo
  {author} {\bibfnamefont {H.~B.}\ \bibnamefont {Hadid}},\ }\bibfield  {title}
  {\enquote {\bibinfo {title} {Selective control of
  {P}oiseuille--{R}ayleigh--{B}{\'e}nard instabilities by a spanwise magnetic
  field},}\ }\href@noop {} {\bibfield  {journal} {\bibinfo  {journal} {Phys.\
  Fluids}\ }\textbf {\bibinfo {volume} {22}},\ \bibinfo {pages} {034103}
  (\bibinfo {year} {2010})}\BibitemShut {NoStop}%
\bibitem [{\citenamefont {Krasnov}\ \emph {et~al.}(2012)\citenamefont
  {Krasnov}, \citenamefont {Zikanov},\ and\ \citenamefont
  {Boeck}}]{krasnov2012numerical}%
  \BibitemOpen
  \bibfield  {author} {\bibinfo {author} {\bibfnamefont {D.}~\bibnamefont
  {Krasnov}}, \bibinfo {author} {\bibfnamefont {O.}~\bibnamefont {Zikanov}}, \
  and\ \bibinfo {author} {\bibfnamefont {T.}~\bibnamefont {Boeck}},\ }\bibfield
   {title} {\enquote {\bibinfo {title} {Numerical study of magnetohydrodynamic
  duct flow at high {R}eynolds and {H}artmann numbers},}\ }\href@noop {}
  {\bibfield  {journal} {\bibinfo  {journal} {J.\,Fluid Mech.}\ }\textbf
  {\bibinfo {volume} {704}},\ \bibinfo {pages} {421--446} (\bibinfo {year}
  {2012})}\BibitemShut {NoStop}%
\bibitem [{\citenamefont {Priede}\ \emph {et~al.}(2012)\citenamefont {Priede},
  \citenamefont {Aleksandrova},\ and\ \citenamefont
  {Molokov}}]{priede2012linear}%
  \BibitemOpen
  \bibfield  {author} {\bibinfo {author} {\bibfnamefont {J.}~\bibnamefont
  {Priede}}, \bibinfo {author} {\bibfnamefont {S.}~\bibnamefont
  {Aleksandrova}}, \ and\ \bibinfo {author} {\bibfnamefont {S.}~\bibnamefont
  {Molokov}},\ }\bibfield  {title} {\enquote {\bibinfo {title} {Linear
  stability of magnetohydrodynamic flow in a perfectly conducting rectangular
  duct},}\ }\href@noop {} {\bibfield  {journal} {\bibinfo  {journal} {J.\,Fluid
  Mech.}\ }\textbf {\bibinfo {volume} {708}},\ \bibinfo {pages} {111--127}
  (\bibinfo {year} {2012})}\BibitemShut {NoStop}%
\bibitem [{\citenamefont {Yoon}\ \emph {et~al.}(2004)\citenamefont {Yoon},
  \citenamefont {Chun}, \citenamefont {Ha},\ and\ \citenamefont
  {Lee}}]{yoon2004numerical}%
  \BibitemOpen
  \bibfield  {author} {\bibinfo {author} {\bibfnamefont {H.~S.}\ \bibnamefont
  {Yoon}}, \bibinfo {author} {\bibfnamefont {H.~H.}\ \bibnamefont {Chun}},
  \bibinfo {author} {\bibfnamefont {M.~Y.}\ \bibnamefont {Ha}}, \ and\ \bibinfo
  {author} {\bibfnamefont {H.~G.}\ \bibnamefont {Lee}},\ }\bibfield  {title}
  {\enquote {\bibinfo {title} {A numerical study on the fluid flow and heat
  transfer around a circular cylinder in an aligned magnetic field},}\
  }\href@noop {} {\bibfield  {journal} {\bibinfo  {journal} {Int.\ J.\ Heat
  Mass Trans.}\ }\textbf {\bibinfo {volume} {47}},\ \bibinfo {pages}
  {4075--4087} (\bibinfo {year} {2004})}\BibitemShut {NoStop}%
\bibitem [{\citenamefont {Hussam}\ \emph {et~al.}(2011)\citenamefont {Hussam},
  \citenamefont {Thompson},\ and\ \citenamefont {Sheard}}]{hussam2011dynamics}%
  \BibitemOpen
  \bibfield  {author} {\bibinfo {author} {\bibfnamefont {W.~K.}\ \bibnamefont
  {Hussam}}, \bibinfo {author} {\bibfnamefont {M.~C.}\ \bibnamefont
  {Thompson}}, \ and\ \bibinfo {author} {\bibfnamefont {G.~J.}\ \bibnamefont
  {Sheard}},\ }\bibfield  {title} {\enquote {\bibinfo {title} {Dynamics and
  heat transfer in a quasi-two-dimensional {MHD} flow past a circular cylinder
  in a duct at high {H}artmann number},}\ }\href@noop {} {\bibfield  {journal}
  {\bibinfo  {journal} {Int.\ J.\ Heat Mass Trans.}\ }\textbf {\bibinfo
  {volume} {54}},\ \bibinfo {pages} {1091--1100} (\bibinfo {year}
  {2011})}\BibitemShut {NoStop}%
\bibitem [{\citenamefont {Cassells}\ \emph {et~al.}(2016)\citenamefont
  {Cassells}, \citenamefont {Hussam},\ and\ \citenamefont
  {Sheard}}]{cassells2016heat}%
  \BibitemOpen
  \bibfield  {author} {\bibinfo {author} {\bibfnamefont {O.~G.~W.}\
  \bibnamefont {Cassells}}, \bibinfo {author} {\bibfnamefont {W.~K.}\
  \bibnamefont {Hussam}}, \ and\ \bibinfo {author} {\bibfnamefont {G.~J.}\
  \bibnamefont {Sheard}},\ }\bibfield  {title} {\enquote {\bibinfo {title}
  {Heat transfer enhancement using rectangular vortex promoters in confined
  quasi-two-dimensional magnetohydrodynamic flows},}\ }\href@noop {} {\bibfield
   {journal} {\bibinfo  {journal} {Int.\ J.\ Heat Mass Trans.}\ }\textbf
  {\bibinfo {volume} {93}},\ \bibinfo {pages} {186--199} (\bibinfo {year}
  {2016})}\BibitemShut {NoStop}%
\bibitem [{\citenamefont {Hamid}\ \emph {et~al.}(2016)\citenamefont {Hamid},
  \citenamefont {Hussam},\ and\ \citenamefont {Sheard}}]{hamid2016combining}%
  \BibitemOpen
  \bibfield  {author} {\bibinfo {author} {\bibfnamefont {A.~H.~A.}\
  \bibnamefont {Hamid}}, \bibinfo {author} {\bibfnamefont {W.~K.}\ \bibnamefont
  {Hussam}}, \ and\ \bibinfo {author} {\bibfnamefont {G.~J.}\ \bibnamefont
  {Sheard}},\ }\bibfield  {title} {\enquote {\bibinfo {title} {Combining an
  obstacle and electrically driven vortices to enhance heat transfer in a
  quasi-two-dimensional {MHD} duct flow},}\ }\href@noop {} {\bibfield
  {journal} {\bibinfo  {journal} {J.\,Fluid Mech.}\ }\textbf {\bibinfo {volume}
  {792}},\ \bibinfo {pages} {364--396} (\bibinfo {year} {2016})}\BibitemShut
  {NoStop}%
\bibitem [{\citenamefont {Fr{\"u}h}\ and\ \citenamefont
  {Nielsen}(2003)}]{fruh2003origin}%
  \BibitemOpen
  \bibfield  {author} {\bibinfo {author} {\bibfnamefont {W.~G.}\ \bibnamefont
  {Fr{\"u}h}}\ and\ \bibinfo {author} {\bibfnamefont {A.~H.}\ \bibnamefont
  {Nielsen}},\ }\bibfield  {title} {\enquote {\bibinfo {title} {On the origin
  of time-dependent behaviour in a barotropically unstable shear layer},}\
  }\href@noop {} {\bibfield  {journal} {\bibinfo  {journal} {Nonlinear Proc.\
  Geoph.}\ }\textbf {\bibinfo {volume} {10}},\ \bibinfo {pages} {289--302}
  (\bibinfo {year} {2003})}\BibitemShut {NoStop}%
\bibitem [{\citenamefont {Vo}\ \emph {et~al.}(2015)\citenamefont {Vo},
  \citenamefont {Montabone}, \citenamefont {Read},\ and\ \citenamefont
  {Sheard}}]{vo2015non}%
  \BibitemOpen
  \bibfield  {author} {\bibinfo {author} {\bibfnamefont {Tony}\ \bibnamefont
  {Vo}}, \bibinfo {author} {\bibfnamefont {Luca}\ \bibnamefont {Montabone}},
  \bibinfo {author} {\bibfnamefont {Peter~L}\ \bibnamefont {Read}}, \ and\
  \bibinfo {author} {\bibfnamefont {Gregory~J}\ \bibnamefont {Sheard}},\
  }\bibfield  {title} {\enquote {\bibinfo {title} {Non-axisymmetric flows in a
  differential-disk rotating system},}\ }\href@noop {} {\bibfield  {journal}
  {\bibinfo  {journal} {Journal of Fluid Mechanics}\ }\textbf {\bibinfo
  {volume} {775}},\ \bibinfo {pages} {349--386} (\bibinfo {year}
  {2015})}\BibitemShut {NoStop}%
\bibitem [{\citenamefont {Duran-{M}atute}\ \emph {et~al.}(2010)\citenamefont
  {Duran-{M}atute}, \citenamefont {Albagnac}, \citenamefont {Kamp},\ and\
  \citenamefont {Van~Heijst}}]{duran2010dynamics}%
  \BibitemOpen
  \bibfield  {author} {\bibinfo {author} {\bibfnamefont {M.}~\bibnamefont
  {Duran-{M}atute}}, \bibinfo {author} {\bibfnamefont {J.}~\bibnamefont
  {Albagnac}}, \bibinfo {author} {\bibfnamefont {L.~P.~J.}\ \bibnamefont
  {Kamp}}, \ and\ \bibinfo {author} {\bibfnamefont {G.~J.~F.}\ \bibnamefont
  {Van~Heijst}},\ }\bibfield  {title} {\enquote {\bibinfo {title} {Dynamics and
  structure of decaying shallow dipolar vortices},}\ }\href@noop {} {\ \textbf
  {\bibinfo {volume} {22}},\ \bibinfo {pages} {116606} (\bibinfo {year}
  {2010})}\BibitemShut {NoStop}%
\bibitem [{\citenamefont {Krasnov}\ \emph {et~al.}(2008)\citenamefont
  {Krasnov}, \citenamefont {Rossi}, \citenamefont {Zikanov},\ and\
  \citenamefont {Boeck}}]{krasnov2008optimal}%
  \BibitemOpen
  \bibfield  {author} {\bibinfo {author} {\bibfnamefont {D.}~\bibnamefont
  {Krasnov}}, \bibinfo {author} {\bibfnamefont {M.}~\bibnamefont {Rossi}},
  \bibinfo {author} {\bibfnamefont {O.}~\bibnamefont {Zikanov}}, \ and\
  \bibinfo {author} {\bibfnamefont {T.}~\bibnamefont {Boeck}},\ }\bibfield
  {title} {\enquote {\bibinfo {title} {Optimal growth and transition to
  turbulence in channel flow with spanwise magnetic field},}\ }\href@noop {}
  {\bibfield  {journal} {\bibinfo  {journal} {J.\,Fluid Mech.}\ }\textbf
  {\bibinfo {volume} {596}},\ \bibinfo {pages} {73} (\bibinfo {year}
  {2008})}\BibitemShut {NoStop}%
\bibitem [{\citenamefont {Zikanov}\ \emph {et~al.}(2013)\citenamefont
  {Zikanov}, \citenamefont {Listratov},\ and\ \citenamefont
  {Sviridov}}]{zikanov2013natural}%
  \BibitemOpen
  \bibfield  {author} {\bibinfo {author} {\bibfnamefont {O.}~\bibnamefont
  {Zikanov}}, \bibinfo {author} {\bibfnamefont {Y.~I.}\ \bibnamefont
  {Listratov}}, \ and\ \bibinfo {author} {\bibfnamefont {V.~G.}\ \bibnamefont
  {Sviridov}},\ }\bibfield  {title} {\enquote {\bibinfo {title} {Natural
  convection in horizontal pipe flow with a strong transverse magnetic
  field},}\ }\href@noop {} {\bibfield  {journal} {\bibinfo  {journal}
  {J.\,Fluid Mech.}\ }\textbf {\bibinfo {volume} {720}},\ \bibinfo {pages}
  {486--516} (\bibinfo {year} {2013})}\BibitemShut {NoStop}%
\bibitem [{\citenamefont {Genin}\ \emph {et~al.}(2011)\citenamefont {Genin},
  \citenamefont {Zhilin}, \citenamefont {Ivochkin}, \citenamefont {Razuvanov},
  \citenamefont {Belyaev}, \citenamefont {Listratov},\ and\ \citenamefont
  {Sviridov}}]{genin2011temperature}%
  \BibitemOpen
  \bibfield  {author} {\bibinfo {author} {\bibfnamefont {L.~G.}\ \bibnamefont
  {Genin}}, \bibinfo {author} {\bibfnamefont {V.~G.}\ \bibnamefont {Zhilin}},
  \bibinfo {author} {\bibfnamefont {Y.~P.}\ \bibnamefont {Ivochkin}}, \bibinfo
  {author} {\bibfnamefont {N.~G.}\ \bibnamefont {Razuvanov}}, \bibinfo {author}
  {\bibfnamefont {I.~A.}\ \bibnamefont {Belyaev}}, \bibinfo {author}
  {\bibfnamefont {Y.~I.}\ \bibnamefont {Listratov}}, \ and\ \bibinfo {author}
  {\bibfnamefont {V.~G.}\ \bibnamefont {Sviridov}},\ }\bibfield  {title}
  {\enquote {\bibinfo {title} {Temperature fluctuations in a heated horizontal
  tube affected by transverse magnetic field},}\ }in\ \href@noop {} {\emph
  {\bibinfo {booktitle} {Proc. Fundamental and Applied {MHD}, 8th International
  {PAMIR} Conference, Borgo, Corsica}}}\ (\bibinfo {year} {2011})\ pp.\
  \bibinfo {pages} {37--41}\BibitemShut {NoStop}%
\bibitem [{\citenamefont {Belyaev}\ \emph {et~al.}(2015)\citenamefont
  {Belyaev}, \citenamefont {Ivochkin}, \citenamefont {Listratov}, \citenamefont
  {Razuvanov},\ and\ \citenamefont {Sviridov}}]{belyaev2015temperature}%
  \BibitemOpen
  \bibfield  {author} {\bibinfo {author} {\bibfnamefont {I.~A.}\ \bibnamefont
  {Belyaev}}, \bibinfo {author} {\bibfnamefont {Y.~P.}\ \bibnamefont
  {Ivochkin}}, \bibinfo {author} {\bibfnamefont {Y.~I.}\ \bibnamefont
  {Listratov}}, \bibinfo {author} {\bibfnamefont {N.~G.}\ \bibnamefont
  {Razuvanov}}, \ and\ \bibinfo {author} {\bibfnamefont {V.~G.}\ \bibnamefont
  {Sviridov}},\ }\bibfield  {title} {\enquote {\bibinfo {title} {Temperature
  fluctuations in a liquid metal {MHD}-flow in a horizontal inhomogeneously
  heated tube},}\ }\href@noop {} {\bibfield  {journal} {\bibinfo  {journal}
  {High Temperature}\ }\textbf {\bibinfo {volume} {53}},\ \bibinfo {pages}
  {734--741} (\bibinfo {year} {2015})}\BibitemShut {NoStop}%
\bibitem [{\citenamefont {Vetcha}\ \emph {et~al.}(2013)\citenamefont {Vetcha},
  \citenamefont {Smolentsev}, \citenamefont {Abdou},\ and\ \citenamefont
  {Moreau}}]{vetcha2013study}%
  \BibitemOpen
  \bibfield  {author} {\bibinfo {author} {\bibfnamefont {N.}~\bibnamefont
  {Vetcha}}, \bibinfo {author} {\bibfnamefont {S.}~\bibnamefont {Smolentsev}},
  \bibinfo {author} {\bibfnamefont {M.}~\bibnamefont {Abdou}}, \ and\ \bibinfo
  {author} {\bibfnamefont {R.}~\bibnamefont {Moreau}},\ }\bibfield  {title}
  {\enquote {\bibinfo {title} {Study of instabilities and quasi-two-dimensional
  turbulence in volumetrically heated magnetohydrodynamic flows in a vertical
  rectangular duct},}\ }\href@noop {} {\ \textbf {\bibinfo {volume} {25}},\
  \bibinfo {pages} {024102} (\bibinfo {year} {2013})}\BibitemShut {NoStop}%
\bibitem [{\citenamefont {Zhang}\ and\ \citenamefont
  {Zikanov}(2014)}]{zhang2014mixed}%
  \BibitemOpen
  \bibfield  {author} {\bibinfo {author} {\bibfnamefont {X.}~\bibnamefont
  {Zhang}}\ and\ \bibinfo {author} {\bibfnamefont {O.}~\bibnamefont
  {Zikanov}},\ }\bibfield  {title} {\enquote {\bibinfo {title} {Mixed
  convection in a horizontal duct with bottom heating and strong transverse
  magnetic field},}\ }\href@noop {} {\bibfield  {journal} {\bibinfo  {journal}
  {J.\,Fluid Mech.}\ }\textbf {\bibinfo {volume} {757}},\ \bibinfo {pages}
  {33--56} (\bibinfo {year} {2014})}\BibitemShut {NoStop}%
\bibitem [{\citenamefont {Takashima}(1996)}]{takashima1996stability}%
  \BibitemOpen
  \bibfield  {author} {\bibinfo {author} {\bibfnamefont {M.}~\bibnamefont
  {Takashima}},\ }\bibfield  {title} {\enquote {\bibinfo {title} {The stability
  of the modified plane {P}oiseuille flow in the presence of a transverse
  magnetic field},}\ }\href@noop {} {\bibfield  {journal} {\bibinfo  {journal}
  {Fluid Dyn.\ Res.}\ }\textbf {\bibinfo {volume} {17}},\ \bibinfo {pages}
  {293} (\bibinfo {year} {1996})}\BibitemShut {NoStop}%
\bibitem [{\citenamefont {Lock}()}]{lock1955stability}%
  \BibitemOpen
  \bibfield  {author} {\bibinfo {author} {\bibfnamefont {R.~C.}\ \bibnamefont
  {Lock}},\ }\bibfield  {title} {\enquote {\bibinfo {title} {The stability of
  the flow of an electrically conducting fluid between parallel planes under a
  transverse magnetic field},}\ }in\ \href@noop {} {\emph {\bibinfo {booktitle}
  {P.\ Roy.\ Soc.\ Lond.\ A Mat.}}}\BibitemShut {Stop}%
\bibitem [{\citenamefont {Moresco}\ and\ \citenamefont
  {Alboussi\`{e}re}(2004)}]{moresco2004experimental}%
  \BibitemOpen
  \bibfield  {author} {\bibinfo {author} {\bibfnamefont {P.}~\bibnamefont
  {Moresco}}\ and\ \bibinfo {author} {\bibfnamefont {T.}~\bibnamefont
  {Alboussi\`{e}re}},\ }\bibfield  {title} {\enquote {\bibinfo {title}
  {Experimental study of the instability of the {H}artmann layer},}\
  }\href@noop {} {\bibfield  {journal} {\bibinfo  {journal} {J.\,Fluid Mech.}\
  }\textbf {\bibinfo {volume} {504}},\ \bibinfo {pages} {167--181} (\bibinfo
  {year} {2004})}\BibitemShut {NoStop}%
\bibitem [{\citenamefont {Poth{\'e}rat}\ \emph {et~al.}(2005)\citenamefont
  {Poth{\'e}rat}, \citenamefont {Sommeria},\ and\ \citenamefont
  {Moreau}}]{potherat2005numerical}%
  \BibitemOpen
  \bibfield  {author} {\bibinfo {author} {\bibfnamefont {A.}~\bibnamefont
  {Poth{\'e}rat}}, \bibinfo {author} {\bibfnamefont {J.}~\bibnamefont
  {Sommeria}}, \ and\ \bibinfo {author} {\bibfnamefont {R.}~\bibnamefont
  {Moreau}},\ }\bibfield  {title} {\enquote {\bibinfo {title} {Numerical
  simulations of an effective two-dimensional model for flows with a transverse
  magnetic field},}\ }\href@noop {} {\bibfield  {journal} {\bibinfo  {journal}
  {J.\,Fluid Mech.}\ }\textbf {\bibinfo {volume} {534}},\ \bibinfo {pages}
  {115--143} (\bibinfo {year} {2005})}\BibitemShut {NoStop}%
\bibitem [{\citenamefont {Morley}\ \emph {et~al.}(2008)\citenamefont {Morley},
  \citenamefont {Burris}, \citenamefont {Cadwallader},\ and\ \citenamefont
  {Nornberg}}]{morley2008gainsn}%
  \BibitemOpen
  \bibfield  {author} {\bibinfo {author} {\bibfnamefont {N.~B.}\ \bibnamefont
  {Morley}}, \bibinfo {author} {\bibfnamefont {J.}~\bibnamefont {Burris}},
  \bibinfo {author} {\bibfnamefont {L.~C.}\ \bibnamefont {Cadwallader}}, \ and\
  \bibinfo {author} {\bibfnamefont {M.~D.}\ \bibnamefont {Nornberg}},\
  }\bibfield  {title} {\enquote {\bibinfo {title} {{GaInSn} usage in the
  research laboratory},}\ }\href@noop {} {\bibfield  {journal} {\bibinfo
  {journal} {Rev.\ Sci.\ Instrum.}\ }\textbf {\bibinfo {volume} {79}},\
  \bibinfo {pages} {056107} (\bibinfo {year} {2008})}\BibitemShut {NoStop}%
\bibitem [{\citenamefont {Sommeria}(1988)}]{sommeria1988electrically}%
  \BibitemOpen
  \bibfield  {author} {\bibinfo {author} {\bibfnamefont {J.}~\bibnamefont
  {Sommeria}},\ }\bibfield  {title} {\enquote {\bibinfo {title} {Electrically
  driven vortices in a strong magnetic field},}\ }\href@noop {} {\bibfield
  {journal} {\bibinfo  {journal} {J.\,Fluid Mech.}\ }\textbf {\bibinfo {volume}
  {189}},\ \bibinfo {pages} {553--569} (\bibinfo {year} {1988})}\BibitemShut
  {NoStop}%
\bibitem [{\citenamefont {Krasnov}\ \emph {et~al.}(2004)\citenamefont
  {Krasnov}, \citenamefont {Zienicke}, \citenamefont {Zikanov}, \citenamefont
  {Boeck},\ and\ \citenamefont {Thess}}]{krasnov2004numerical}%
  \BibitemOpen
  \bibfield  {author} {\bibinfo {author} {\bibfnamefont {D.~S.}\ \bibnamefont
  {Krasnov}}, \bibinfo {author} {\bibfnamefont {E.}~\bibnamefont {Zienicke}},
  \bibinfo {author} {\bibfnamefont {O.}~\bibnamefont {Zikanov}}, \bibinfo
  {author} {\bibfnamefont {T.}~\bibnamefont {Boeck}}, \ and\ \bibinfo {author}
  {\bibfnamefont {A.}~\bibnamefont {Thess}},\ }\bibfield  {title} {\enquote
  {\bibinfo {title} {Numerical study of the instability of the {H}artmann
  layer},}\ }\href@noop {} {\bibfield  {journal} {\bibinfo  {journal}
  {J.\,Fluid Mech.}\ }\textbf {\bibinfo {volume} {504}},\ \bibinfo {pages}
  {183--211} (\bibinfo {year} {2004})}\BibitemShut {NoStop}%
\bibitem [{\citenamefont {Tsai}\ \emph {et~al.}(2016)\citenamefont {Tsai},
  \citenamefont {Hussam}, \citenamefont {Fouras},\ and\ \citenamefont
  {Sheard}}]{tsai2016origin}%
  \BibitemOpen
  \bibfield  {author} {\bibinfo {author} {\bibfnamefont {T.}~\bibnamefont
  {Tsai}}, \bibinfo {author} {\bibfnamefont {W.~K.}\ \bibnamefont {Hussam}},
  \bibinfo {author} {\bibfnamefont {A.}~\bibnamefont {Fouras}}, \ and\ \bibinfo
  {author} {\bibfnamefont {G.~J.}\ \bibnamefont {Sheard}},\ }\bibfield  {title}
  {\enquote {\bibinfo {title} {The origin of instability in enclosed
  horizontally driven convection},}\ }\href@noop {} {\bibfield  {journal}
  {\bibinfo  {journal} {Int.\ J.\ Heat Mass Trans.}\ }\textbf {\bibinfo
  {volume} {94}},\ \bibinfo {pages} {509--515} (\bibinfo {year}
  {2016})}\BibitemShut {NoStop}%
\bibitem [{\citenamefont {Trefethen}(2000)}]{trefethen2000spectral}%
  \BibitemOpen
  \bibfield  {author} {\bibinfo {author} {\bibfnamefont {L.~N.}\ \bibnamefont
  {Trefethen}},\ }\href@noop {} {\emph {\bibinfo {title} {Spectral methods in
  {MATLAB}}}},\ Vol.~\bibinfo {volume} {10}\ (\bibinfo  {publisher} {Siam},\
  \bibinfo {year} {2000})\BibitemShut {NoStop}%
\bibitem [{\citenamefont {Weideman}\ and\ \citenamefont
  {Reddy}(2000)}]{weideman2000matlab}%
  \BibitemOpen
  \bibfield  {author} {\bibinfo {author} {\bibfnamefont {J.~A.}\ \bibnamefont
  {Weideman}}\ and\ \bibinfo {author} {\bibfnamefont {S.~C.}\ \bibnamefont
  {Reddy}},\ }\bibfield  {title} {\enquote {\bibinfo {title} {A {MATLAB}
  differentiation matrix suite},}\ }\href@noop {} {\bibfield  {journal}
  {\bibinfo  {journal} {ACM Transactions on Mathematical Software (TOMS)}\
  }\textbf {\bibinfo {volume} {26}},\ \bibinfo {pages} {465--519} (\bibinfo
  {year} {2000})}\BibitemShut {NoStop}%
\bibitem [{\citenamefont {Schmid}\ and\ \citenamefont
  {Henningson}(2001)}]{schmid2001stability}%
  \BibitemOpen
  \bibfield  {author} {\bibinfo {author} {\bibfnamefont {P.~J.}\ \bibnamefont
  {Schmid}}\ and\ \bibinfo {author} {\bibfnamefont {D.~S.}\ \bibnamefont
  {Henningson}},\ }\href@noop {} {\emph {\bibinfo {title} {Stability and
  transition in shear flows}}},\ Vol.\ \bibinfo {volume} {142}\ (\bibinfo
  {publisher} {Springer Verlag},\ \bibinfo {year} {2001})\BibitemShut {NoStop}%
\bibitem [{\citenamefont {McBain}\ \emph {et~al.}(2009)\citenamefont {McBain},
  \citenamefont {Chubb},\ and\ \citenamefont {Armfield}}]{mcbain2009numerical}%
  \BibitemOpen
  \bibfield  {author} {\bibinfo {author} {\bibfnamefont {G.~D.}\ \bibnamefont
  {McBain}}, \bibinfo {author} {\bibfnamefont {T.~H.}\ \bibnamefont {Chubb}}, \
  and\ \bibinfo {author} {\bibfnamefont {S.~W.}\ \bibnamefont {Armfield}},\
  }\bibfield  {title} {\enquote {\bibinfo {title} {Numerical solution of the
  {O}rr--{S}ommerfeld equation using the viscous {G}reen function and
  split-{G}aussian quadrature},}\ }\href@noop {} {\bibfield  {journal}
  {\bibinfo  {journal} {J.\,Comput.\ Appl.\ Math.}\ }\textbf {\bibinfo {volume}
  {224}},\ \bibinfo {pages} {397--404} (\bibinfo {year} {2009})}\BibitemShut
  {NoStop}%
\bibitem [{\citenamefont {Mack}(1976)}]{mack1976numerical}%
  \BibitemOpen
  \bibfield  {author} {\bibinfo {author} {\bibfnamefont {L.~M.}\ \bibnamefont
  {Mack}},\ }\bibfield  {title} {\enquote {\bibinfo {title} {A numerical study
  of the temporal eigenvalue spectrum of the {B}lasius boundary layer},}\
  }\href@noop {} {\bibfield  {journal} {\bibinfo  {journal} {J.\,Fluid Mech.}\
  }\textbf {\bibinfo {volume} {73}},\ \bibinfo {pages} {497--520} (\bibinfo
  {year} {1976})}\BibitemShut {NoStop}%
\bibitem [{\citenamefont {Reid}\ and\ \citenamefont
  {Harris}(1958)}]{reid1958some}%
  \BibitemOpen
  \bibfield  {author} {\bibinfo {author} {\bibfnamefont {W.~H.}\ \bibnamefont
  {Reid}}\ and\ \bibinfo {author} {\bibfnamefont {D.~L.}\ \bibnamefont
  {Harris}},\ }\bibfield  {title} {\enquote {\bibinfo {title} {Some further
  results on the {B}{\'e}nard problem},}\ }\href@noop {} {\ \textbf {\bibinfo
  {volume} {1}},\ \bibinfo {pages} {102--110} (\bibinfo {year}
  {1958})}\BibitemShut {NoStop}%
\bibitem [{\citenamefont
  {Chandrasekhar}(1961)}]{chandrasekhar1961hydrodynamics}%
  \BibitemOpen
  \bibfield  {author} {\bibinfo {author} {\bibfnamefont {S.}~\bibnamefont
  {Chandrasekhar}},\ }\href@noop {} {\emph {\bibinfo {title} {Hydrodynamic and
  hydromagnetic stability}}}\ (\bibinfo  {publisher} {Oxford University},\
  \bibinfo {year} {1961})\BibitemShut {NoStop}%
\bibitem [{\citenamefont {Burr}\ and\ \citenamefont
  {M{\"u}ller}(2002)}]{burr2002rayleigh}%
  \BibitemOpen
  \bibfield  {author} {\bibinfo {author} {\bibfnamefont {U.}~\bibnamefont
  {Burr}}\ and\ \bibinfo {author} {\bibfnamefont {U.}~\bibnamefont
  {M{\"u}ller}},\ }\bibfield  {title} {\enquote {\bibinfo {title}
  {Rayleigh--{B}{\'e}nard convection in liquid metal layers under the influence
  of a horizontal magnetic field},}\ }\href@noop {} {\bibfield  {journal}
  {\bibinfo  {journal} {J.\,Fluid Mech.}\ }\textbf {\bibinfo {volume} {453}},\
  \bibinfo {pages} {345--369} (\bibinfo {year} {2002})}\BibitemShut {NoStop}%
\bibitem [{\citenamefont {Mistrangelo}\ and\ \citenamefont
  {B{\"u}hler}(2016)}]{mistrangelo2016magneto}%
  \BibitemOpen
  \bibfield  {author} {\bibinfo {author} {\bibfnamefont {Chiara}\ \bibnamefont
  {Mistrangelo}}\ and\ \bibinfo {author} {\bibfnamefont {Leo}\ \bibnamefont
  {B{\"u}hler}},\ }\bibfield  {title} {\enquote {\bibinfo {title}
  {Magneto-convective instabilities in horizontal cavities},}\ }\href@noop {}
  {\ \textbf {\bibinfo {volume} {28}},\ \bibinfo {pages} {024104} (\bibinfo
  {year} {2016})}\BibitemShut {NoStop}%
\bibitem [{\citenamefont {Andreev}\ \emph {et~al.}(2003)\citenamefont
  {Andreev}, \citenamefont {Thess},\ and\ \citenamefont
  {Haberstroh}}]{andreev2003visualization}%
  \BibitemOpen
  \bibfield  {author} {\bibinfo {author} {\bibfnamefont {O.}~\bibnamefont
  {Andreev}}, \bibinfo {author} {\bibfnamefont {A.}~\bibnamefont {Thess}}, \
  and\ \bibinfo {author} {\bibfnamefont {Ch.}\ \bibnamefont {Haberstroh}},\
  }\bibfield  {title} {\enquote {\bibinfo {title} {Visualization of
  magnetoconvection},}\ }\href@noop {} {\bibfield  {journal} {\bibinfo
  {journal} {Phys.\ Fluids}\ }\textbf {\bibinfo {volume} {15}},\ \bibinfo
  {pages} {3886--3889} (\bibinfo {year} {2003})}\BibitemShut {NoStop}%
\bibitem [{\citenamefont {Hattori}\ \emph {et~al.}(2015)\citenamefont
  {Hattori}, \citenamefont {Patterson},\ and\ \citenamefont
  {Lei}}]{hattori2015stability}%
  \BibitemOpen
  \bibfield  {author} {\bibinfo {author} {\bibfnamefont {T.}~\bibnamefont
  {Hattori}}, \bibinfo {author} {\bibfnamefont {J.~C.}\ \bibnamefont
  {Patterson}}, \ and\ \bibinfo {author} {\bibfnamefont {C.}~\bibnamefont
  {Lei}},\ }\bibfield  {title} {\enquote {\bibinfo {title} {On the stability of
  internally heated natural convection due to the absorption of radiation in a
  laterally confined fluid layer with a horizontal throughflow},}\ }\href@noop
  {} {\bibfield  {journal} {\bibinfo  {journal} {Int.\ J.\ Heat Mass Trans.}\
  }\textbf {\bibinfo {volume} {81}},\ \bibinfo {pages} {846--861} (\bibinfo
  {year} {2015})}\BibitemShut {NoStop}%
\bibitem [{\citenamefont {Aujogue}\ \emph {et~al.}(2015)\citenamefont
  {Aujogue}, \citenamefont {Poth{\'e}rat},\ and\ \citenamefont
  {Sreenivasan}}]{aujogue2015onset}%
  \BibitemOpen
  \bibfield  {author} {\bibinfo {author} {\bibfnamefont {K.}~\bibnamefont
  {Aujogue}}, \bibinfo {author} {\bibfnamefont {A.}~\bibnamefont
  {Poth{\'e}rat}}, \ and\ \bibinfo {author} {\bibfnamefont {B.}~\bibnamefont
  {Sreenivasan}},\ }\bibfield  {title} {\enquote {\bibinfo {title} {Onset of
  plane layer magnetoconvection at low {E}kman number},}\ }\href@noop {}
  {\bibfield  {journal} {\bibinfo  {journal} {Phys.\ Fluids}\ }\textbf
  {\bibinfo {volume} {27}},\ \bibinfo {pages} {106602} (\bibinfo {year}
  {2015})}\BibitemShut {NoStop}%
\bibitem [{\citenamefont {Nakagawa}()}]{nakagawa1957experiments}%
  \BibitemOpen
  \bibfield  {author} {\bibinfo {author} {\bibfnamefont {Y.}~\bibnamefont
  {Nakagawa}},\ }\bibfield  {title} {\enquote {\bibinfo {title} {Experiments on
  the instability of a layer of mercury heated from below and subject to the
  simultaneous action of a magnetic field and rotation},}\ }in\ \href@noop {}
  {\emph {\bibinfo {booktitle} {P.\ Roy.\ Soc.\ Lond.\ A Mat.}}}\BibitemShut
  {Stop}%
\bibitem [{\citenamefont {Aujogue}\ \emph {et~al.}(2016)\citenamefont
  {Aujogue}, \citenamefont {Poth{\'e}rat}, \citenamefont {Bates}, \citenamefont
  {Debray},\ and\ \citenamefont {Sreenivasan}}]{aujogue2016little}%
  \BibitemOpen
  \bibfield  {author} {\bibinfo {author} {\bibfnamefont {K.}~\bibnamefont
  {Aujogue}}, \bibinfo {author} {\bibfnamefont {A.}~\bibnamefont
  {Poth{\'e}rat}}, \bibinfo {author} {\bibfnamefont {I.}~\bibnamefont {Bates}},
  \bibinfo {author} {\bibfnamefont {F.}~\bibnamefont {Debray}}, \ and\ \bibinfo
  {author} {\bibfnamefont {B.}~\bibnamefont {Sreenivasan}},\ }\bibfield
  {title} {\enquote {\bibinfo {title} {Little {Earth} experiment: an instrument
  to model planetary cores},}\ }\href@noop {} {\bibfield  {journal} {\bibinfo
  {journal} {Rev.\ Sci.\ Instrum.}\ }\textbf {\bibinfo {volume} {87}},\
  \bibinfo {eid} {084502} (\bibinfo {year} {2016})}\BibitemShut {NoStop}%
\bibitem [{\citenamefont {Nakagawa}(1955)}]{nakagawa1955experiment}%
  \BibitemOpen
  \bibfield  {author} {\bibinfo {author} {\bibfnamefont {Y.}~\bibnamefont
  {Nakagawa}},\ }\bibfield  {title} {\enquote {\bibinfo {title} {An experiment
  on the inhibition of thermal convection by a magnetic field},}\ }\href@noop
  {} {\bibfield  {journal} {\bibinfo  {journal} {Nature}\ }\textbf {\bibinfo
  {volume} {175}},\ \bibinfo {pages} {417--419} (\bibinfo {year}
  {1955})}\BibitemShut {NoStop}%
\end{thebibliography}

%

\end{document}